\documentclass[aps,prd,twocolumn,groupedaddress,asmsymb,amsmath,amssymb,longbibliography]{revtex4-1}

\usepackage{graphicx}
\usepackage{calligra}
\usepackage{upgreek}
\usepackage{color}
\usepackage{caption}
\usepackage{stackrel}
\usepackage[usenames,dvipsnames]{xcolor}

\numberwithin{equation}{section}

\DeclareMathAlphabet{\mathpzc}{OT1}{pzc}{m}{it}

\newcommand{\be}{\begin{equation}}
\newcommand{\ee}{\end{equation}}
\newcommand{\bea}{\begin{eqnarray}}
\newcommand{\eea}{\end{eqnarray}}
\newcommand{\lb}{\label}

\newcommand{\bj}{{\mbox{\boldmath $\dot{\j}$}}}
\newcommand{\ofs}{{\overline{f}_s}}
\newcommand{\orhos}{{\overline{\rho}_s}}
\newcommand{\oeps}{{\overline{\epsilon}_s}}
\newcommand{\obv}{{\overline{\bf v}}}
\newcommand{\ov}{{\overline{v}}}

\newcommand{\bv}{{\bf v}}
\newcommand{\bu}{{\bf u}}
\newcommand{\tbus}{\widetilde{{\bf u}}_s}
\newcommand{\bw}{{\bf w}}
\newcommand{\bz}{{\bf z}}
\newcommand{\bk}{{\bf k}}
\newcommand{\bp}{{\bf p}}
\newcommand{\bx}{{\bf x}}
\newcommand{\obx}{{\overline{\bf x}}}
\newcommand{\obp}{{\overline{\bf p}}}
\newcommand{\obP}{{\overline{\bf P}}}
\newcommand{\orho}{{\overline{\rho}}}
\newcommand{\otau}{{\overline{\tau}}}
\newcommand{\br}{{\bf r}}
\newcommand{\bE}{{\bf E}}
\newcommand{\bB}{{\bf B}}
\newcommand{\bI}{{\bf I}}
\newcommand{\bJ}{{\bf J}}

\newcommand{\bP}{{\bf P}}
\newcommand{\bR}{{\bf R}}

\newcommand{\bX}{{\bf X}}

\newcommand{\boeps}{{\mbox{\boldmath $\varepsilon$}}}

\newcommand{\btau}{{\mbox{\boldmath $\tau$}}}

\newcommand{\bPi}{{\mbox{\boldmath $\Pi$}}}
\newcommand{\grad}{{\mbox{\boldmath $\nabla$}}}
\newcommand{\bdot}{{\mbox{\boldmath $\cdot$}}}
\newcommand{\bdots}{{\mbox{\boldmath $:$}}}
\newcommand{\bzed}{{\mbox{\boldmath $0$}}}
\newcommand{\btimes}{{\mbox{\boldmath $\times$}}}

\newcommand{\obj}{{\overline{\mbox{\boldmath $\j$}}}}
\newcommand{\Dlim}{{{\mathcal D}\mbox{-}\lim}}

\newcommand{\of}{\overline{f}}

\newcommand{\hbvs}{\widehat{{\bf v}}_s}
\newcommand{\hbw}{\widehat{{\bf w}}}
\newcommand{\hbE}{\widehat{{\bf E}}}
\newcommand{\hbB}{\widehat{{\bf B}}}
\newcommand{\tbE}{\widetilde{{\bf E}}}
\newcommand{\tbB}{\widetilde{{\bf B}}}
\newcommand{\ttau}{\widetilde{\tau}}
\newcommand{\htau}{\widehat{\tau}}
\newcommand{\obE}{\overline{{\bf E}}}
\newcommand{\obB}{\overline{{\bf B}}}
\newcommand{\calQ}{{\mathcal Q}}

\newcommand{\wtc}{\frac{1}{c}}
\newcommand{\uit}{\mathpzc{u}}
\newcommand{\tit}{\mathpzc{t}}
\newcommand{\btimescom}{{\stackbin{\rm \btimes}{,}}}
\newcommand{\odots}{{\stackbin{\circ}{\text{\scriptsize{$\circ$}}}}}
\newcommand{\bodots}{{\mbox{\boldmath $\odots$}}}
\newcommand{\bcirc}{{\mbox{\boldmath $\circ$}}}
\newcommand{\sit}{\mbox{{\Large {\calligra s}}}\,}

\newcommand{\magc}[1]{\textcolor{black}{#1}}
\newcommand{\red}[1]{\textcolor{black}{#1}}
\newcommand{\green}[1]{\textcolor{black}{#1}}
\newcommand{\black}[1]{\textcolor{black}{#1}}

\begin{document}

%\preprint{}

\title{Cascades and Dissipative Anomalies in Nearly Collisionless Plasma Turbulence}
%and Dissipative Anomalies}


\author{Gregory L. Eyink${\,\!}^{1,2}$}
%\thanks{}
\affiliation{${\,\!}^1$Department of Applied Mathematics \& Statistics, The Johns Hopkins University, Baltimore, MD, USA, 21218}
\affiliation{${\,\!}^2$Department of Physics \& Astronomy, The Johns Hopkins University, Baltimore, MD, USA, 21218}


\date{\today}

\begin{abstract}

We develop first-principles theory of kinetic plasma turbulence governed by the Vlasov-Maxwell-Landau equations 
in the limit of vanishing collision rates. Following an exact renormalization-group approach pioneered by Onsager, 
we demonstrate the existence of a ``collisionless range'' of scales (lengths and velocities) in 1-particle phase space 
where the ideal Vlasov-Maxwell equations are satisfied in a ``coarse-grained sense''. Entropy conservation may 
nevertheless be violated in that range by a ``dissipative anomaly'' due to nonlinear entropy cascade. 
We derive ``4/5th-law'' type expressions for the entropy flux, which allow us to characterize the singularities 
(structure-function scaling exponents) required for its non-vanishing. Conservation laws of mass, momentum and energy 
are not afflicted with anomalous transfers in the collisionless limit. In a subsequent limit of small gyroradii, however, 
anomalous contributions to inertial-range energy balance may appear due both to cascade of bulk energy and to turbulent redistribution 
of internal energy in phase space. In that same limit the ``generalized Ohm's law" derived from the particle momentum 
balances reduces to an ``ideal Ohm's law'', but only in a coarse-grained sense that does not imply magnetic flux-freezing 
and that permits magnetic reconnection at all inertial-range scales. We compare our results with prior theory based on the gyrokinetic 
(high gyro-frequency) limit, with numerical simulations, and with spacecraft measurements of the solar wind and 
terrestrial magnetosphere.

%x
%We investigate dissipative anomalies in a turbulent fluid governed by the compressible Navier-Stokes 

%x
%equation. We follow an exact approach pioneered by Onsager, which we explain as a non-perturbative application 

%x
%of the principle of renormalization-group invariance. In the limit of high Reynolds and P\'eclet numbers, 

%x
%the flow realizations are found to be described as distributional or ``coarse-grained'' solutions of 

%x
%the compressible Euler equations, with standard conservation laws broken by turbulent anomalies. 

%x
%The anomalous dissipation of kinetic energy is shown to be due not only to local cascade, but also 

%x
%to a distinct mechanism called pressure-work defect. Irreversible heating in stationary, planar shocks 

%x
%with an ideal-gas equation of state exemplifies the second mechanism. Entropy conservation anomalies 

%x
%are also found to occur by two mechanisms: an anomalous input of negative entropy (negentropy) 

%x
%by pressure-work and a cascade of negentropy to small scales. We derive  ``4/5th-law''-type expressions 

%x
%for the anomalies, which allow us to characterize the singularities (structure-function scaling exponents) 

%x
%required to sustain the cascades. We compare our approach with alternative theories and empirical 

%x
%evidence. It is argued that the ``Big Power-Law in the Sky'' observed in electron density scintillations in 

%x
%the interstellar medium is a manifestation of a forward negentropy cascade, or an inverse cascade of usual 

%x
%thermodynamic entropy. 
\end{abstract}

\pacs{?????}
%\keywords{}

\maketitle

%\section{}
%\subsection{}

\section{Introduction}\label{sec:I}

In turbulent plasmas at very high temperatures and low densities the collisions of 
constituent particles are so infrequent that fluid models assuming small mean-free path 
lengths are invalid and the plasma must be described by kinetic equations for the 
particle distribution functions \cite{lifshitz1981physical}. A prime example of societal importance is 
magnetic-confinement fusion \cite{ongena2016magnetic}, where turbulent transport 
essentially limits performance but where mean free path lengths at typical operating 
conditions are $\sim 10$ km, much larger than the size of the device. 
The solar wind is one of the best-studied examples of a turbulent plasma in Nature, with 
a wealth of {\it in situ} spacecraft measurements showing turbulent-like spectra down to lengths  
of order a kilometer, but the mean-free-path for electron-ion collisions in the near-Earth 
solar wind is $\sim 1$ AU \cite{marsch2006kinetic}. The terrestrial magnetosphere is likewise a 
nearly collisionless plasma with turbulence occurring either typically (magnetosheath) or sporadically (magnetopause)
\cite{zimbardo2010magnetic}. The Magnetospheric Multiscale mission 
\cite{burch2016magnetospheric,burch2016electron} 
is currently measuring proton and electron velocity distribution functions in this environment 
at high phase-space resolution and cadence. Exploration of this velocity-space has been described as the 
``next frontier" of kinetic heliophysics \cite{howes2017prospectus}. More generally, turbulent, nearly collisionless 
plasma environments are ubiquitous in astrophysics. The interstellar medium exhibits 
an approximately Kolmogorov spectrum of electron density over $\sim 13$ orders of magnitude,
the so-called ``Big Power Law in the Sky" 
\cite{armstrong1981density,armstrong1995electron,chepurnov2010extending} 
but almost a third of this range lies below the ion mean-free path length $\sim 10^7$ km.

%Motivated mainly by applications to turbulent plasmas in space physics and astrophysics, 
%but should also have implications for turbulence in terrestrial plasmas, e.g. fusion devices.
%Range of validity in solar wind: all scales less than Coulomb mean-free-path length.  

% Plasma Physics: Confinement, Transport and Collective Effects
% edited by Andreas Dinklage, Thomas Klinger, Gerrit Marx, Lutz Schweikhard
% mean-free path for electrons and deuterons at T=15 keV and n=2.5 10^{20} /m^3
% about 10 km 

These diverse physics challenges call for fundamental theory of kinetic plasma turbulence. 
Recently, a first-principles paradigm has emerged for fluid turbulence, based upon a
mathematical analysis pioneered by Onsager for incompressible fluids
\cite{onsager1945distribution,onsager1949statistical,eyink2006onsager}. In this ``ideal turbulence'' theory,
the dissipative anomaly---or non-vanishing dissipation of kinetic energy in the inviscid limit---
is explained as a consequence of nonlinear energy cascade for ``coarse-grained'' or ``distributional''
solutions of the incompressible Euler equations. Onsager's analysis can be understood
as an exact, non-perturbative application of the principle of renormalization-group (RG)
invariance \cite{gross1976applications,eyink2015turbulent,eyink2018review},  
and it predicts the fluid H\"older singularities necessary for 
turbulent energy cascade. This ``ideal turbulence" theory for incompressible fluids has recently 
been supported by rigorous mathematical developments following from the Nash-Kuiper 
theorem and Gromov's $h$-principle \cite{delellis2012h,delellis2013continuous}.  The 
physical domain of the Onsager theory has also been extended recently to compressible fluids,
both non-relativistic \cite{aluie2013scale,eyink2017cascades1} and relativistic \cite{eyink2017cascades2}, with 
cascade of thermodynamic entropy and anomalous entropy production as central concepts. 

The purpose of this paper is to develop a similar exact theory for kinetic turbulence of nearly 
collisionless plasmas. It was suggested already some time ago by Krommes \& Hu 
\cite{krommes1994role,krommes1999thermostatted} that collisional production of kinetic entropy should remain 
non-zero in plasmas with vanishing collision rates. Empirical evidence for such ``anomalous entropy 
production'' has since been obtained by numerical simulations of gyrokinetic turbulence, both forced 
and decaying (see section \ref{sec:VIIIb} for a review). In the gyrokinetic formulation, 
Schekochihin et al. \cite{schekochihin2009astrophysical, schekochihin2008gyrokinetic} made 
an explicit analogy with the inertial-range of incompressible turbulence and proposed a 
gyrokinetic ``entropy cascade" through a range of scales in phase-space where collisions 
can be neglected. We here derive this picture for the full Vlasov-Maxwell-Landau equations of 
a weakly-coupled, multi-species plasma, by extending Onsager's exact, non-perturbative RG 
analysis to phase-space. The ``collisionless range'' of scales shall be shown to be governed by 
``coarse-grained'' or ``distributional'' solutions of the Vlasov-Maxwell kinetic equations, 
with an entropy-production anomaly due to a nonlinear entropy cascade. We derive 
expressions for the entropy flux through phase-space scales that are analogous to the ``4/5th-law" 
of Kolmogorov \cite{kolmogorov1941dissipation,uriel1995turbulence} 
for energy flux in incompressible turbulence and we exploit them 
to deduce the singularities of particle distributions and electromagnetic fields that are required 
in order to sustain the cascade of entropy. 

Such a careful, systematic mathematical framework for kinetic plasma turbulence %, we shall argue at length below, 
is valuable not only for its predictive power and conceptual clarity,
but also is necessary to avoid inconsistencies and apparent contradictions that arise 
from naive, informal discussions. There is an analogy with the theory of collisional
transport in plasmas which, prior to the  systematic derivation by Braginskii \cite{braginskii1965transport}, 
led frequently to ``paradoxes which have been the source of various errors and ambiguities'' 
(\cite{braginskii1965transport}, p.213). An even closer analogy is the situation in elementary 
particle physics prior to the discovery of the axial anomaly in quantum gauge theories. 
Naively, both the vector and axial-vector currents are conserved in massless spinor electrodynamics, 
but the simultaneous assumption of both conservation laws leads to the Veltman-Sutherland ``paradox" 
and the ``forbidden'' soft pion decay $\pi^0\rightarrow\gamma\gamma$.  As is well-known, 
this paradox is resolved by the chiral anomaly, which modifies the naive conservation 
of axial current and which accounts for the experimentally observed neutral-pion 
decay in a pseudovector coupling calculation \cite{adler2005anomalies,ioffe2006axial}. 
The origin of the axial anomaly lies in the ultraviolet divergences that appear when quantum fields,
which exist only as distributions or generalized functions, are naively multiplied pointwise. 
Careful regularization of these divergences, e.g. by  gauge-invariant ``point-splitting" of the spinor fields, 
yields the anomaly in axial charge conservation. The turbulent dissipative anomaly in the naive 
conservation of kinetic energy arises in a very similar manner, as stressed by Polyakov 
\cite{polyakov1993theory,polyakov1992conformal} and already understood long ago by Onsager 
\cite{onsager1945distribution,onsager1949statistical}. 

The need for sophistication in treating kinetic plasma turbulence is quite clear from the fact that the 
``collisionless range" of scales is governed by the Vlasov-Maxwell equations, in a certain sense, 
but entropy is nevertheless not conserved, as it would be for smooth solutions of the 
Vlasov-Maxwell equations in the standard sense. Similar cautionary remarks apply not just to entropy
conservation in a turbulent plasma, but also to other quantities which are naively conserved.  For 
example, it is true that the ideal Ohm's law is valid (in a certain sense) in the inertial range
of the solar wind, at scales much larger than the ion gyroradius, but this law does not hold 
in a manner that implies conservation of magnetic flux at those scales, as is frequently asserted
\cite{schekochihin2009astrophysical,kunz2015inertial,loureiro2017role}
\footnote{For example, in \cite{schekochihin2009astrophysical}, section 8.1 
on inertial-range solar wind turbulence, the authors wrote: ``In a plasma such as the solar wind ...
for $k_\perp\uprho_i\ll 1$, these fluctuations are rigorously described by the RMHD equations. 
The magnetic flux is frozen into the ion motions...'' %, so displacing a parcel of plasma should produce 
%a matching (Alfv\'enic) perturbation of the magnetic field line and vice versa.''  
RMHD indeed governs the 
shear-Alfv\'en modes in the inertial-range of the solar wind, under reasonable assumptions, but not in 
a sense that implies magnetic flux-freezing to the ion flow at those scales. See section \ref{sec:VIIIc}.}. 
This fact has important implications for the problem of magnetic reconnection in a turbulent plasma 
\cite{eyink2015turbulent,lalescu2015inertial}. We shall treat this 
problem here in the framework of the Vlasov-Maxwell-Landau kinetic theory, by considering 
the momentum conservation of the various charged particle species and the ``generalized Ohm's law" 
derived from them. We shall also discuss the energy balances of the particle species and of 
the electromagnetic fields,  in order to investigate the possibility of energy cascades in kinetic plasma turbulence. 
Energy and momentum in totality (particles $+$ fields) are conserved in the Vlasov-Maxwell-Landau 
model, so that no dissipative anomaly of total energy or momentum is possible. There can, however, be anomalous transfers 
between different components of energy (electromagnetic, kinetic \magc{energy of bulk} velocities, kinetic energy 
of fluctuation velocities) and also in phase-space. We investigate this possibility in the limit 
of vanishing collision rates and also in subsidiary limits, such as vanishingly small gyroradii.  

A notable aspect of the analysis presented here is that it involves almost no discussion 
of the rich array of linear waves supported by a plasma (shear-Alfv\'en waves, 
slow/fast magnetosonic waves, ion acoustic waves, kinetic Alfv\'en waves, whistler waves, etc.). This contrasts 
with the vast majority  of works, where plasma turbulence is regarded by default as an array of 
interacting linear waves. The dominance of this wave point of view is due in part to its great empirical success,
with the imprint of linear waves, such as their dispersion relations and eigenmodes, often 
clearly observed even in strongly interacting turbulent plasmas. On the other hand, it is also true that    
the mathematics of linear plasma waves is very familiar and well-developed \cite{stix1992waves}, 
whereas exact nonlinear theory of kinetic plasma turbulence is less straightforward and far 
fewer works are devoted to it \magc{\cite{matthaeus2014nonlinear,coburn2015third}}. In their recent 
discussion of the wave-turbulence dichotomy, Coburn et al. \cite{coburn2015third} have remarked that: 
\begin{quotation} 
``For most of the space age our view of solar wind fluctuations (magnetic, velocity, density, etc.) has been based on the theory of plasma waves. 
Attempts to incorporate turbulence concepts into this thinking have often been treated as little more than an afterthought that is either a secondary 
dynamic or a concept in direct conflict with the wave interpretation.'' --- \cite{coburn2015third}, p.1.
\end{quotation} 
It is the main purpose of the present paper to satisfy this need and to provide exact, first-principles theory
of the nonlinear cascades in kinetic turbulence of nearly collisonless plasmas. We shall remain mostly silent on 
the linear wave aspects, but this involves no rejection of their importance. A complete theory of kinetic plasma 
turbulence will certainly require a full synthesis of the linear wave and nonlinear cascade points of view. 

%CONTENTS: 
%II: Onsager theory, III: VML equations, IV: Phase-space coarse-graining, V: Coarse-grained VM equations
%(estimates in Appendix A), 
%VI: Entropy Cascade, VII: Mass, Momentum, Energy, VIII: Prior works: gyrokinetics, empirical studies,
%turbulent magnetic reconnection, IX: Conclusions

\section{Vlasov-Maxwell-Landau Equations}\label{sec:III} 

The theory of kinetic plasma turbulence in the present paper will be developed within the 
framework of the Vlasov-Maxwell-Landau equations for a weakly-coupled plasma,  
with a large Debye number or plasma parameter, $\Lambda=n \lambda_D^3\gg 1$ (where $\lambda_D$ is 
the Debye length).  In order to provide background and to set notations, we briefly describe 
this system and its basic properties, the dimensionless number groups which characterize 
its solutions, and important prior work on the collisionless limit. 

\subsection{Basic Equations}\lb{sec:IIIa} 

The Vlasov-Maxwell-Landau equations describe the evolution of the 
distribution functions $f_s(\bx,\bv,t)$ in 1-particle phase-space of $S$ species of particles with 
charges $q_s$ and masses $m_s,$ $s=1,...,S,$ and of the smoothed electromagnetic fields $\bE(\bx,t),$
$\bB(\bx,t),$ conditionally averaged over microscopic molecular states with given 
particle distributions $f_s,$ $s=1,...,S$; see e.g. 
\cite{lifshitz1981physical,ecker1972theory,klimontovich1982kinetic}   
These equations in the non-relativistic case
\footnote{This may be termed the ``semi-relativistic Vlasov-Maxwell-Landau system'', since 
the full relativistic Maxwell equations are retained for the fields, but the velocities of particles of 
all species $s$ are approximated by $\bv\doteq \bp/m_s$ under the assumption that 
$|\bp|\ll m_s c$ for all momenta in the support of the distribution functions $f_s(\bx,\bp,t)$
for $s=1,...,S$}
have the form of a Boltzmann-type kinetic equation for each species
\be \partial_t f_s + \bv\bdot\grad_\bx  f_s
+q_s \bE_* \bdot \grad_\bp f_s = C_s(f) \lb{III1} \ee
or 
\be \partial_t f_s + \grad_\bx \bdot (\bv f_s)
+\grad_\bp \bdot (q_s \bE_* f_s )= C_s(f), \lb{III2} \ee
for $s=1,...,S$ and the conditionally-averaged Maxwell equations
\bea 
&& \grad_\bx\bdot\bE = 4\pi \sum_s q_s n_s \cr
&&  \grad_\bx\btimes\bB -\wtc \partial_t\bE = \frac{4\pi}{c}\bj \cr
&&  \grad_\bx\btimes\bE + \wtc \partial_t\bB = \bzed, \quad \grad_\bx\bdot\bB = 0. \quad 
\lb{III3} \eea
with electric field in the rest frame of the particle population with velocity $\bv$ given by: 
\be \bE_*=\bE+\frac{1}{c}\bv\btimes \bB, \lb{III4} \ee  
with particle number density: 
\be n_s(\bx,t) = \int d^3 v \, f_s(\bx,\bv,t),  \lb{III5}\ee 
mass density $\rho_s=m_s n_s,$  and momentum density: 
\be \rho_s(\bx,t)\bu_s(\bx,t)=\int d^3v\ m_s \bv\,f_s(\bx,\bv,t) \lb{III6} \ee
for $s=1,...,S$ and with total electric current density:  
\be \bj(\bx,t) = \sum_s q_s n_s(\bx,t)\bu_s(\bx,t). \lb{III7} \ee
The equations \eqref{III1} and \eqref{III2} are equivalent because the vector-field $(\bv,q_s\bE_*)$ is Hamiltonian 
and has zero phase-space divergence $\grad_\bx\bdot\bv+\grad_\bp\bdot (q_s\bE_*)=0.$ Note that we avoid 
additional factors of $m_s$ in the equations by introducing the momentum variable $\bp=m_s\bv$
for each species. To complete the description, we need to specify 
the collision operator for species $s,$ given by  
\be C_s(f)=\sum_{s'} C_{ss'}(f_s,f_{s'}) \lb{III8} \ee
summed over collisions with species $s'.$ Here we shall consider the Landau collision operator
\cite{landau1936kinetische}: 
\bea 
&& C_{ss'}(f_s,f_{s'}) = 2\pi q_s^2 q_{s'}^2\ln \Lambda \cr
&& \hspace{20pt} \times 
\grad_\bp\bdot\left[ \int d^3v' \frac{\bPi_{\bv-\bv'}}{|\bv-\bv'|}
\left(\grad_\bp-\grad_{\bp'}\right)(f_sf_{s'})\right],\cr
&&  \lb{III9} \eea  
where $f_s=f_s(\bx,\bv,t),$ $f_{s'}=f_{s'}(\bx,\bv',t),$ where 
$\bPi_\bw=\bI-\bw\bw/|\bw|^2$ is the projection orthogonal to $\bw,$ 
and where the plasma parameter $\Lambda$ arises as a cut-off 
in the collision integral for impact factors greater than the Debye length (and, 
in principle, depends upon $s,s'$ pairs).  Although this is a standard kinetic 
model for a plasma \cite{lifshitz1981physical}, 
it has never been rigorously derived from a microscopic description 
\cite{villani2002review,golse2012mean} and global existence of (strong) solutions 
is an open problem \cite{arsenio2013solutions}.  Physically, alternative collision-integrals 
such as that of Balescu-Lenard \cite{lenard1960bogoliubov,balescu1960irreversible}
might give improved accuracy when large-velocity bumps or tails develop in the distribution 
functions \cite{radu1997statistical}. However, so long as these improved collision 
integrals satisfy an $H$-theorem and have similar differential form  
as the Landau operator (albeit with higher-order nonlinearity), then the analysis 
of the present paper will carry over. 

\subsection{Conservation Laws and $H$-Theorem}\lb{sec:IIIb} 

Essential properties of the kinetic equations are the local conservation laws 
for the various quantities preserved by collisions.  The mass for each particle species 
is conserved when collisions do not transform one species to another, so that 
$\int d^3v\,C_{ss'}=0$,  and the $m_s$-moment of \eqref{III1} in integration over velocity 
$\bv$ then gives 
\be \partial_t\rho_s+\grad_\bx\bdot(\rho_s\bu_s)=0. \lb{III10} \ee 
Momentum balance for species $s$ is obtained from the first moment of \eqref{III1} with $m_s\bv$, or: 
\be \partial_t (\rho_s\bu_s) +\grad_\bx\bdot(\rho_s\bu_s\bu_s+\bP_s) = q_s n_s \bE_{*s}+ \bR_s
\lb{III11} \ee
where the pressure tensor is 
\be \bP_s=\int d^3v\, m_s (\bv-\bu_s)(\bv-\bu_s) f_s, \lb{III12} \ee
the electric field in the bulk rest-frame of species $s$ is 
\be \bE_{*s} = \bE +\frac{1}{c}\bu_s\btimes \bB, \lb{III13} \ee 
and the drag force on species $s$ is 
\be \bR_s = \sum_{s'} \int d^3v\ m_s\bv \, C_{ss'}. \lb{III14} \ee  
When $\sum_s \bR_s=\bzed,$ the total momentum of the particles and fields is conserved. 
Finally, taking the moment of \eqref{III1} with $(1/2)m_s|\bv|^2$ gives kinetic energy balances for each species $s$:
\be \partial_t E_s +\grad_\bx \bdot(E_s\bu_s +\bP_s\bdot\bu_s+{\bf q}_s) =
\bj_s\bdot\bE +\bR_s\bdot\bu_s+Q_s \lb{III15} \ee
with $\bj_s=q_s n_s\bu_s$ the partial electric current of species $s,$ with kinetic energy density
\be E_s = \int d^3v\, \frac{1}{2}m_s|\bv|^2\, f_s, \lb{III16}  \ee
with heat flux, 
\be {\bf q}_s = \int d^3v\, \frac{1}{2}m_s|\bv-\bu_s|^2\,(\bv-\bu_s)\,f_s \lb{III17} \ee 
and with collisional heat exchange with other species
\be Q_s = \sum_{s'} \int d^3v\, \frac{1}{2}m_s|\bv-\bu_s|^2\,\, C_{ss'}. \lb{III18} \ee  
Total energy of the particles and fields is conserved when collisions are elastic and 
$\sum_s (\bR_s\bdot\bu_s+Q_s)=\sum_{ss'}\int d^3v\, (1/2)m_s|\bv|^2\, C_{ss'}=0.$ The 
Landau operator \eqref{III9}, as well known, has all of these properties. 

One can further subdivide the energy density $E_s$ of species $s$ into a bulk kinetic energy 
density $(1/2)\rho_s|\bu_s|^2$ and an ``internal'' \footnote{The terminology ``internal energy'' as used 
in equilibrium thermodynamics is really appropriate only when the distribution function $f_s$ is a 
local Maxwellian, whereas generally the fluctuational energy may reside in high-velocity tails or  
correspond to a very complex distribution in phase-space} or fluctuation energy density
\be \epsilon_s:= \frac{1}{2}{\rm Tr}\,\bP_s=\int d^3v\,  \frac{1}{2}m_s|\bv-\bu_s|^2\, f_s. \lb{III19} \ee
Note that $\epsilon_s=(3/2)p_s$ if the pressure tensor is decomposed into a scalar pressure 
$p_s$ and a traceless, anisotropic pressure tensor $\stackrel{\circ}{\bP}_s,$ as $\bP_s
=p_s{\bf I}+\stackrel{\circ}{\bP}_s.$ It is easy using \eqref{III10},\eqref{III11} to derive 
the balance equation for bulk kinetic energy of species $s$: 
\bea 
&& \partial_t (\frac{1}{2}\rho_s|\bu_s|^2) 
+\grad_\bx \bdot\left(\frac{1}{2}\rho_s|\bu_s|^2\bu_s +\bP_s\bdot\bu_s\right) \cr
&& \hspace{50pt} 
=\bP_s\bdots\grad_\bx\bu_s+\bj_s\bdot\bE +\bR_s\bdot\bu_s, \lb{III20} \eea
and then by subtracting \eqref{III20} from \eqref{III15} to obtain the balance equation for 
internal/fluctuational energy: 
\be \partial_t \epsilon_s +\grad_\bx \bdot(\epsilon_s\bu_s +{\bf q}_s) = -\bP_s\bdots\grad_\bx\bu_s+
Q_s \lb{III21} \ee
It is notable that fields directly exchange energy only with the bulk flows, via the Ohmic 
term $\bj_s\bdot\bE$, and subsequently energy is transferred between bulk  flows and fluctuations by the 
pressure-strain term $\bP_s\bdots\grad_\bx\bu_s$ \cite{yang2017Aenergy}. 

Very fundamental to our theory of kinetic plasma turbulence is the phase-space
{\it entropy density} of species $s$: 
\be \sit(f_s) = -f_s \ln f_s. \lb{III22} \ee
As well-known, this quantity simply counts the number of microstates of particle species $s$ compatible 
with the given macroscopic distribution $f_s$ \cite{boltzmann1877beziehung}. Using \eqref{III1}, the density 
$\sit(f_s)$ is easily shown to satisfy the phase-space balance equation:  
\bea 
&&\partial_t \sit(f_s) +\grad_\bx\bdot(\sit(f_s)\bv) + \grad_\bp\bdot(q_s\bE_* \sit(f_s)) \cr 
&& \hspace{50pt} = -(\ln f_s+1)C_s(f). \lb{III23} \eea
When integrated over $\bv$ and summed over $s,$ this gives the balance of total particle entropy  
density in space
\be s_{tot}(f) =\sum_s \int d^3v\ \sit(f_s) \lb{III24} \ee 
of the form 
\be \partial_t s_{tot}+\grad_\bx \bdot {\bf J}_S =\sigma, \lb{III25} \ee
with spatial entropy current density 
\be \bJ_S = \sum_s \int d^3v\, \bv\, \sit(f_s) \lb{III26} \ee
and local entropy production rate
\bea 
&& \sigma(\bx,t):= -\sum_s \int d^3v\, \ln f_s \, C_{s} \cr
&&=\sum_{ss'}\frac{\Gamma_{ss'}}{2} 
\int d^3v\int d^3v'\frac{|\bPi_{\bv-\bv'}\left(\grad_\bp-\grad_{\bp'}\right)(f_sf_{s'})|^2}{f_sf_{s'}|\bv-\bv'|} \cr
&&\hspace{100pt} \geq 0. 
\lb{III27} \eea 
Here we have introduced the shorthand notation $\Gamma_{ss'}=q_s^2q_{s'}^2\ln \Lambda.$
The non-negativity of the entropy production in \eqref{III27} is the statement of the $H$-theorem 
for the Landau collision operator. 

\subsection{Dimensionless Quantities}\lb{sec:IIIc} 

%Non-collisional: Liouville theorem/circulation
We now consider the Vlasov-Maxwell-Landau equations in a dimensionless form. For each species 
$s=1,...,S,$ we take as characteristic length the largest scale of variation $L_s$ of the distribution 
function of species $s.$ The characteristic velocity for species $s$ will be taken to be its thermal 
velocity $v_{th,s}$ and the characteristic time to be $\tau_s=L_s/v_{th,s}.$ The characteristic 
magnitude of $f_s$ will be taken as $\langle n_s\rangle/v_{th,s}^3,$ where $\langle n_s\rangle$
is the mean density of species $s.$ We thus introduce dimensionless variables:
\be \hat{x}=x/L_s, \,\, \hat{t}=t/\tau_s, \,\, \hat{v}=v/v_{th,s}, 
\,\, \hat{f}_s=v_{th,s}^3f_s/\langle n_s\rangle \lb{III28} \ee
for each separate species $s=1,...,S.$ In order to non-dimensionalize electromagnetic variables 
we introduce an effective density $n_0$ and length-scale $L_0$ so that typical field 
magnitudes are $E_0\sim B_0\sim e n_0 L_0$
%\footnote{We need not assume that $E_*\sim B_*,$and in that case the particle-field term 
%in the rescaled kinetic equation (???) will have electric 
%field term proportional to $1/\beta_s^E$ for ``electric beta parameter'' 
%$\beta_s^E=m_s v_{th,s}^2 < n_s >/(E_*^2/4\pi)$} 
and we then take 
\be \hat{x}=x/L_0, \,\, \hat{t}=ct/L_0, \,\, \hat{\bE}=\bE/E_0, 
\,\, \hat{\bB} = \bB/B_0 \lb{III29} \ee
Using $q_s=Z_se,$ the inhomogeneous Maxwell equations in these rescaled variables become
\bea 
&& \grad_{\hat{\bx}}\bdot\hat{\bE} = 4\pi \sum_s \frac{\langle n_s\rangle}{n_0} Z_s\hat{n}_s\cr
&&  \grad_{\hat{\bx}}\btimes\hat{\bB} -\partial_{\hat{t}}\hat{\bE} 
= 4\pi \sum_s \frac{\langle n_s\rangle}{n_0} \frac{v_{th,s}}{c} Z_s\hat{n}_s\hat{\bu}_s
\label{III30} \eea
while the homogeneous Maxwell equations are unchanged in form. Note that the length-scale
$L_0$ drops out of the rescaled equations \eqref{III30} and one of the factors $n_0,$ $L_0$ 
can be chosen as desired, e.g. to be a typical magnitude of $\langle n_s\rangle$ or of $L_s,$ 
if these are of similar orders of magnitude for all $s=1,...,S.$ With the rescaled variables in (\ref{III28})
and the rescaled field-strengths in (\ref{III29}), one then obtains the dimensionless kinetic equation 
for species $s$ as 
\bea 
&& \partial_{\hat{t}} \hat{f}_s + \hat{\bv}\bdot\grad_{\hat{\bx}}  \hat{f}_s
+ (Z_s/\beta_{0s}) \hat{\bE}_* \bdot \grad_{\hat \bp} \hat{f}_s \cr 
&& \hspace{40pt} = \sum_{s'} \hat{\Gamma}_{ss'} \, \hat{C}_s(\hat{f}_s,\hat{f}_{s'}) \lb{III31} \eea
where 
\be \hat{\bE}_* = \hat{\bE} + \frac{v_{th,s}}{c} \hat{\bv} \btimes \hat{\bB}, \ee 
\be \beta_{0s} = \frac{m_s v_{th,s}^2}{e B_0 L_s}= \frac{m_s v_{th,s}^2\langle n_s\rangle}{B_0^2}
\left(\frac{n_0}{\langle n_s\rangle}\right)\left(\frac{L_0}{L_s}\right),  
\lb{III32} \ee 
\be \hat{\Gamma}_{ss'}=\frac{2\pi q_s^2 q_{s'}^2 \langle n_{s'}\rangle \ln\Lambda}{m_s^2v_{th,s'}^3}\tau_s 
\lb{III33} \ee
Note that the standard beta parameter for species $s$ is $\beta_s=m_s v_{th,s}^2 \langle n_s\rangle/(B^2_0/4\pi)$ 
and is nearly the same as the quantity $\beta_{0s}$ defined in (\ref{III32}). The meaning of the constants $\hat{\Gamma}_{ss'}$ 
is elucidated by recalling that the Spitzer-Harm collision rate \cite{spitzer1956physics,helander2005collisional}
for particle pair $s,$ $s'$ is:
\be \nu_{ss'}=\frac{2\pi q_s^2q_{s'}^2 n_{s'}\ln \Lambda}{\mu_{s,s'}^2 (v^{rel}_{s,s'})^3} \lb{III34} \ee
up to a prefactor of order unity, where $\mu_{s,s'}$ is the reduced mass for pairs $s,$ $s'$ given by 
$1/\mu_{s,s'}=(1/m_s)+(1/m_{s'})$ and where $v^{rel}_{s,s'}$ is the typical relative velocity of particles 
of species $s,$ $s',$ or $\max\{v_{th,s},v_{th,s'}\}$ on order of magnitude. Thus, the quantity $\hat{\Gamma}_{ss'}$ 
defined in (\ref{III33}) is essentially equal to $\nu_{ss'}\tau_s,$ or the ratio of the characteristic time $\tau_s$ of species $s$ and 
and the mean-free-time for its collisions with species $s'$. We follow 
\cite{schekochihin2009astrophysical,schekochihin2008gyrokinetic} in 
referring to $Do_{ss'}=1/\hat{\Gamma}_{ss'}$ as the {\it Dorland number} for the pair $s,s'$ 
\footnote{\magc{More properly, this might be termed the ``global Dorland number'', since it refers to the 
large length-scale $L_s,$ whereas the original references \cite{schekochihin2009astrophysical,schekochihin2008gyrokinetic}
introduced an ``ion-scale Dorland number'' which refers instead to the ion gyroradius $\uprho_i$}}
In terms of the mean-free-path $\ell_{s,s'}=v_{th,s}/\nu_{s,s'}$ for collisions of species $s$ with $s'$
we can also write the Dorland number as $\ell_{s,s'}/L_s.$ We thus 
see that $Do_{ss'}$ is a measure of the collisionality of the plasma, with the plasma being nearly 
collisionless when $1/\hat{\Gamma}=Do:=\min_{s,s'} Do_{ss'}\gg 1.$ 
Hereafter we consider this weakly collisional regime with all other dimensionless parameters 
($\beta_{0s}$, $\langle n_s\rangle/n_0,$ etc.) assumed to have magnitudes of order unity.  
We remove hats $\hat{(\cdot)}$ on all variables for simplicity of notations. 

\subsection{Collisionless Limit and Dissipative Anomaly}\lb{sec:IIId} 

The collisional terms in the kinetic equation \eqref{III1} formally disappear in the limit 
$\Gamma:=\max_{ss'}\Gamma_{ss'}\to 0$ and its solutions may be expected to converge, in a certain sense,
to solutions of the collisionless Vlasov-Maxwell equations. Naively, the entropy production \eqref{III27} 
also vanishes in this limit because the prefactors $\Gamma_{ss'}\to 0.$ However, this need 
not be the case if the velocity-gradients of the distribution functions that appear in the collision 
integral diverge in the same limit. The simplest mechanism for producing large velocity-gradients 
is the free-streaming or ballistic advection of spatial structure, \magc{which underlies 
linear Landau damping \cite{landau1946vibrations}} 
and which has long been known to produce 
``velocity-space filamentation" in collisionless Vlasov simulations (e.g. see the review \cite{klimas1987method}
with many earlier references). In the papers of 
Krommes \& Hu \cite{krommes1994role} and Krommes \cite{krommes1999thermostatted} 
it was pointed out that entropy production rates in a long-time statistical steady-state of a plasma 
obtained by taking the limit $t\to\infty$ first are determined entirely by the forcing and thus must remain 
constant in the subsequent limit $\Gamma\to 0.$ 
The papers \cite{krommes1994role,krommes1999thermostatted} argued 
that the required fine structure in velocity space could be produced by ballistic streaming
and drew an explicit analogy with non-vanishing viscous dissipation of kinetic energy for fluid turbulence 
in the high-Reynolds number limit, or what is \black{called} the ``dissipative anomaly'' \cite{eyink2018review}. 
In following work of Schekochihin et al. \cite{schekochihin2008gyrokinetic,schekochihin2009astrophysical} 
within the gyrokinetic approach to plasma 
turbulence, it was pointed out that the analogue of a high Reynolds-number ``inertial-range'' can exist 
at sub-ion scales in position and velocity space for high Dorland-number plasma turbulence,  
with ion entropy cascading through that range by a nonlinear perpendicular phase-mixing mechanism
\footnote{To prevent possible confusion, we note that the sub-ion scales $\ell\ll\rho_i$ in the papers 
\cite{schekochihin2008gyrokinetic,schekochihin2009astrophysical} 
are described in the traditional terminology of space physics as the 
``dissipation range'', whereas the scales $\ell\gg\rho_i$ are referred to as the ``inertial range''. However, 
refs.\cite{schekochihin2008gyrokinetic,schekochihin2009astrophysical} 
made an analogy, just as we do, of the ``dissipation range'' of kinetic 
plasma turbulence with the ``inertial-range'' of hydrodynamic turbulence in neutral fluids.}. 
Employing phenomenological arguments,  the authors of 
\cite{schekochihin2008gyrokinetic,schekochihin2009astrophysical} 
argued that small-scales in velocity-space are produced more efficiently by nonlinear entropy cascade than 
by the simpler ballistic phase-mixing mechanism. 
%In a subsequent numerical study of gyrokinetic turbulence 
%resolving scales between the ion and electron gyroradii, Howes et al. (2011) obtained magnetic 
%energy spectra in qualitative agreement with the predictions of their phenomenological theory and, 
%furthermore, presented some empirical evidence in support of an ion entropy cascade. 

In the present paper we shall further develop the connection between high Reynolds-number turbulence
and nearly collisionless (high Dorland-number) plasma kinetics, but without making the more 
restrictive assumptions necessary for validity of a gyrokinetic description (i.e. without assuming 
evolution time-scales for any species $s$ long compared with its gyrofrequency). We shall show 
that existence of a turbulent cascade of entropy emerges as a natural consequence of the conjecture 
of \cite{krommes1994role,krommes1999thermostatted,schekochihin2008gyrokinetic,schekochihin2009astrophysical}
that collisional entropy production persists in the collisionless limit. 
We formalize the latter conjecture as the precise hypothesis 
that the entropy production \eqref{III27} converges in the collisionless limit
\be \lim_{\Gamma\to 0}\sigma(\bx,t)=\sigma_\star(\bx,t) \lb{III36} \ee 
as a measure in $\bx$-space for each $t$. This formulation is motivated by the analogy with energy dissipation 
in incompressible fluid turbulence \cite{eyink2018review} and also by the case of 
compressible fluids where, for shock solutions, the entropy production converges in exactly this fashion 
in the infinite Reynold-number limit \cite{eyink2017cascades1}.  There is, however, a strengthened 
version of the hypothesis which is also natural and which involves the collisional 
entropy-production density in the 2-particle phase-space, or 
\be \varsigma(\bx,\bv,\bv',t):=\sum_{ss'}\frac{\Gamma_{ss'}}{2} 
\frac{|\bPi_{\bv-\bv'}\left(\grad_\bp-\grad_{\bp'}\right)(f_sf_{s'})|^2}{f_sf_{s'}|\bv-\bv'|}, \lb{III37} \ee
so that $\sigma(\bx,t)=\int d^3v\int d^3v' \varsigma(\bx,\bv,\bv',t).$
This density involves only a single position variable $\bx,$ since a pair of particles must pass 
through the same space point (to within a Debye radius) in order to experience an unscreened Coulomb collision. 
As obvious from the definition \eqref{III37}, this phase-space density involves only velocity-gradients 
of the particle distributions and not space-gradients. It may therefore be expected to remain 
a continuous function of $\bx$ in the limit $\Gamma\to 0,$ if the particle distributions likewise 
remain continuous in $\bx$ and $\bv$ (e.g. as gyrokinetic theory suggests; see section \ref{sec:VIIIa}). 
In that case, it is reasonable to make the stronger hypothesis that the 2-particle phase-space density 
of entropy production converges 
\be \lim_{\Gamma\to 0}\varsigma(\bx,\bv,\bv',t)=\varsigma_\star(\bx,\bv,\bv',t) \lb{III38} \ee 
as a finite measure in $(\bv,\bv')$-space for every $(\bx,t).$ Of course, this assumption implies that 
in \eqref{III36}, but now even pointwise in $\bx$ rather than simply as a measure.  The validity 
of both these hypotheses can be explored in numerical simulations of the Vlasov-Maxwell-Landau 
system, similarly as in \cite{pezzi2016collisional,pezzi2017solar}.  In the present paper we explore their 
theoretical consequences.  As we shall see,  the Onsager ``ideal turbulence theory'' 
\cite{eyink2018review} carries over under these assumptions to plasma kinetics and predicts properties 
of the collisionless limit of Vlasov-Maxwell-Landau (VML) solutions with anomalous entropy production. 
This analysis leads to the concept of  ``weak'' or ``coarse-grained'' solutions of the Vlasov-Maxwell (VM) 
equations with irreversible entropy production by nonlinear entropy cascade in phase-space. 

\section{Phase-Space Coarse-Graining}\label{sec:IV} 

The most obvious requirement for non-vanishing of the entropy production  as in \eqref{III36} or 
\eqref{III38} is divergence of velocity-gradients of the particle distribution functions in the limit $\Gamma\to 0,$
or an ``ultraviolet divergence'' at small-scales in velocity space. One should furthermore 
expect that space-gradients of the particle distribution functions will diverge as well 
in the collisionless limit.  Note that the characteristic curves of the VM equation are 
the Hamiltonian particle motions in an electromagnetic field and, for non-trivial fields, these will generally 
lead to large space-gradients as well as to large velocity-gradients. Such divergences make it 
impossible to interpret the VML equations naively in this limit and pursuit of a dynamical description 
which can remain valid requires a suitable regularization. We shall here follow closely the discussion 
for hydrodynamic turbulence in \cite{eyink2018review} and make use of a similar ``coarse-graining'' or 
``block-spin'' regularization in the 1-particle phase space.  

\subsection{Definition of Coarse-Graining}\label{sec:IVa}

For any time-dependent function $a(\bx,\bv,t)$ on the 1-particle phase-space, we define its 
coarse-graining \footnote{In principle, one may also coarse-grain in time with a kernel 
$K_\tit(\tau)=(1/\tit)K(\tau/\tit),$ keeping only frequencies $\omega<1/\tit$ for a time-resolution scale $\tit.$ We 
do not coarse-grain temporally in this work, since it is not necessary mathematically to derive any of our results.
Coarse-graining in phase-space alone is already enough to regularize the time-derivatives of $f_s$, 
without the need for any additional coarse-graining. This follows because of the constraint 
of $f_s$ being a solution of the kinetic equation \eqref{III1}, which is first-order in time. This 
fact implies that $\partial_t \ofs(\obx,\obv,t)$ exists as a smooth function defined by the other terms 
in the eq.\eqref{V1}, which are all regularized. Similar remarks hold for $\bE,$ $\bB$ fields 
which are solutions of the Maxwell equations \eqref{III3}. Empirical measurements have always 
a finite time-resolution, of course, and, to model these, it would be appropriate to coarse-grain 
in time. This may be easily done and modifies little our analysis. Expressions for cumulants 
such as \eqref{IV7} in that case involve also time-increments and scaling exponents defined similarly 
to $\sigma_p^a$ in \eqref{VI33} will characterize space-time regularity. The two sets 
of exponents, for space regularity and space-time regularity, can be easily related 
\cite{isett2013regularity,drivas2018onsager}}
at position resolution $\ell$ and velocity resolution $\uit$ by  
\be \overline{a}(\obx,\obv,t)= \int d^3r \,G_\ell(\br) \int d^3w\, H_\uit(\bw) \, 
a(\obx+\br,\obv+\bw,t) \lb{IV1} \ee
where $H_\uit(\bw)=\uit^{-3}H(\bw/\uit)$ for a kernel $H$ satisfying the properties: 
\bea 
&& \qquad H(\bw)\geq 0 \quad \mbox{ (non-negative) }\cr
&& \int d^3w\ H(\bw)=1 \quad \mbox{ (normalized) } \cr
&& \int d^3w\ \bw\, H(\bw)=\bzed  \quad \mbox{ (centered) } \cr
&& \int d^3w\ |\bw|^2\, H(\bw)=1 \quad \mbox{ (unit variance) }.  
\lb{IV2} \eea 
We also assume that $H$ is smooth and rapidly decaying, e.g. $H\in C^\infty_c({\mathbb R}^3),$
and for convenience assume isotropy, or $H=H(w)$ with $w=|\bw|,$ so that
$\int d^3w\ w_i w_j \, H(\bw)=(1/3)\delta_{ij}.$ 
In the same manner, $G_\ell(\br)=\ell^{-3} G(\br/\ell)$ for a kernel $G$ satisfying the analogous 
properties. It is sometimes useful to rewrite the definition 
\eqref{IV1} as 
\be \overline{a}(\obx,\obv,t)= \langle a(\obx+\br,\obv+\bw,t) \rangle_{\ell,\uit} \lb{IV3} \ee 
where the local average $\langle\cdot\rangle_{\ell,\uit}$ is over displacements $\br,$ $\bw$\\
with respect to the distribution $G_\ell(\br)H_\uit(\bw).$ 
In our discussion below we shall also sometimes employ coarse-graining only with respect to position 
or only with respect to velocity, which we denote by 
\bea 
\overline{a}_\ell(\obx,\bv,t) &=& \int d^3r \,G_\ell(\br) a(\obx+\br,\bv,t) \cr 
                                          &=& \langle a(\obx+\br,\bv,t)\rangle_\ell \cr 
\overline{a}_\uit (\bx,\obv,t)&=& \int d^3w\, H_\uit(\bw) \, a(\bx,\obv+\bw,t) \cr
                                          &=& \langle a(\bx,\obv+\bw,t)\rangle_\uit 
\lb{IV4} \eea
There is consistency between these various notions of coarse-graining if a phase-space function
lacks dependence on one variable. For example, if $b=b(\bx,t)$ is independent of $\bv$, 
then $\overline{b}=\overline{b}_\ell$ and we need not distinguish these two quantities. Likewise, if 
$c=c(\bv,t)$ is independent of $\bx,$ then $\overline{c}=\overline{c}_\uit.$

One more 
concept that we shall employ extensively in our analysis below is that of {\it coarse-graining cumulants}
$\otau(f_1,...,f_p).$ These are defined as usual \cite{huang1987statistical,germano1992turbulence} 
through the iterative expansion of coarse-grained products into finite sums of cumulants: 
\be \overline{a_1\cdots a_n} =  \sum_I \prod_{r=1}^{r_I}\otau(a_{i_1^{(r)}},...,a_{i^{(r)}_{{p_r}}}) 
\lb{IV5} \ee
where the sum is over all distinct partitions $I$ of $\{1, ..., n\}$ into $r_I$ disjoint subsets 
$\{i^{(r)}_1 , ..., i^{(r)}_{{p_r}} \}$  of $p_r$ members each, $r = 1, ..., r_I$ , so that
$\sum_{r=1}^{r_I} p_r = n$ for each partition $I$. By solving the iterated expansions for 
cumulants in terms of coarse-grained products one obtains, for example, 
\bea
&& \hspace{30pt} \otau(a_1)=\overline{a}_1, \quad \otau(a_1,a_2)=\overline{a_1a_2}-\overline{a}_1\overline{a}_2, \cr
&& \otau(a_1,a_2,a_3)=\overline{a_1a_2a_3}-\overline{a_1a_2}\ \overline{a}_3
-\overline{a_1a_3}\ \overline{a}_2-\overline{a_2a_3}\ \overline{a}_1 \cr
&& \hspace{120pt} + 2\, \overline{a}_1\ \overline{a}_2\ \overline{a}_3 
\lb{IV6} \eea
and so forth for cumulants of higher order. A relation that is crucial to our analysis is
\be \otau(a_1,a_2) = \langle \delta a_1\,\delta a_2\rangle-\langle \delta a_1\rangle\langle\delta a_2\rangle
\lb{IV7} \ee
where $\delta_{\br,\bw}a(\bx,\bv,t)=a(\bx+\br,\bv+\bw,t)-a(\bx,\bv,t)$ is the increment for a phase-space
displacement $(\br,\bw)$ \cite{eyink2015turbulent,eyink2018review}. 
A similar result holds for the 2nd-order cumulant $\otau_\ell(b_1,b_2)$ defined 
with respect to the average $\langle \cdot \rangle_\br$ over $\br$ and with the increment taken to be
$\delta_\br b.$ The same remark holds for $\otau_\uit(c_1,c_2),$ average $\langle \cdot\rangle_\uit$ over
$\bw,$ and increment $\delta_\bw c.$ In fact, expressions for higher-order cumulants 
in terms of increments hold as well, completely analogous to \eqref{IV7} for 2nd-order cumulants  
\cite{eyink2015turbulent,eyink2018review}.  

The phase-space coarse-graining operation \eqref{IV1} clearly regularizes 
all gradients, so that $\grad_\obx \overline{a}$ and $\grad_\obv\overline{a}$ are finite and smooth,
even if quantity $a$ exists only as a distribution on phase-space. Moreover, one can derive 
expressions for these gradients in terms of increments:
\bea 
&& \grad_\obx \overline{a}(\obx,\obv,t)  = \cr
&& -\frac{1}{\ell} \int d^3r \,(\grad G)_\ell(\br) \int d^3w\, H_\uit(\bw)
(\delta_\br a)(\obx,\obv+\bw,t)\cr
&& \lb{IV8} \eea 
and
\bea 
&& \grad_\obv \overline{a}(\obx,\obv,t)  = \cr
&& -\frac{1}{\uit} \int d^3r \,G_\ell(\br) \int d^3w\, (\grad H)_\uit(\bw) 
(\delta_\bw a)(\obx+\br,\obv,t),\cr
&& \lb{IV9} \eea 
by exploiting $\int d^3r\ (\grad G)_\ell(\br)=\int d^3w\ (\grad H)_\uit(\bw)=\bzed.$ These formulas
permit one to estimate the order of magnitude of the coarse-grained gradients. We emphasize 
that the length scale $\ell$ and velocity scale $\uit$ introduced by our coarse-graining 
regularization are completely arbitrary. No objective physical fact can depend upon their precise 
values.  The coarse-graining  \eqref{IV1} is a purely passive operation which corresponds to 
observing a given phase-space function $a(\bx,\bv,t)$ with some chosen resolutions $\ell$ in position 
and $\uit$ in velocity.  As we see below, the arbitrariness of these regularization scales can be 
exploited to deduce exact consequences, analogous to RG-invariance in quantum field-theory 
and statistical physics \cite{gross1976applications} and analogous to Onsager's ``ideal turbulence'' 
theory for incompressible fluid turbulence \cite{eyink2018review}.  

\subsection{Phase-Space Favre Average}\label{sec:IVb}   

In the theory of compressible fluid turbulence, a mass-density weighted average was introduced 
by Favre \cite{favre1969statistical} 
within a statistical ensemble approach to compressible fluid turbulence. Density-weighting 
may be employed also for coarse-graining averages, e.g. \cite{garnier2009large,aluie2013scale,eyink2017cascades1}. 
It should be emphasized that the use of density-weighting is not obligatory, but has the advantage that
it reduces the number of terms in coarse-grained equations and generally provides each 
term with an intuitive physical interpretation. Therefore, we employ weighted coarse-graining here 
as well, but with the novelty that coarse-graining averages are weighted by the phase-space 
particle distributions rather than by mass-densities. For a field $a=a(\bx,\bv,t)$ we thus define 
its {\it phase-space Favre average} at scales $\ell,$ $\uit$ weighted by the 
particle distribution of species $s$ as 
\be \widehat{a}_s:=\overline{a f_s}/\of_s. \lb{IV10} \ee 
We contrast this with the traditional {\it physical-space Favre average} 
at scale $\ell$ for a field $b=b(\bx,t)$ with no $\bv$-dependence, which is weighted 
by the mass-density of species $s$ so that 
\be \widetilde{b}_s:=\overline{b\, \rho_s}/\orho_s=\overline{b\, n_s}/\overline{n}_s. \lb{IV11} \ee 
Even for a purely spatial field $b=b(\bx,t)$ with no $\bv$-dependence, these two averages do not agree, 
\be \widehat{b}_s(\obx,\obv,t)\neq \widetilde{b}_s(\obx,t), \lb{IV12} \ee
because the correlations between positions and velocities in the distribution function 
$f_s(\bx,\bv,t)$ induce a nontrivial $\obv$-dependence in $\widehat{b}_s$. There is, however, an easily 
derived consistency relation 
\be \int d^3\ov\ \widehat{b}_s \of_s= \overline{b\, n}_s = \widetilde{b}_s\overline{n}_s
=\widetilde{b}_s \int d^3\ov\ \of_s, \lb{IV13} \ee 
which holds for any $b=b(\bx,t).$ 

Just as for unweighted coarse-graining, one may define {\it phase-space Favre cumulants} 
$\htau_s(a_1,...,a_n)$ through the iterative decompositions 
\be \widehat{(a_1\cdots a_n)}_s =  \sum_I \prod_{r=1}^{r_I}\htau_s(a_{i_1^{(r)}},...,a_{i^{(r)}_{{p_r}}}) 
\lb{IV14} \ee
for $n=1,2,3,...$ Likwise, one may define physical-space Favre cumulants $\ttau_s(b_1,...,b_n)$ 
with respect to the standard Favre average for $b_i=b_i(\bx,t),$ $i=1,2,3,...$ Since 
Favre-averaging is just a convenience, one may always express Favre cumulants in terms of 
unweighted cumulants, e.g. for $\htau_s(a)=\widehat{a}_s$
\be
\widehat{a}_s = \overline{a} + \frac{1}{\ofs}\otau(a,f_s), 
\lb{IV15} \ee
also 
\bea
&& \htau_s(a_1,a_2)= \otau(a_1,a_2)+ \frac{1}{\ofs}\otau(a_1,a_2,f_s) \cr
&& \hspace{60pt} -\frac{1}{\ofs^2}\otau(a_1,f_s)\otau(a_2,f_s)
\lb{IV16} \eea
and so forth. Because the unweighted cumulants $\otau(a_1,...,a_n)$ can be expressed in terms of 
increments $\delta a_i$ $i=1,...,n$ via relations such as \eqref{IV7}, it follows that the 
Favre cumulants $\htau(a_1,...,a_n)$ can be expressed in terms of increments $\delta f_s$ and $\delta a_i,$ 
$i=1,...,n$.  

\subsection{Coarse-Grained Distribution}\label{sec:IVc}   

Basic dynamical objects for the coarse-graining regularization are the {\it coarse-grained distributions} 
$\ofs(\obx,\obv,t)$ for each particle species $s=1,...,S.$ Before we consider their evolution, however, we 
note some simple properties of the coarse-grained distributions that follow directly from their definition. 
First, one easily obtains the velocity moments up to quadratic order as 
\be \int d^3\ov\ m_s \ofs(\obx,\obv,t)=\orhos(\obx,t) \lb{IV17} \ee
\be \int d^3\ov\ m_s\obv\,\ofs(\obx,\obv,t)=\overline{\rho_s\bu_s}(\obx,t) \lb{IV18} \ee
%\be \int d^3\ov\ \obv\,\obv\,\ofs(\obx,\obv,t)=\overline{\left(\rho_s\bu_s\bu_s+\bP_s+\frac{1}{3}\uit^2\rho_s\bI\right)}(\obx,t) \ee
\be \int d^3\ov\ m_s\obv\,\obv\,\ofs(\obx,\obv,t)=\overline{\left(\rho_s\bu_s\bu_s+\bP_s+\frac{1}{3}\rho_s\uit^2\bI\right)}
(\obx,t) \lb{IV19} \ee
where to obtain the last two relations we used $\int d^3w\, \bw\, H_\uit(\bw)=\bzed$ and 
$\int d^3w\, \bw\,\bw\, H_\uit(\bw)=(1/3)\uit^2\bI.$ 
Simple consequences of the above three moment conditions are then 
\be \int d^3\ov\ \obv\,\ofs\Big/ \int d^3\ov\,\ofs =\tbus, \lb{IV20} \ee 
\be \int d^3\ov\ \frac{1}{2}m_s|\obv|^2\,\ofs=\overline{E}_s+\frac{1}{2}\orhos\uit^2, \lb{IV21} \ee
and 
\be \int d^3\ov\ m_s(\obv-\tbus)(\obv-\tbus)\,\ofs =\orhos \ttau(\bu_s,\bu_s)+\obP_s+\frac{1}{3}\orho_s\uit^2\bI. \lb{IV22} \ee

To interpret the last three results, note that $\ofs(\obx,\obv,t)$ represents an imperfectly measured distribution 
function for particle species $s,$ \magc{observed} with resolution $\ell$ in positions and resolution $\uit$ in velocities. The 
relation \eqref{IV20} states that the bulk flow velocity for the measured distribution coincides with the 
Favre-average of the true bulk velocity. Likewise, the relations \eqref{IV21} and \eqref{IV22} give 
the resolved energy density and resolved pressure tensor calculated from the measured distribution. 
Aside from the extra isotropic term $(1/3)\orhos\uit^2\bI,$ the resolved pressure tensor is given by 
\be \obP_s^*=\obP_s+\orhos \ttau(\bu_s,\bu_s), \lb{IV23} \ee
which we call the {\it intrinsic resolved pressure tensor}. Note that no calculation involving 
only the measured distribution function $\ofs(\obx,\obv,t)$ can yield separately the coarse-grained 
pressure tensor $\obP_s$ or the subscale stress tensor $\orhos \ttau(\bu_s,\bu_s)$ and only the 
combination is intrinsically defined for the measured distribution. This is similar to the concept 
of ``intrinsic resolved internal energy'' that was introduced in \cite{eyink2017cascades1} for a 
turbulent compressible fluid, which is likewise the only internal energy that be obtained from 
coarse-grained observations of the basic fluid variables. In kinetic theory, we may define
the {\it intrinsic resolved internal energy} by $\oeps^*=(1/2){\rm tr}\,(\bP_s^*),$ or 
\be \oeps^*=\oeps+\frac{1}{2}\orhos \ttau(\bu_s\bf{;}\bu_s), \lb{IV24} \ee 
using the short-hand notation $\ttau({\bf b}{\bf ;} {\bf b}')=\sum_{i=1}^3 \ttau(b_i,b_i').$ 
We then see that $\overline{E}_s=(1/2)\orhos|\tbus|^2+\oeps^*.$ The quantity $\oeps^*$ in 
\eqref{IV24} is the only internal/fluctuational energy that can be obtained from the imperfectly measured distribution 
function $\ofs(\obx,\obv,t),$ for which energy in \magc{kinetic} fluctuations $\epsilon_s$ and energy in unresolved, 
turbulent fluctuations of the bulk velocity $\tbus$ are indistinguishable.  

Finally, we note one of the most important properties of the coarse-grained distributions. 
Because the phase-space entropy density $\sit(f_s)$ is concave in $f_s,$ one has the basic inequality 
\be \sit(\ofs)\geq \overline{\sit(f_s)}. \ee 
Thus, as is well-known (e.g. \cite{gibbs1902elementary}, Chapter XII) the entropy of each species $s$ 
can only increase under coarse-graining:
\bea 
S(\of_s):&=&\int d^3\overline{x}\int d^3\ov\ \sit(\ofs)\cr
&\geq& \int d^3x\int d^3v\ \sit(f_s)=S(f_s). 
\eea 
This result implies that, if increase of total particle entropy $S_{tot}(f):=\sum_s S(f_s)$ is persistent in the collisionless 
limit $\Gamma\to 0,$ then an observer with only coarse-grained measurements of the phase-space 
distribution functions at finite resolutions $\ell,$ $\uit$ will also observe an increase in 
$S_{tot}(\of)=\sum_s S(\ofs)$. As we show now, however, the entropy production observed at fixed scales 
$\ell,$ $\uit$ is not due to the direct effect of collisions in the limit $\Gamma\to 0.$

\section{Coarse-Grained Vlasov-Maxwell Equations}\label{sec:V}

The coarse-grained particle distribution functions and coarse-grained electromagnetic fields may 
have a well-defined dynamics in the collisionless limit, as all of their gradients necessarily remain finite. 
The dynamics at fixed resolutions $\ell,$ $\uit$ in fact is governed by a coarse-grained version of the 
collisionless Vlasov-Maxwell equations, valid for very large (but finite) Dorland number. 

\subsection{Negligibility of Collisions}\label{sec:Va}

The equations for the particle distribution functions coarse-grained at scales $\ell,$ $\uit$ are 
\be \partial_t \of_s + \grad_\obx \bdot (\overline{\bv f_s})
+\grad_\obp \bdot (\overline{q_s \bE_* f_s})= \overline{C_s(f)}, \lb{V1} \ee
since the coarse-graining operation commutes with all partial derivatives. The coarse-grained 
collision operator is given by $\overline{C}_s=\sum_{s'}\overline{C}_{ss'}$ with 
\bea 
&& \overline{C}_{ss'}(\obx,\obv,t)\cr
&& =\int d^3r\, G_\ell(\br)
\int d^3v\, H_\uit(\bv-\obv) \, C_{ss'}(\obx+\br,\bv,t) \cr
&& = -\frac{\Gamma_{ss'}}{m_s\uit} \int d^3r\, G_\ell(\br)
\int d^3v\, (\grad H)_\uit(\bv-\obv)\bdot \cr
&& \hspace{25pt} \int d^3v' \frac{\bPi_{\bv-\bv'}}{|\bv-\bv'|}\bdot
\left(\grad_\bp-\grad_{\bp'}\right)(f_s f_{s'}).
\lb{V2} \eea
Here we have used the specific form of the Landau collision integral \eqref{III9} and integrated 
by parts once to move the $\grad_\bv$ derivative to the kernel $H_\uit.$ In the final 
expression in \eqref{V2}, $f_s=f_s(\bx+\br,\bv,t),$ $f_{s'}=f_{s'}(\bx+\br,\bv',t).$

We now show that $\overline{C}_{ss'}\to 0$ as $\Gamma\to 0,$ by deriving an appropriate 
upper bound. We first factorize the integrand in \eqref{V2} into a product of two terms to give 
\bea 
&& \overline{C}_{ss'}(\obx,\obv,t)\cr
&& \hspace{15pt} = -\frac{\Gamma_{ss'}}{m_s\uit} \int d^3r \int d^3v  \int d^3v'  \cr 
&& \hspace{25pt} G_\ell^{1/2}(\br)(\grad H)_\uit(\bv-\obv) \left(\frac{f_sf_{s'}}{|\bv-\bv'|}\right)^{1/2}\cr
&& \hspace{25pt} \bdot\frac{G_\ell^{1/2}(\br)\bPi_{\bv-\bv'}}{(f_sf_{s'}|\bv-\bv'|)^{1/2}}
\left(\grad_\bp-\grad_{\bp'}\right)(f_sf_{s'})
\lb{V3} \eea
and then apply Cauchy-Schwartz inequality to obtain
\bea 
&& |\overline{C}_{ss'}(\obx,\obv,t)|\leq  \frac{\Gamma_{ss'}}{m_s\uit} \times\cr 
&& \sqrt{\int d^3r \int d^3v  \int d^3v'  \,
G_\ell(\br)|(\grad H)_\uit(\bv-\obv)|^2 \frac{f_sf_{s'}}{|\bv-\bv'|}}\times \cr
&& \sqrt{\int d^3r \int d^3v  \int d^3v'  \,
\frac{G_\ell(\br)|\bPi_{\bv-\bv'}\left(\grad_\bp-\grad_{\bp'}\right)(f_sf_{s'})|^2}{f_sf_{s'}|\bv-\bv'|}}\cr
&&
\lb{V4} \eea
The integral under the first square-root contains a factor $1/|\bv-\bv'|$ in its integrand diverging 
as $\bv'\to\bv,$ but this is an integrable singularity in 3D. It is not hard to show under reasonable 
assumptions on the particle distributions that this integral remains finite as $\Gamma\to 0$ 
(Appendix \ref{app:A1}).  The integral under the second square root is, 
to within a factor, the spatial coarse-graining of the $s,s'$ term in the local entropy 
production defined in \eqref{III27}. We therefore obtain an upper bound, with $C_{f,\ell,\uit}$ independent of $\Gamma,$
\be  |\overline{C}_{ss'}(\obx,\obv,t)|\leq C_{f,\ell,\uit}(\obx,\obv)\sqrt{\Gamma_{ss'}\overline{\sigma}(\obx,t)} 
\lb{V5} \ee
which is vanishing in the limit $\Gamma\to 0$ with $\ell,$ $\uit$ fixed. Since it is the coarse-grained 
entropy-production which appears in this bound, we only need to assume that $\sigma(\bx,t)\to 
\sigma_\star(\bx,t)$ as a measure (eq.\eqref{III36}) and not pointwise in $\bx$ or in any stronger 
sense (e.g. \eqref{III38}). 

%Balescu-Lenard:
%\bea 
%&& C_{ss'}(f_s,f_{s'}) = 2\pi q_s^2 q_{s'}^2 \cr
%&& \hspace{20pt} \times 
%\grad_\bp\bdot\left[ \int d^3v' {\bf C}(\bv-\bv',f)
%\left(\grad_\bp-\grad_{\bp'}\right)(f_sf_{s'})\right],\cr
%&&  \lb{III9} \eea  
%Balescu (1997), Strain (2007)

The conclusion of this argument is that for any fixed scales $\ell,$ $\uit$ then for sufficiently 
large (but finite) Dorland numbers, the fields $\ofs,$ $s=1,...,S$ and $\obE,$ $\obB$ will 
satisfy, to any desired degree of accuracy, the {\it coarse-grained Vlasov-Maxwell equations}:
\bea 
&& \partial_t \of_s + \grad_\obx \bdot (\overline{\bv f_s}) 
+\grad_\obp \bdot(\overline{q_s \bE_* f_s})= 0, \cr 
&& \hspace{180pt} s=1,..,S \cr 
&& \grad_\obx\bdot\obE= 4\pi \sum_s q_s \overline{n}_s, \cr
&&  \grad_\obx\btimes\obB -\wtc \partial_t\obE = \frac{4\pi}{c}\obj, \cr
&&  \grad_\obx\btimes\obE + \wtc \partial_t\obB = \bzed, \quad \grad_\obx\bdot\obB = 0. \quad 
\lb{V6} \eea
The validity of the coarse-grained Maxwell equations is immediate, of course, because of the linearity of the 
Maxwell equations in $f_s,$ $s=1,..,S$ and $\bE,$ $\bB.$ For any fixed value of the Dorland number $Do\gg 1,$
the range of scales $\ell,$ $\uit$ where collisions have no direct effect and where the above ``coarse-grained 
VM equations'' are well-satisfied shall be called the ``collisionless range'' of kinetic turbulence. 
This concept is completely analogous to the ``inertial-range'' of hydrodynamic turbulence, 
where likewise viscosity has no direct effect and ``coarse-grained Euler equations'' are valid. This is essentially 
the same analogy suggested in \cite{schekochihin2008gyrokinetic,schekochihin2009astrophysical} 
but now derived and interpreted in a precise fashion. 

\magc{Explicit estimates of the cutoff scales $\ell_c,$ $\uit_c$ where collisions become important 
can be obtained from our analysis. Since the derivation involves material in later sections of the paper 
and is somewhat out of logical order, we present the details in \ref{app:C}. Here we 
just remark briefly that estimate \eqref{V5} can be improved, to:  
\be 
\overline{C}_{ss'}(\obx,\obv,t) \leq C'' \sqrt{\nu_{ss'}\,\overline{\varsigma}_{s,\ell,\uit}(\obx,\obv,t)\, 
\ofs(\obx,\obv,t)}\times \frac{v_{th,ss'}}{\uit},
\lb{V6a} \ee
where $\overline{\varsigma}_{s,\ell,\uit}(\obx,\obv,t)$ is a coarse-grained collisional entropy production rate
of particle species $s$ per phase-space volume, $v_{th,ss'}=\max\{v_{th,s},v_{th,s'}\}$, and $\nu_{ss'}$
is the Spitzer-Harm collision rate \eqref{III34} for particles of species $s,s'.$
By making the stronger hypothesis \eqref{III38} on non-vanishing entropy production, one can 
infer that $\overline{\varsigma}_{s,\ell,\uit}(\obx,\obv,t)$ remains finite in the limit $Do\to\infty,$
so that estimate \eqref{V6a} also implies that collisions can be neglected at fixed $\ell,$ $\uit$ 
in the limit. Furthermore, from \eqref{V6a} one can infer the following condition  
to determine cutoff scales $\ell_c,$ $\uit_c:$  
\be \frac{\left(\omega^{eddy}_{s,\ell,\uit}\right)^2}{\omega^{diss}_{s,\ell,\uit}} \simeq \nu_{ss'} 
\left(\frac{v_{th,ss'}}{\uit}\right)^2. \lb{V6b}\ee 
where $\omega^{eddy}_{s,\ell,\uit}(\obx,\obv,t)$ is a suitably defined ``eddy-turnover \green{rate}'' and 
where $\omega^{diss}_{s,\ell,\uit}(\obx,\obv,t)$ is a coarse-grained collisional ``dissipation \green{rate}'', at scales $\ell,$ $\uit$
in phase-space. See  Appendix \ref{app:C}. When $\omega^{eddy}_{s,\ell,\uit}\sim \omega^{diss}_{s,\ell,\uit}$
the condition \eqref{V6b} essentially coincides with the heuristic criterion proposed in the gyrokinetic literature 
(see \cite{schekochihin2008gyrokinetic}, section 2 and \cite{schekochihin2009astrophysical}, eq.(251)) 
but now derived locally in phase-space and thus consistent with possible intermittency.}

\magc{Since \eqref{V6b} imposes only a single condition on two parameters $\ell,$ $\uit,$ an additional relation 
is required to completely determine $\ell_c,$ $\uit_c$. In gyrokinetic turbulence theory this has been taken 
to be a relation $\uit/v_{th,s}\sim\ell/\rho_s$ that connects scaling in position space and velocity space, 
with $\uprho_s$ the gyroradius for species $s.$ See eq.(17) in \cite{schekochihin2008gyrokinetic} and eq.(252) 
in \cite{schekochihin2009astrophysical}. From the renormalization-group point of view, however, $\ell,$ $\uit$ 
are two independent regularization scales determined by completely arbitrary resolutions of observations
\cite{gross1976applications}. One can thus impose any additional constraint whatsoever, such as 
\be \ell \sim \uprho_s (\uit/v_{th,s})^\beta,  \qquad \beta>0, \lb{V6c} \ee 
so long as $\ell,$ $\uit$ vanish together. The scales $\ell_c^{(\beta)},$ $\uit_c^{(\beta)}$ where collisions 
first become non-negligible in the coarse-grained VM eqs.\eqref{V6} will necessarily be $\beta$-dependent, 
but  no {\it objective} physical statement can depend upon which value of $\beta$ is adopted in \eqref{V6c}.  
There may, on the other hand, be a ``natural choice'' which makes the description simpler (just as any curvilinear 
coordinate system may be adopted to describe a given physics problem, but some coordinate choices are far more 
convenient). In particular, for the case of kinetic turbulence, there may be a physical relation between the 
scales of phase-space ``eddies'' in position-space $\ell$ and velocity-space $\uit,$ which determines 
a natural choice of $\beta$ and which removes this freedom in the definition of $\ell_c,$ $\uit_c.$} 

If suitable (strong) limits of the VML solutions exist \footnote{We here use the standard terminology of analysis,
with ``strong convergence'' denoting convergence in norms such as the $L^p$ norms and ``weak convergence''
denoting convergence after smearing with an element of the dual space, e.g. \cite{kolmogorov1975introductory}.
A remarkable result of the DiPerna-Lions theory \cite{diperna1989global} is that, even for sequences of 
solutions $f_s^{(n)},$ $\bE^{(n)},$ $\bB^{(n)}$ of the Vlasov-Maxwell system that converge only weakly 
to limits $f_s,$ $\bE,$ $\bB,$ nevertheless the nonlinear wave-particle interaction term 
$\grad_\bv \bdot[ (\bE^{(n)}+(\bv/c)\btimes\bB^{n)}) f_s^{(n)})$ also converges (distributionally) to 
$\grad_\bv \bdot[ (\bE+(\bv/c)\btimes\bB) f_s].$ The corresponding statement is not true 
of the advective nonlinear term $\grad_\bx\bdot[\bu\bu]$ for the incompressible Euler equation, nor for most
nonlinear PDE's,  but instead depends upon special features of the Vlasov-Maxwell equations. Because of this 
special ``stability'' property of VM solutions, it is quite likely that strong limits as $Do\to\infty $ of the VML 
solutions need not be assumed in order to obtain (distributional) solutions of VM. On the other hand, the same 
remarkable convergence statements need not hold for other nonlinear functions, such as the phase-space 
entropy densities $\sit(f_s).$ Thus, some of our key conclusions, such as the anomalous entropy balance 
\eqref{VI11}, may require strong convergence},  $f_s,\,\bE,\, \bB\, \to \, f_{\star\,s},\bE_\star,\bB_\star$
as $Do\to \infty,$ then the coarse-grained VM equations \eqref{V6} will hold for those limit fields 
with any choice of $\ell,$ $\uit.$ This is equivalent to the statement that the limit fields $f_{\star\,s},\bE_\star,\bB_\star$
are ``weak'' or ``distributional'' solutions of the Vlasov-Maxwell equations 
(Propositions 1 \& 2 in \cite{drivas2018onsager}). In other words, the limit fields will satisfy 
\bea 
&& \partial_t f_{\star\, s} + \grad_\bx \bdot (\bv f_{\star\, s}) 
+\grad_\bp \bdot( q_s (\bE_\star)_* f_{\star\, s})= 0, \cr 
&& \hspace{180pt} s=1,..,S \cr 
&& \grad_\bx\bdot\bE_\star = 4\pi \sum_s q_s n_{\star\, s}, \cr
&&  \grad_\bx\btimes\bB_\star -\wtc \partial_t\bE_\star = \frac{4\pi}{c}\bj_\star, \cr
&&  \grad_\bx\btimes\bE_\star + \wtc \partial_t\bB_\star = \bzed, \quad \grad_\bx\bdot\bB_\star = 0. \quad 
\lb{V7} \eea
in the sense of distributions. Here we may note that there is rigorous mathematical theory 
on global existence of weak solutions of the VM equations, the state of the art of which
is represented essentially by the work of DiPerna \& Lions \cite{diperna1989global}. Those authors 
prove that, for any initial data $f_{0s},$ $s=1,..,S$ and $\bE_0,$ $\bB_0$ which satisfy the conditions   
\bea 
&& \int d^3x \int d^3v (1+|\bv|^2) f_{0s} <\infty, \quad \int d^3x \int d^3v f_{0s}^2<\infty \cr
&& \hspace{180pt} s=1,..,S \cr 
&& \hspace{30pt} \grad_\bx\bdot \bE_0=\sum_s q_s\int d^3v\, f_{0s}, \quad \grad_\bx\bdot \bB_0=0 \cr 
&& \hspace{60pt} \int d^3x [|\bE_0|^2 + |\bB_0|^2] <\infty,
\lb{V8} \eea 
then weak/distributional solutions of the VM equations with these initial conditions exist globally 
in time (but may not be unique). We shall discuss some properties of these known weak solutions 
further below. We note here only that the weak solutions in the DiPerna-Lions theory \cite{diperna1989global} 
are {\it not} obtained as collisionless limits of solutions of the VML equations or other Boltzmann-type equations, 
and that such limits have not to date been mathematically proved (or disproved) to exist \footnote{Of course, 
this is not an unusual situation in physics. There are many examples of mathematical objects whose existence is 
supported by almost overwhelming empirical evidence and theoretical arguments, but which have never been rigorously 
proved to exist from first principles.These examples include crystalline solid phases at sufficiently low temperatures
\cite{farmer2017crystallization} or the non-Gaussian renormalization group fixed point believed to describe the critical 
properties of all equilbrium systems in the universality class of the 3D Ising model \cite{gallavotti2014renormalization}. 
Needless to say, the development of solid state physics did not have to wait for the mathematical proof from many-body 
quantum theory  (still unavailable) that crystals exist!}. Better mathematical understanding of the collisionless 
limit would provide important new concepts and tools for the theory of kinetic plasma turbulence. We emphasize,
however, that we do not need to assume in this work that limits $f_s,\,\bE,\, \bB\, \to \, f_{\star\,s},\bE_\star,\bB_\star$
must exist for $Do\to \infty.$ Our principal conclusions are independent of this hypothesis. 

\subsection{Eddy-Drift and Effective Fields}\label{sec:Vb}

Although the ``coarse-grained VM equations'' hold \magc{to any desired accuracy} 
for fixed $\ell,$ $\uit$ when $Do\gg 1,$
this does not mean that the VM equations in the naive sense hold for the coarse-grained 
fields $\ofs,$ $s=1,...,S$ and $\obE,$ $\obB$. To explain this point clearly, we shall 
write the equations \eqref{V6} in a form as close as possible to the ordinary VM equations. 
This can be done in a simple way by using the concept of phase-space Favre average 
introduced in section \ref{sec:IVb} to write $\overline{\bv f_s}=\hbvs\of_s$ and 
$\overline{\bE_* f_s}=\hbE_{*s}\of_s$ so that the 
``coarse-grained Vlasov equation'' becomes:  
\be \partial_t \of_s + \grad_\obx \bdot(\hbvs\of_s) 
+\grad_\obp \bdot(q_s\hbE_{*s}\of_s)=0. \lb{V9} \ee
If the effective fields $\hbvs$, $\hbE_{*s}$ introduced in this fashion were the same 
as $\obv$ and $\obE+(\obv/c)\btimes \obB$ then the coarse-grained quantities 
$\ofs,$ $s=1,...,S$ and $\obE,$ $\obB$ would satisfy the VM equations in the conventional sense.
In the example of hydrodynamic turbulence, however, $\overline{\bv\bv}=\obv\,\obv+\btau\neq \obv\,\obv,$
because of the additional ``subscale'' or ``turbulent'' stress $\btau$ that was 
introduced by integrating out small eddies. In the same manner, we shall show now that
$\hbvs$, $\hbE_{*s}$ do not coincide with $\obv,$ $\obE+(\obv/c)\btimes \obB$ but contain 
additional contributions because of the elimination of ``small eddies'' in the phase-space. 

We note first directly from the definition of Favre average that 
\be 
\hbvs=\obv+\hbw_s(\obx,\obv,t) 
\lb{V10} \ee
with an {\it eddy-drift velocity} given by 
\bea \hbw_s :&=&\frac{1}{\ofs} \langle \bw f_s(\obx+\br,\obv+\bw,t)\rangle_{\ell,\uit} \cr 
&=& \frac{1}{\of_s}
\langle \bw\,\delta_\bw \overline{f}_{s,\ell}(\obx,\obv) \rangle_{\uit} 
\lb{V11} \eea 
The second expression is obtained by performing first the $\langle\cdot\rangle_\ell$-average 
over $\br$ and then using the property $\langle \bw\rangle_\uit=\bzed.$ This expression shall be useful 
in making estimates of the magnitude of $\hbw_s$. The physical meaning of this ``eddy-drift'' 
is that the local mean velocity of the population of particles within distances $\ell,$ $\uit$ 
of the phase point $(\obx,\obv)$ does not coincide with $\obv,$ and $\hbw_s$ is the 
average drift velocity of this population relative to $\obv$ itself. 

One can likewise derive for the effective fields in \eqref{V9} the expressions 
\be \hbE_{*s} = \hbE_s + \wtc \obv\btimes \hbB_s + \wtc \widehat{(\bw\btimes\bB)}_s 
\lb{V12} \ee 
with 
\bea \hbE_s(\obx,\obv,t) &=& \obE(\obx,t) + \frac{1}{\ofs} \otau(\bE,f_s) \cr
&=& \obE(\obx,t) + \frac{1}{\of_s} \otau_\ell(\bE,\overline{f}_{s,\uit}),  
\lb{V13} \eea
also 
\bea 
\hbB_s(\obx,\obv,t) &=& \obB(\obx,t) + 
 \frac{1}{\of_s}
\otau(\bB,f_s) \cr
&=& \obB(\obx,t) + \frac{1}{\of_s}
\otau_\ell(\bB,\overline{f}_{s,\uit}), \cr
&& 
\lb{V14} \eea
and 
\bea 
&& \widehat{(\bw\btimes\bB)}_s(\obx,\obv,t) \cr 
&& \hspace{50pt} =\frac{1}{\ofs} \langle \bw\btimes \bB(\obx+\br,t) f_s(\obx+\br,\obv+\bw,t)\rangle_{\ell,\uit}\cr
&& \hspace{50pt}=\frac{1}{\of_s}
\langle \bw\btimes\bB(\obx+\br,t)\,\delta_\bw f_s(\obx+\br,\obv) \rangle_{\ell,\uit} \cr
&&
\lb{V15} \eea
These results are again direct consequences of the definition of Favre coarse-graining. 
The derivation of \eqref{V15} is quite similar to that of \eqref{V11}.
The first lines in \eqref{V13},\eqref{V14} follow by the general relation \eqref{IV15}
between Favre and unweighted coarse-graining, and the second lines in \eqref{V13},\eqref{V14} 
follow from the $\bv$-independence of $\bE,$ $\bB,$ which allows the 
$\langle\cdot\rangle_\uit$-average over $\bw$ to be performed first.  

Notice that the Favre-averaged fields $\hbE_s$, $\hbB_s$ become velocity-dependent 
due to the terms $\otau_\ell(\bE,\overline{f}_{s,\uit})$, $\otau_\ell(\bB,\overline{f}_{s,\uit})$,
which account for the fine-scale correlations of particles and fields. This is similar to the 
velocity-dependence of conditionally-averaged fields in the derivation of 
the Vlasov-Maxwell system from the BBGKY hierarchy, except that the latter dependence 
arises from multi-particle statistical correlations and disappears when molecular chaos holds 
(e.g. \cite{ecker1972theory}, section III.1.1). 
In the ``collisionless range'' of kinetic turbulence, on the other hand, the correlations arise from turbulent 
fluctuations in the phase space and they do not vanish under any physically plausible assumptions. 
As we shall see, these correlations are a major contributor to kinetic turbulent cascades. \red{Similar 
correlations arise microscopically at the next order in the expansion in the plasma parameter, 
leading to the collision integral expressed in the form $C_s(f)=-q_s\grad_\bp\bdot \langle \delta\bE_*\delta f_s\rangle,$
where the average here is over statistics of the individual ions (e.g. see \cite{klimontovich1982kinetic}, eq.(26.13)). 
Thus, the contributions in \eqref{V9} which arise from the correlation terms $\otau_\ell(\bE,\overline{f}_{s,\uit})$, 
$\otau_\ell(\bB,\overline{f}_{s,\uit})$ in $ \hbE_{*s}$ represent ``collisions'' of turbulent eddies.} It is interesting 
that in the exact theory presented here at the level of the VML description, these nonlinear wave-particle 
interaction terms can explicitly drive a cascade in velocity space. In the gyrokinetic approximation 
there is no corresponding term which can create phase-space fine-scale structure by direct  
``advection'' in velocity space and the necessary fine-structure for persistent entropy dissipation
arises instead from the velocity-dependence of ring-averages (\cite{schekochihin2009astrophysical} , p.345). 

Using the second lines of each of the formulas \eqref{V11},\eqref{V13}--\eqref{V15}, we 
can estimate the magnitudes of all of the contributions to $\hbw_s$ and $\hbE_{*s}$ in 
\eqref{V10},\eqref{V12}: 
\be \hbw_s(\obx,\obv,t) =O(\uit\,\delta_\uit \! f_s/f_s),  \lb{V16} \ee 
\be \hbE_s(\obx,\obv,t) =\obE(\obx,t) + O(\delta_\ell E\,\delta_\ell \! f_s/f_s),  
\lb{V17} \ee 
\be
\obv\btimes\hbB_s(\obx,\obv,t) =
\obv\btimes\obB(\obx,t)+ O(\overline{v}\,\delta_\ell B\,\delta_\ell \! f_s/f_s),  
\lb{V18} \ee
\be \widehat{(\bw\btimes\bB)}_s(\obx,\obv,t) =O(\uit\,B\,\delta_\uit \! f_s/f_s),  
\lb{V19} \ee
Here we use the short-hand notations 
\be \delta_\ell f_s:= \sup_{|\br|<\ell}|\delta_\br f_s|, \quad \delta_\uit f_s:= \sup_{|\bw|<\uit}|\delta_\bw f_s| \ee  
and likewise for all other quantities. The estimates \eqref{V16}-\eqref{V19} are all exact upper bounds,
but can also be taken as order-of-magnitude estimates of the terms 
\eqref{V11},\eqref{V13}--\eqref{V15}, if one assumes that there are no significant cancellations
in the local phase-space averages defining those terms. (As we shall discuss later, this is probably 
a dubious assumption.) We see explicitly from \eqref{V16}--\eqref{V19} that the 
quantities $\hbvs$, $\hbE_{*s}$ appearing in the ``coarse-grained Vlasov equations'' are 
different from $\obv,$ $\obE+(\obv/c)\btimes \obB$ and, thus, $\ofs,$ $s=1,...,S$ and $\obE,$ $\obB$
do not satisfy the VM equations in the conventional sense.  

From a conceptual point of view, the quantities $\hbvs$, $\hbE_{*s}$ are scale-dependent 
``renormalizations'' of the ``bare'' quantities $\bv,$ $\bE_*$  that appear in the ``fine-grained'' 
VML equations \eqref{III1}-\eqref{III3}. The particle distribution functions measured in any real experiment 
will always have some finite  resolutions $\ell,$ $\uit$ in position- and velocity-space and thus correspond 
to the coarse-grained distributions $\ofs(\obx,\obv,t)$ and not to the fine-grained distributions $f_s(\bx,\bv,t)$ 
that exactly satisfy the Vlasov-Landau equation \eqref{III1}. At \black{sufficiently} large but finite $Do$ 
and with fixed resolutions $\ell,$ $\uit,$ these measured distributions $\ofs$ will satisfy 
\black{to any desired degree} the renormalized equation \eqref{V9}, which is only 
equivalent to a Vlasov equation in the ``coarse-grained sense" \eqref{V6}. By contrast, any
fine-grained distributions $f_{\star s}(\bx,\bv,t)$ obtained in the strong limit $Do\to\infty$ 
exactly satisfy the collisionless Vlasov equation \eqref{V7}, 
but only in a distributional sense. The limits $f_{\star s}$ are singular Vlasov solutions with 
non-differentiable dependence on position and velocity, which can never be strictly observed in Nature.
They are idealized mathematical objects which are approached better and better by the smooth 
VML solutions as $Do$ increases and as the fine-grained distributions $f_s$ become 
more and more nearly singular. 

\section{Entropy Cascade in Phase Space}\label{sec:VI}

The results in the previous section resolve the ``paradox'' that the Vlasov-Maxwell equations are 
valid at fixed scales $\ell,$ $\uit$ as $\Gamma\to 0,$ in the sense of eq.\eqref{V6}, and yet entropy 
$S_{tot}(\of)$ increases at those scales, even without any direct contribution from collisions. As we now show, the 
entropy production in the coarse-grained description at fixed resolutions $\ell,$ $\uit$ is due to a
nonlinear entropy cascade through phase-space scales, in exact analogy to the kinetic-energy cascade
in incompressible fluid turbulence. 

\subsection{Coarse-Grained Entropy Balance}\label{sec:VIa}

The first important observation is that the ``coarse-grained Vlasov equation'' in \eqref{V6} or 
\eqref{V9} satisfies no Liouville theorem, so that $\ofs$ is not conserved along characteristic 
curves of $\hbvs$, $\hbE_s$. Instead, direct calculation yields along characteristics that 
\bea 
&& \partial_t \of_s + \hbvs \bdot \grad_\obx \of_s 
+q_s\hbE_{*s}\bdot \grad_\obp \of_s \cr
&& \qquad \qquad = -(\grad_\obx\bdot\hbvs + q_s\grad_\obp\bdot\hbE_{*s})\ofs
\lb{VI1} \eea
with generally $\grad_\obx\bdot\hbvs + q_s\grad_\obp\bdot\hbE_{*s}\neq 0.$ 
Below we give explicit expressions for this phase-space divergence which show clearly 
that it need not vanish. As a simple consequence of \eqref{VI1}, one obtains the 
following phase-space balance equation satisfied by the entropy density of the 
coarse-grained distribution for species $s$:
\bea
&& \partial_t \sit[\of_s] + \grad_\obx \bdot(\hbvs \sit[\of_s]) 
+\grad_\obp \bdot(q_s\hbE_{*s} \sit[\of_s]) \cr
&&\qquad\qquad =\varsigma^{flux,s}_{\ell,\uit}(\obx,\obv,t)
\lb{VI2} \eea
where 
\be
\varsigma^{flux,s}_{\ell,\uit}(\obx,\obv,t):=
(\grad_\obx\bdot\hbvs + q_s\grad_\obp\bdot\hbE_{*s})\of_s 
\lb{VI3} \ee
The quantity $\varsigma^{flux,s}_{\ell,\uit}(\obx,\obv,t)$ represents rate of transfer of entropy 
of species $s$ from unresolved scales $<\ell,$ $\uit$ in the phase-space, where it is created 
by collisions, up to the resolved scales $>\ell,$ $\uit,$ locally for each phase-space point 
$(\obx,\obv)$ \footnote{The phase-space contraction rate, or violation
of the Liouville theorem, is well-known in other contexts to be related to the rate of entropy production,
e.g. in molecular dynamics theory of non-equilibrium statistical mechanics and transport behavior, 
cf. \cite{evans1990statistical}, section 10.1}. It is exactly analogous to the local 
energy flux $\Pi_\ell(\bx,t)$ for incompressible fluid turbulence (\cite{eyink2018review},(III.8)), 
except for a change in sign. Because of the sign-difference, $\varsigma^{flux,s}_{\ell,\uit}$ is better 
regarded as a flux of {\it negentropy}, or negative entropy, to small-scales in phase-space, which 
is there dissipated by collisions. We recall here that the ``generalized energy'' in gyrokinetics is 
the electromagnetic field energy {\it minus} the entropy of particles (see 
\cite{schekochihin2008gyrokinetic,schekochihin2009astrophysical} 
and the discussion in section \ref{sec:VIIIa}). Negentropy also plays a central role in 
the ``ideal turbulence theory'' for compressible fluids 
\cite{eyink2017cascades1,eyink2017cascades2,drivas2018onsager}. 
  
The sign of $\varsigma^{flux,s}_{\ell,\uit}(\obx,\obv,t)$ will vary from point to point in phase-space
and also with scales $\ell,$ $\uit.$ However, its integral over velocity and summation over $s$
\be \sigma^{flux}_{\ell,\uit}(\obx,t):=\sum_s \int d^3\ov\ \varsigma^{flux,s}_{\ell,\uit}(\obx,\obv,t) \lb{VI4} \ee
must be positive on average. Indeed, velocity integration of \eqref{VI2} and summation over $s$ yields
\be \partial_t s_{tot}(\of) + \grad_\obx\bdot \bJ_{S,\ell\,\uit}^{res} = \sigma^{flux}_{\ell,\uit}, \lb{VI5} \ee 
with space-density of total resolved entropy 
\be s_{tot}(\of):= \sum_s \int d^3\ov \ \sit(\ofs) \lb{VI6} \ee
and with resolved entropy current density
\bea \bJ_{S,\ell\,\uit}^{res}:&=& \sum_s \int d^3\ov\ \hbvs \, \sit(\ofs) \cr
&=& -\sum_s \int d^3\ov\ \overline{\bv f_s}\, \ln \ofs \lb{VI7} \eea
Averaging \eqref{III25} over space, we first choose $Do$ sufficiently large so that 
\be \frac{d}{dt}\langle s_{tot}(f) \rangle=\langle \sigma\rangle\doteq \langle\sigma_\star\rangle>0, 
\lb{VI8} \ee
with $\langle\cdot\rangle$ representing the space-average. 
We then subsequently choose $\ell,$ $\uit$ sufficiently small so that the average of \eqref{VI5} 
over space gives 
\be \langle \sigma^{flux}_{\ell,\uit}\rangle = \frac{d}{dt}\langle s_{tot}(\of) \rangle\doteq
\frac{d}{dt}\langle s_{tot}(f) \rangle. \lb{VI9} \ee
Comparing the two expressions for $(d/dt)\langle s_{tot}(f) \rangle$ in \eqref{VI8} and \eqref{VI9}, 
one concludes that for $Do\gg 1$ there is a range of sufficiently small $\ell,$ $\uit$
such that 
\be \langle \sigma^{flux}_{\ell,\uit}\rangle \doteq \langle \sigma_\star \rangle>0. \lb{VI10} \ee
Thus, there is a range of nearly constant negentropy flux which, furthermore, is positive,
corresponding to a forward cascade of negentropy or an inverse cascade of the standard entropy 
\footnote{Identities $-(\grad_\obv\bdot  {\bE}_{*s}^{\!\!\!\widehat{}\,\,\,} )\ofs=-\overline{\grad_\bv\bdot (\bE_* f_s)}+
({\bE}_{*s}^{\!\!\!\widehat{}\,\,\,} \bdot\grad_\obv)\ofs$  and $-(\grad_\obx\bdot  {\bw}_{s}^{\!\!\!\widehat{}\,\,\,} )\ofs
=-\overline{\grad_\bx\bdot (\bv f_s)}+({\bw}_{s}^{\!\!\!\widehat{}\,\,\,} \bdot\grad_\obx)\ofs$ 
show that positive entropy production results at a phase-space point when the coarse-grained 
rate of spreading of the fine-grained distribution is greater there than the rate of spreading 
of the coarse-grained distribution. It is worth recalling that the particle-field interaction term can be written as 
$(q_s/m_s)\bE_*\bdot\grad_\bv f_s=(q_s \bE/m_s)\bdot \grad_\bv f_s-(q_s\bB/m_sc)\bdot (\bv\btimes\grad_\bv)f_s,$
with the first term corresponding over a time-increment $dt$ to a translation of the particle 
distribution in velocity space by $dt (q_s \bE/m_s)$ and the second term corresponding to a rotation of the distribution 
in velocity space  by rotation vector $-dt(q_s\bB/m_sc).$ Of course, the free-streaming term 
$-(\bv\bdot \grad_\bx) f_s$ corresponds to translation of the particle distribution in position space
by displacement $\bv\, dt$ over time $dt$}. 

We can derive a more general result if we assume that (strong) limits exist $f_s\to f_{\star\, s}$
as $Do\to\infty.$ In that case, one has the limiting entropy balance
\be \partial_t s_{tot}(f_\star) + \grad_\bx \bdot\bJ_{S\star} =\sigma_\star \lb{VI11} \ee 
in the sense of distributions, directly from \eqref{III25}. Furthermore, one has in the limit $\ell,$ $\uit\to 0$ that 
$s_{tot}(\of_\star)\to s_{tot}(f_\star)$ in the sense of distributions 
for the total entropy defined in \eqref{III24} and likewise as $\ell,$ $\uit\to 0$
\bea \bJ_{S\star,\ell\,\uit}^{res}&=& -\sum_s \int d^3\ov\ \overline{\bv f_{\star\,s}}\, \ln \of_{\star s}\cr
&\to& -\sum_s \int d^3v\ \bv f_{\star s}\, \ln f_{\star s} \, = \, \bJ_{S\star}
\lb{VI12} \eea 
in the sense of distributions, for the entropy current density defined in \eqref{III26}. Because 
the eq.\eqref{VI5} follows for $\of_{\star\,s},$ $s=1,...,S$ as a consequence of eq.\eqref{V7}, one 
can also conclude that 
\bea \lim_{\ell,\uit\to 0} \sigma^{flux}_{\star,\ell,\uit} 
&=& \lim_{\ell,\uit\to 0}\left[\partial_t s_{tot}(\of_\star)+ \grad_\bx\bdot\bJ_{S\star,\ell\,\uit}^{res}\right] \cr
& =& \partial_t s_{tot}(f_\star)+ \grad\bdot\bJ_{S\star}\cr
&=& \sigma_\star
\lb{VI13} \eea 
in the sense of distributions, where \eqref{VI11} was used in the last step. Equation \eqref{VI13} is 
equivalent to the statement that, for any smooth, compactly-supported function on space-time,
$\varphi(\bx,t)\geq 0$ with $\int d^3x\int dt\, \varphi=1,$ then, for the local space-time average defined by $\varphi,$
%\newpage 
\bea 
&& \lim_{\ell,\uit\to 0} \int d^3x\int dt \ \varphi(\bx,t)\,\sigma^{flux}_{\star,\ell,\uit}(\bx,t)\cr
&& \hspace{40pt} =
\int d^3x\int dt \ \varphi(\bx,t)\, \sigma_\star(\bx,t). 
\lb{VI14} \eea 
This is obviously a stronger statement than \eqref{VI10}, which required a global space average.
The result \eqref{VI13} or \eqref{VI14} is analogous to the local relation (in the sense of distributions) between
kinetic energy flux and viscous energy dissipation derived for incompressible fluid turbulence 
by Duchon \& Robert \cite{duchon2000inertial}. 

The balance for total entropy obtained in \eqref{VI11} as $Do\to\infty$, with $f_{\star s},$ 
$s=1,...,S$ a set of weak or distributional solutions of the Vlasov-Maxwell equations \eqref{V7} is an 
example of what is called an ``anomalous balance'' in quantum field-theory and condensed-matter 
physics \cite{polyakov1993theory,polyakov1992conformal,eyink2018review}. 
A positive source term $\sigma_*>0$ implies increasing total entropy for the ``weak" solutions, 
whereas total entropy is conserved for smooth solutions of the Vlasov-Maxwell equations.
The non-vanishing entropy-production $\sigma_\star>0$ is an example of a ``dissipative anomaly'', 
like that predicted by Onsager \cite{onsager1949statistical,eyink2018review} for incompressible 
Euler solutions describing hydrodynamic turbulence as $Re\to\infty.$ As in the fluid case, 
such anomalies are possible only if the solutions are sufficiently ``singular'' or ``rough''.
We next derive the analogue of ``4/5th-laws'' which express the entropy flux \eqref{VI3} 
in terms of increments of particle distributions and fields and which allow us to establish exact 
constraints on the degree of singularity/rugosity required for the turbulent solutions 
to sustain a non-vanishing negentropy flux to small scales in phase-space. 

\subsection{$4/5^{th}$ Laws for Entropy Fux}\label{sec:VIb}

The formula \eqref{VI3} for the entropy flux through scales in phase-space can be 
further evaluated with the expressions for $\hbvs,$ $\hbE_s$ given in \eqref{V10}-\eqref{V15}. 
The net contribution of $\obv$ and $\obE+(\obv/c)\btimes\obB$ to the divergence in 
\eqref{VI3} is clearly zero, and the non-vanishing contributions arise from the subscale 
correlation terms. \black{From \eqref{V11}-\eqref{V15}, these quantities all have 
the general form ${\bf A}/\ofs,$ where ${\bf A}$ is an expression for the sub-scale correlation.
Since $\grad\bdot({\bf A}/\ofs)\ofs=\grad\bdot{\bf A}-{\bf A}\bdot \grad\log\ofs$, the contributions
to the entropy flux $\varsigma^{flux,s}_{\ell,\uit}$ consist generally of a total divergence term 
$\grad\bdot{\bf A}$ and a second term proportional to $\grad\ofs.$ More precisely,
\be (\grad_\obx\bdot\hbw_s)\, \of_s = \grad_\obx \bdot(\hbw_s\, \of_s) -\hbw_s \bdot\grad_\obx\of_s 
\lb{VI15a} \ee 
and 
\be q_s(\grad_\obp\bdot\hbE_s)\,\of_s = -\grad_\obp \bdot \bk_S^{*s} + \bk_S^{*s}\bdot\grad_\obp\log\ofs 
\lb{VI16} \ee
with 
\bea \bk_S^{*s}&:=& -q_s\Big[\otau_\ell(\bE,\overline{f}_{s,\uit})+\wtc\obv\btimes \otau_\ell(\bB,\overline{f}_{s,\uit})
+\wtc \widehat{(\bw\btimes\bB)}_s\ofs \Big]\cr
&& \, 
\lb{VI17} \eea 
We now make an important observation, that ``flux terms'' in coarse-grained balance equations are generally 
defined pointwise in phase space only up to total divergences, which may be considered as contributions 
to transport in phase-space rather than as transfer between scales. In this spirit, the quantity $\bk_S^{*s}$ 
defined in \eqref{VI17} may be 
taken to represent a {\it turbulent transport of entropy in momentum-space.} Likewise, the quantity
\be \bj_S^{*s} = -\hbw_s\, \of_s =-\langle \bw\, f_s\rangle_{\ell,\uit} \ee
may be considered to be {\it turbulent transport of entropy in position-space}. Using these definitions, we 
may now rewrite the coarse-grained entropy balance \eqref{VI2} as
\bea
&& \partial_t \sit[\of_s] + \grad_\obx \bdot(\hbvs \sit[\of_s]+\bj_S^{*s}) 
+\grad_\obp \bdot(q_s\hbE_{*s} \sit[\of_s]+\bk_S^{*s}) \cr
&&\qquad\qquad =\varsigma^{*flux,s}_{\ell,\uit}(\obx,\obv,t)
\lb{VI18} \eea
where the source term on the righthand side 
\be
\varsigma^{*flux,s}_{\ell,\uit}(\obx,\obv,t):=
\bj_S^{*s}\bdot\grad_\obx\log\ofs+ \bk_S^{*s}\bdot\grad_\obp\log\ofs,  
\lb{VI19} \ee
is another possible representation of {\it entropy flux} across scales $\ell,$ $\uit$ in phase space,
alternative to \eqref{VI3}} \footnote{\magc{It is worth pointing out that the same expression \eqref{VI19}
for entropy flux can be obtained without using Favre averaging, by instead treating particle distributions 
$f_s$ as ``advected scalars'' for the incompressible flow in phase-space generated by the 
Hamiltonian equations of a charged particle in an EM field. Standard derivations of 4/5th-type laws 
for advected scalars in incompressible fluid turbulence using 
spatial coarse-graining, e.g. eqs.(29)-(33) in \cite{eyink1996intermittency}, recover \eqref{VI19}}}. 

\black{
This expression for entropy flux has an intuitive physical interpretation when expressed in terms of 
\be \lambda[f_s]:=\delta S[f]/\delta f_s(\bx,\bv) = -(\log f_s+1), \lb{VI20} \ee
the potential ``entropically conjugate'' to $f_s$. Turbulent entropy production is obviously positive 
whenever the turbulent transport vectors $\bj_S^{*s},$ $\bk_S^{*s}$ are anti-aligned with the 
corresponding gradients $\grad_\obx \lambda[\ofs],$ $\grad_\obp \lambda[\ofs].$ The sign need 
not be positive everywhere in phase space, of course, but may often be negative. However, the 
considerations in the previous section \ref{sec:VIa} on the sign of $\sigma^{flux}_{\ell,\uit}$ all 
carry over to the corresponding quantity
\be \sigma^{*flux}_{\ell,\uit}(\obx,t):=\sum_s \int d^3\ov\ \varsigma^{*flux,s}_{\ell,\uit}(\obx,\obv,t). 
\lb{VI21} \ee
This is obvious for the space-average, because the two quantities differ only by a divergence term 
and thus $\langle\sigma^{*flux}_{\ell,\uit}\rangle=\langle\sigma^{flux}_{\ell,\uit}\rangle.$ Furthermore, 
the pointwise distributional limits of these two quantities must also coincide, taking first $Do\to \infty$ and then   
\be \lim_{\ell,\uit\to 0} \sigma^{*flux}_{\star,\ell,\uit} = \sigma_\star\geq 0, \lb{VI22} \ee
where $\sigma_\star$ is the same quantity that appears in \eqref{VI13} as the distributional limit of 
$\sigma^{flux}_{\star,\ell,\uit}.$  More generally, distributional limits of $\varsigma^{*flux,s}_{\ell,\uit}$
and $\varsigma^{flux,s}_{\ell,\uit}$ must coincide. This follows again because of the fact that these
quantities differ only by terms of the form $\grad\bdot{\bf A}.$ The gradient $\grad$ can always be shifted 
after smearing in phase space to the test function $\varphi(\bx,\bp,t),$ via an integration by parts, whereas 
estimates \eqref{V16}-\eqref{V19} of the correlation terms 
${\bf A}$ show that each of these vanishes as $\ell,$ $\uit\to 0$ under very 
mild assumptions, e,g. continuity of the limiting solutions $\bE_\star,$ $\bB_\star,$ $f_{\star s},$ $s=1,..,S.$
\footnote{\black{In fact, a standard ``density argument'' from real analysis shows that the terms 
$\grad\bdot{\bf A}\to 0$ distributionally as $\ell,$ $\uit\to 0$ even under the much weaker assumption that 
$\bE_\star,$ $\bB_\star,$ $f_{\star s},$ $s=1,..,S$ are $L^2$ functions, which is almost the minimal 
regularity required for the Vlasov-Maxwell equations to make sense.}}}

\black{
The most compelling reason to prefer the modified quantity $\varsigma^{*flux,s}_{\ell,\uit}$ in \eqref{VI19} as a measure 
of ``entropy flux'' is that the original definition $\varsigma^{flux,s}_{\ell,\uit}$ in \eqref{VI3} suffers large cancellations when integrated 
over phase space, and the net contribution to the entropy cascade in fact arises from the much smaller 
quantity $\varsigma^{*flux,s}_{\ell,\uit}.$ Indeed, the contributions to $\varsigma^{flux,s}_{\ell,\uit}$ from 
the $\grad\bdot{\bf A}$ terms are quadratic in increments, like typical turbulent transport terms in space, 
whereas all of the contributions to $\varsigma^{*flux,s}_{\ell,\uit}$ are cubic in increments, like typical 
turbulent fluxes, and thus generally smaller in magnitude. Specifically, the entropy flux defined in \eqref{VI19}
consists of four contributions 
\bea
&&\varsigma^{*flux,s}_{\ell,\uit}= - \hbw_s\bdot\grad_\obx\of_s 
-(q_s/m_s) \otau_\ell(\bE,\overline{f}_{s,\uit})\bdot\grad_\obv\of_s/\of_s \cr 
&& \hspace{40pt} +(q_s/m_s c) \otau_\ell(\bB,\overline{f}_{s,\uit})\bdot(\obv\btimes\grad_\obv)\of_s/\of_s\cr 
&& \hspace{40pt} -(q_s/m_s c) \widehat{(\bw\btimes\bB)}_s \bdot\grad_\obv\of_s. 
\lb{VI15} \eea 
These four quantities can all be expressed in terms of phase-space increments of the VML solutions 
$f_s,$ $s=1,...,S$ and $\bE,$ $\bB$ by means of the general relation \eqref{IV7} for the correlation terms
$\otau_\ell(\bE,\overline{f}_{s,\uit}),$ $\otau_\ell(\bB,\overline{f}_{s,\uit})$,  the identities \eqref{V11},\eqref{V15}
for $\hbw_s,$ $\widehat{(\bw\btimes\bB)}_s,$ and the equations  \eqref{IV8}-\eqref{IV9} for the gradients 
$\grad_\obx\ofs,$ $\grad_\obv\ofs.$  
\magc{These  expressions provide exact ``4/5th-laws'' for entropy cascade in kinetic turbulence 
(see Appendix \ref{app:D} for explicit formulas and further discussion), which have previously 
been obtained only in 2D gyrokinetic turbulence \cite{plunk2009theory,plunk2010two}. Exploiting them,}
we can make order-of-magnitude estimates of each of the four terms contributing to the phase-space entropy 
flux in eq.\eqref{VI15}: 
\be
- \hbw_s\bdot\grad_\obx\of_s  = O\left(\frac{\uit\,(\delta_\uit f_s) (\delta_\ell\! f_s)}{\ell f_s}\right),  
\lb{VI23} \ee
\be
-\frac{q_s}{m_s}\otau_\ell(\bE,\overline{f}_{s,\uit})\bdot \frac{\grad_\obv\of_s}{\of_s} 
 = O\left(\frac{q_s(\delta_\ell E)(\delta_\ell f_{s})(\delta_\uit f_s)}{m_s\uit f_s}\right),  
\lb{VI24} \ee
\be
\frac{q_s}{m_s c} \otau_\ell(\bB,\overline{f}_{s,\uit})\bdot \frac{(\obv\btimes\grad_\obv)\of_s}{\of_s}
= O\left(\frac{\ov\,q_s(\delta_\ell B)\,(\delta_\uit f_s)(\delta_\ell f_{s})}{c\,m_s \uit f_s} \right),  \lb{VI25} \ee 
\be
-\frac{q_s}{m_s c} \widehat{(\bw\btimes\bB)}_s \bdot\grad_\obv\of_s
=O\left(\frac{q_sB\,(\delta_\uit \! f_s)^2}{m_s c\, f_s}\right)
\lb{VI26} \ee
These estimates all hold as exact upper bounds.
One can already see from these estimates the possibility to have a non-vanishing 
entropy flux as $\ell,$ $\uit\to 0$, because the diverging factors $1/\ell,$ $1/\uit$
in \eqref{VI23}-\eqref{VI25} that arose from gradients in space and velocity may compensate 
for the vanishing increment factors. Note that there is an exact cancellation 
$\uit/\uit=1$ in the estimate \eqref{VI26}, which implies that there can be no such compensation 
for this particular term, which vanishes whenever the particle distributions $f_s,$ $s=1,...,S$ 
remain continuous as $Do\to\infty  .$ A persistent entropy flux in that limit is therefore expected 
to arise only from the first three contributions \eqref{VI23}-\eqref{VI25} to the modified entropy 
flux $\varsigma^{*flux,s}_{\ell,\uit}$ in \eqref{VI19}.} 

\magc{Each of the three contributions to entropy flux has a clear physical significance. The two terms 
\eqref{VI24}-\eqref{VI25} are entropy transfer due to nonlinear wave-particle interactions, arising 
from turbulent fluctuations of electric and magnetic fields, respectively. The term \eqref{VI23}
represents instead entropy transfer due to phase-mixing arising from linear advection. 
In the theory of Landau damping \cite{landau1946vibrations},  linear phase-mixing is well recognized 
as a mechanism that can transfer entropy to small scales in velocity-space, both in the physics 
\cite{krommes1994role, krommes1999thermostatted,zocco2011reduced,kanekar2015fluctuation} 
and mathematics (\cite{mouhot2011landau}, section 2.7) literatures. To be clear, there is no 
Onsager-type ``entropy dissipation anomaly'' in traditional Landau damping with an initially smooth, 
decaying perturbation of a Vlasov-Maxwell equilibrium, which is an entropy-conserving process. 
Because the particle distribution remains smooth (but with linearly 
growing velocity-gradients), the flux of entropy vanishes at sufficiently small scales in velocity-space.
In a forced, steady-state, on the other hand, the phase-mixing mechanism can  
produce an entropy cascade to arbitrarily small scales
\cite{krommes1994role, krommes1999thermostatted,zocco2011reduced,kanekar2015fluctuation}, 
but this requires an extremely singular particle distribution. In fact, if we impose the gyrokinetic 
relation $\ell/\uprho_s\sim \uit/v_{th,s},$ we see from our eq. \eqref{VI23} that the linear-advection 
contribution to entropy flux is bounded by $(\delta_\uit f_s)^2/f_s$ and hence vanishes as $\uit\to 0,$
whenever the distribution function $f_s$ remains continuous or even square-integrable 
(see footnote [76]) in the collisionless limit. This general result agrees with the linear kinetic 
model calculation in 
\cite{kanekar2015fluctuation}, eq.(4.25) showing that total ``free-energy'' diverges in the limit of   
vanishing collisional damping \footnote{\magc{Note, in fact, that the $m_{||}^{-1/2}$ spectrum 
of parallel Hermite modes derived in \cite{zocco2011reduced,kanekar2015fluctuation} corresponds 
to exponent $\rho_2^{f_s}=-1/2$ as defined in our Eq.\eqref{VI33}. See e.g. \cite{mhaskar2017local}}}.  
Our eq.\eqref{VI23} implies that in nonlinear kinetic turbulence, where particle-distributions are 
expected to remain even H\"older continuous, the linear advection contribution to entropy  
flux will generally be sub-dominant compared to the wave-particle interaction contributions
\eqref{VI24}-\eqref{VI25}, although this conclusion obviously can depend upon the arbitrary 
relation \eqref{V6c} which is adopted between scales $\ell,$ $\uit.$}

As cautioned earlier, the phase-space coarse-graining averages \black{involved in the definitions 
of the four terms in \eqref{VI15}} may involve substantial cancellations. \black{Furthermore, the 
four individual terms are all quantities of indefinite sign---although non-negative when summed together
and averaged---so that additional cancellations will certainly occur in integrating these over phase space.} The bounds 
\eqref{VI23}-\eqref{VI26} on the entropy flux contributions derived above may therefore be considerable 
overestimates. As we shall see in our discussion of the gyrokinetic predictions in sections \ref{sec:VIIIa}-\ref{sec:VIIIb}, there are reasons 
to expect extensive cancellations indeed will occur, which are missed by the above rather 
crude upper bounds.  Despite their giving only upper bounds, the estimates \eqref{VI23}-\eqref{VI26} 
nevertheless suffice to derive non-trivial exact constraints on scaling properties of turbulent 
solutions in order to be consistent with a non-vanishing entropy flux to small scales in phase-space. 

\subsection{Scaling Exponent Constraints}\label{sec:VIc}

The scaling exponents that we discuss are those which appear in the {\it structure functions} 
of (absolute) increments of phase-space variables $a(\bx,\bv),$ which are defined similarly as for the 
hydrodynamic case (\cite{eyink2018review},Eq.(IV.5)), by 
\be S_p^a(\br):=\langle |\delta_\br a|^p\rangle, \quad 
   R_p^a(\bw):=\langle |\delta_\bw a|^p\rangle. \lb{VI27} \ee 
Here the notation $\langle \cdot\rangle $ stands for a local average over some bounded open region $O$
in phase-space, that is,  
\be \langle a\rangle := \frac{1}{|O|}\iint_O d^3x\,d^3v \ a(\bx,\bv) \lb{VI28}\ee
where $|O|$ is the phase-volume of the region $O.$ The average depends, of course, on the 
particular region which is selected. This may be the entire region of phase-space where 
entropy cascade occurs if that has finite phase-volume \footnote{We always assume that the 
only velocities which occur in the solutions $f_s,$ $s=1,...,S$ of the VLM equations have $|\bv|<c,$ 
the speed of light, or otherwise the semi-relativistic model \eqref{III1}-\eqref{III3} becomes 
unphysical and must be replaced by a fully relativistic Vlasov-Maxwell model. We also assume that the 
turbulence occurs in only a bounded region of position-space.} or any bounded, open subregion. 
Our results shall give local conditions for entropy cascade to occur within any chosen such domain 
of phase-space. Note that the structure-functions defined by \eqref{VI27} are directly related to 
local $L_p$-norms in phase-space:
\be S_p^a(\br)=\|\delta_\br a\|_p^p, \quad R_p^a(\bw)=\|\delta_\bw a \|_p^p \lb{VI29}\ee
See e.g. \cite{kolmogorov1975introductory}. Basic properties of such $L_p$-norms will be 
our main analytical tools, in particular the well-known H\"older inequality and 
also the nesting property of the norms, or $\|a\|_p\leq \|a\|_{p'}$ for $p'\geq p.$ Note that we may consider
the above structure-functions as well for variables $b$ that are functions of $\bx$ only, or variables $c$ that 
are functions of $\bv$ only. If the region considered has product form $O=O_x\times O_v$ for 
bounded open subsets $O_x$ and $O_v$ of position-space and velocity-space, respectively, then the
local phase-space structure functions reduce to the corresponding (local) structure-functions in 
position-space or velocity-space. 

We seek conditions that must hold in order for there to be constant entropy flux 
as in \eqref{VI10}, that is, for space-average \black{$\langle \sigma_{\ell,\uit}^{*flux}\rangle
=\langle\sigma_\star\rangle$} in a range of scales $\ell,$ $\uit,$ which extends down 
to $\ell,\uit=0$ for $Do\to\infty.$ In light of \eqref{VI4}, this can occur only  
if for some region $O$ and some $s$\black{
\be \lim_{\ell,\uit\to 0}\langle \varsigma^{*flux,s}_{\ell,\uit} \rangle\neq 0 \lb{VI30} \ee}
As we show now, this condition imposes constraints on the structure-function {\it scaling exponents} 
$\zeta_p^E,$ $\zeta_p^B,$ $\zeta_p^{f_s},$ $\xi_p^{f_s},$ $s=1,...,S$
of the solution variables $a=\bE,$ $\bB,$ $f_s,$ $s=1,...,S.$ For any such variable $a$, we can define 
the exponents by assuming that scaling laws hold of the form 
\be S_p^a(\br)\sim   C_p a_{rms}^p \left(\frac{|\br|}{L_a}\right)^{\zeta^a_p} , \,\,\,
   R_p^a(\bw)\sim D_p a_{rms}^p \left(\frac{|\bw|}{V_a}\right)^{\xi^a_p}
\lb{VI31} \ee
for $|\br|\sim \ell,$ $|\bw|\sim \uit$ in the range of $\ell,$ $\uit$ where non-vanishing 
flux condition \eqref{VI30} holds. Equivalently, and somewhat more conveniently,  we may 
discuss exponents $\sigma_p^E,$ $\sigma_p^B,$ $\sigma_p^{f_s},$ $\rho_p^{f_s},$ $s=1,...,S$
defined by the scaling laws
\be \!\! \|\delta_\br a\|_p\sim   C_p^{1/p} a_{rms} {\left(\frac{|\br|}{L_a}\right),}^{\!\!\!\!\!\!\sigma_p^a} \,\,
   \|\delta_\bw a \|_p \sim D_p^{1/p} a_{rms} \left(\frac{|\bw|}{V_a}\right)^{\!\!\rho_p^a} 
   \lb{VI32} \ee
with $\sigma_p^a=\zeta_p^a/p$ and $\rho_p^a=\xi_p^a/p.$ Although it is natural to assume 
that scaling laws such as \eqref{VI31} or \eqref{VI32} will hold, this assumption is 
not necessary.  If the infinite-$Do$ limit variable $a_\star$ exists and its $p$th-order 
moments $\langle |a_\star|^p\rangle$ are finite, then we can instead take 
\be \sigma_p^{a}=\liminf_{|\br|\to 0}\frac{\log\|\delta_\br a_\star \|_p}{\log |\br|}, \quad
\rho_p^{a}=\liminf_{|\bw|\to 0}\frac{\log\|\delta_\bw a_\star\|_p}{\log |\bw|} \lb{VI33}\ee
where the limit-infimum is guaranteed to exist. The exponents defined by \eqref{VI33} 
coincide with those given by the scaling laws \eqref{VI31} or \eqref{VI32}, whenever 
the latter hold. Otherwise, $\sigma_p^{a}$ and $\rho_p^{a}$ give the (fractional) smoothness 
in position and velocity, respectively, of the phase-space variable $a$ in $L_p$-mean sense, 
or the maximal ``Besov exponents''. See \cite{eyink1995besov,eyink2010notes}
\footnote{One can avoid any assumption of the existence of an infinite-$Do$ limit $a_\star$ by 
defining the $p$th-order exponent of $a_{{\,\!}_{Do}}$ by 
$\sigma_p^a=\sup\{s:\, \sup_{Do}\|a_{{\,\!}_{Do}}\|_{B^{s,\infty}_p}<\infty$\}, which is the 
largest smoothness exponent $s$ for which  $a_{{\,\!}_{Do}}\in B^{s,\infty}_p,$ uniformly in $Do.$
See \cite{drivas2018onsagerincompressible} for an analogous definition in the context 
of the incompressible Navier-Stokes turbulence}

We now show that non-smoothness or ``roughness'' of the solutions $\bE,$ $\bB,$ $f_s,$ $s=1,...,S$
is required in order to permit a non-vanishing flux as in \eqref{VI30}. For this, it is enough to 
obtain bounds on the norms\black{
\be \|\varsigma_{\ell,\uit}^{*flux,s}\|_1 \leq \|\varsigma_{\ell,\uit}^{*flux,s}\|_{p/3}, \quad p\geq 3, 
\lb{VI34} \ee}
\hspace{-10pt} that vanish if the solutions are too smooth. 
By the triangle-inequality we need bounds on the $L_{p/3}$-norms of the three contributions
to entropy flux in \eqref{VI16}-\eqref{VI18} (noting that the fourth contribution \eqref{VI19}
to flux will always vanish as $\ell,$ $\uit\to 0$ when $p$th-moments of $\bB$ and $f_s$ are finite). 
Simple applications of the nesting property and the H\"older inequality give 
%- \hbw_s\bdot\grad_\obx\of_s  = O\left(\frac{\uit\,(\delta_\uit f_s) (\delta_\ell\! f_s)}{\ell f_s}\right),  
%\lb{VI23} \ee
%\be
%-\frac{q_s}{m_s}\otau_\ell(\bE,\overline{f}_{s,\uit})\bdot \frac{\grad_\obv\of_s}{\of_s} 
% = O\left(\frac{q_s(\delta_\ell E)(\delta_\ell f_{s})(\delta_\uit f_s)}{m_s\uit f_s}\right),  
%\lb{VI24} \ee
%\be
%\frac{q_s}{m_s c} \otau_\ell(\bB,\overline{f}_{s,\uit})\bdot \frac{(\obv\btimes\grad_\obv)\of_s}{\of_s}
\black{
\be
\left\| \hbw_s\bdot\grad_\obx\of_s \right\|_{p/3} 
=O\left(\frac{\uit\,\|\delta_\uit f_s\|_p \,\|\delta_\ell\! f_s\|_p}{\ell \min\{f_s\}}\right),  
\lb{VI35}\ee
\bea
&& \left\|\frac{q_s}{m_s}\otau_\ell(\bE,\overline{f}_{s,\uit})\bdot \frac{\grad_\obv\of_s}{\of_s}\right\|_{p/3} \cr 
&& \hspace{80pt} = O\left(\frac{q_s\|\delta_\ell E\|_p\, \|\delta_\ell f_{s}\|_p\,\|\delta_\uit f_s\|_p}{m_s\uit \min\{f_s\}}\right),  \cr
&& \lb{VI36} \eea
\bea 
&& \left\|\frac{q_s}{m_s c} \otau_\ell(\bB,\overline{f}_{s,\uit})\bdot \frac{(\obv\btimes\grad_\obv)\of_s}{\of_s}\right\|_{p/3} \cr
&& \hspace{60pt} =O\left( \frac{\max\{\overline{v}\}q_s\|\delta_\ell B\|_p\,\|\delta_\ell f_{s}\|_p\,\|\delta_\uit f_s\|_p}{c\, m_s \uit \min\{f_s\}} \right).  \cr 
&&
\lb{VI37}\eea} 
Here we defined
\be 
\|\delta_\ell f_{s}\|_p:=\sup_{|\br|<\ell}\|\delta_\br f_s\|_p, \quad 
\|\delta_\uit f_{s}\|_p:=\sup_{|\bw|<\uit}\|\delta_\bw f_s\|_p.  \lb{VI38}\ee  
%and likewise 
%\bea 
%\|\delta_\uit\delta_\ell f_{s}\|_p:&=&\sup_{|\br|<\ell,|\bw|<\uit}\|\delta_\br\delta_\bw f_s\|_p\cr
%&=&O\left(\min\{\|\delta_\ell f_s\|_p,\|\delta_\uit f_s\|_p\} \right)\lb{VI39}\eea 
We have also assumed strict positivity of the distribution, or $\min\{f_s\}=\min_{(\bx,\bv)\in O} f_s(\bx,\bv,t)>0,$
which means that there are no ``perfect holes'' in the distribution function of species $s$ 
where $f_s=0.$ This does not, of course, rule out conventional phase-space holes 
where the density $f_s$ becomes much smaller than the density in surrounding regions 
but remains non-zero \cite{hutchinson2017electron}. 

%Because of the bound \eqref{VI39}, we see that the second terms in the estimates 
%\eqref{VI35}-\eqref{VI37} are always subleading relative to the first terms
%footnote{To avoid misunderstanding, this does {\it not} mean that the cubic terms in the various flux 
%contributions are necessarily negligible as $\ell,$ $\uit\to 0$ compared with the quadratic terms. 
%We are only making a statement here about the upper bounds on the two terms}, so that
%\be \|\grad_\obx\bdot\hbw_s\, \of_s\|_{p/3} =O(\uit\,\|\delta_\uit \delta_\ell\! f_s\|_p/\ell) \lb{VI40}\ee
%\bea
%&& \max\{\|\grad_\obv\bdot\hbE_s\,\of_s\|_{p/3},\|(\obv/c)\bdot\grad_\obv\btimes\hbB_s\,\of_s\|_{p/3}\}\cr
%&& \hspace{60pt} =O\left (\max\{\|\delta_\ell E\|_p,\|\delta_\ell B\|_p\}\,\|\delta_\uit\delta_\ell f_{s}\|_p/\uit \right)\cr
%&& 
%\lb{VI41}\eea
%We have gathered together in \eqref{VI41} the two contributions from the electromagnetic field,
%since they have a very similar character. 
We now try to get the tightest bound on the entropy 
flux by minimizing the sum of the bound \eqref{VI35} \black{on the advective phase-mixing contribution and
the bound on the total field-particle interaction contribution 
\bea
&& \left\|\frac{q_s}{m_s}\left[\otau_\ell(\bE,\overline{f}_{s,\uit})
+\wtc \obv\btimes \otau_\ell(\bB,\overline{f}_{s,\uit})\right]\bdot \frac{\grad_\obv\of_s}{\of_s}\right\|_{p/3} \cr 
&& \hspace{30pt} = O\left(\frac{q_s\max\{\|\delta_\ell E\|_p,\delta_\ell B\|_p\}\,\|\delta_\uit f_s\|_p\, \|\delta_\ell f_{s}\|_p}{m_s\uit \min\{f_s\}}\right),  \cr
&& \lb{VI41} \eea
obtained by combining estimates \eqref{VI36},\eqref{VI37} and by noting that $\max\{\overline{v}\}\leq c.$} 
As we have emphasized throughout this work, there is complete freedom in choosing the two scales $\ell,$ $\uit,$ 
as long as they are sufficiently small. They represent an arbitrary choice of resolution of the 
turbulent cascade process. Hence, we can exploit this arbitrariness and choose $\uit$ to be 
the value which minimizes the sum of the bounds \black{\eqref{VI35}} and \eqref{VI41}, with $\ell$ fixed. 
Elementary calculus gives 
\be 
\uit = \left[ \ell \max\{\|\delta_\ell E\|_p,\|\delta_\ell B\|_p\}\right]^{1/2}=O\left( \ell^{(\sigma_p^F+1)/2}  \right), 
\lb{VI42}\ee 
which also concides with the choice of $\uit$ for which the two bounds \black{\eqref{VI35}} and \eqref{VI41}
are ``balanced'' or have comparable magnitudes. In \eqref{VI42} we have introduced the exponent 
$\sigma_p^F=\min\{\sigma_p^E,\sigma_p^B\}$ which gives the minimal $p$th-order smoothness of the 
electromagnetic field. Putting together all of the previous estimates, then for the choice of $\uit$ 
determined by \eqref{VI42} we have \black{
\bea  
&& \|\varsigma_{\ell,\uit}^{flux,s}\|_{p/3}
= O\left(\frac{\uit}{\ell}\,\|\delta_\ell\! f_s\|_p\, \|\delta_\uit\! f_s\|_p\right)\cr
&& =O\left( \ell^{(\sigma_p^F-1)/2}\cdot \ell^{\sigma_p^{f_s}}\cdot \ell^{\rho_p^{f_s}(\sigma_p^F+1)/2}\right).
\lb{VI43}\eea}
\hspace{-7pt} Clearly the upper bound \eqref{VI43} for $p\geq 3$ will vanish as $\ell,\uit\to 0$ if \black{
\be \frac{1}{2}(\sigma_p^F-1)+\sigma_p^{f_s}+\frac{1}{2}\rho_p^{f_s}(\sigma_p^F+1)> 0. \lb{VI44}\ee} 
We thus arrive at the exponent inequality \black{
\be \sigma_p^F+2\sigma_p^{f_s}+\rho_p^{f_s}(\sigma_p^F+1)\leq 1, \quad p\geq 3, \lb{VI45}\ee} 
\hspace{-7pt} as a necessary condition for non-vanishing entropy cascade to small scales in phase-space. 

If we assume, for simplicity, that $\sigma_p^F=\sigma_p^{f_s}=\rho_p^{f_s}=\sigma_p$ \black{for all fields,  
with some single $\sigma_p$ (``uni-scaling"),} then the above inequality \eqref{VI45} requires that 
\black{$4\sigma_p+\sigma_p^2\leq 1$ or $\sigma_p\leq \sigma_{cr}=\sqrt{5}-2\doteq 0.2361$.}
as the condition for non-vanishing entropy cascade. This result must not be interpreted as 
a prediction that the \black{``mean-field'' value $\sigma_{cr}\doteq 0.2361$} will be the scaling that physically 
occurs. Our result \eqref{VI45} should be compared with the inequality for velocity scaling exponents $\zeta^u_p\leq p/3$ 
or $\sigma_p^u\leq 1/3$ when $p\geq 3,$ which was first derived by Constantin et al. \cite{constantin1994onsager}
(see also \cite{eyink1995local,eyink2018review}) as a necessary condition for kinetic energy 
cascade in incompressible fluid turbulence. Empirical results from experiments and simulations 
in that case indicate that $\sigma_3^u\doteq 1/3$ (just slightly smaller) but that 
$\sigma_p^u$ for $p\gg 3$ is considerably smaller than the Kolmogorov value 1/3. This is due to 
the effect of ``intermittency'' in which the energy cascade rate becomes strongly fluctuating in space 
and time \cite{uriel1995turbulence,eyink2018review}. For very large $p$ values the scaling 
of velocity structure functions is determined by more singular structures with $\sigma_p^u$ much less
than $1/3$. However, these singular structures are also more sporadic and 
thus contribute relatively little to energy cascade.  There are presumably similar phase-space 
intermittency effects in the entropy cascade of kinetic plasma turbulence, \black{
e.g. associated to sheets of strong electric current density \cite{chasapis2018situ}}. Thus,  our 
exponent inequality \eqref{VI45} is probably far from equality for $p\gg 3.$

In gyrokinetic turbulence, we expect that even for $p$ near 3 the physically observed exponents 
$\sigma_p^E,$ $\sigma_p^B,$ $\sigma_p^{f_s},$ $\rho_p^{f_s}$, $s=1,...,S$ will 
satisfy the bound \eqref{VI45} as an inequality, with a sizable gap, rather than as an equality.
As we shall see in section \ref{sec:VIIIa}, the gyrokinetic predictions for scaling exponents in various entropy cascade 
ranges satisfy our bound \eqref{VI45} easily, with a considerable gap. This should be expected, 
because our estimates take into account no physical effects of plasma wave oscillations or 
fast particle gyrations which could lead to strong depletion of nonlinearity. For example, in weak 
wave turbulence, rapid wave oscillations are known to cancel completely all nonlinear
wave interactions except those with resonant wave frequencies 
\cite{zakharov1992kolmogorov,nazarenko2011wave}.  
In general, effects of wave oscillations or particle gyrations will lead 
to large cancellations in the exact expression \black{\eqref{VI15}} for entropy flux,
so that the upper bounds \black{\eqref{VI23}-\eqref{VI26}} will be large overestimates. Because 
of the depletion of nonlinearity, more singular structures must develop to support the 
entropy cascade and the physically occurring exponents will not yield an equality in 
our condition \eqref{VI45}. For the same reason, our equation \eqref{VI42} cannot 
be regarded as a physical relation between position and velocity scales $\ell,$ $\uit$ in 
a gyrokinetic entropy cascade range 
\footnote{\black{A more plausible candidate for such a physical relation is the 
scaling law $\uit/v_{th,s}\sim\ell/\rho_s$ proposed by 
\cite{schekochihin2008gyrokinetic,schekochihin2009astrophysical} within a gyrokinetic description.
However, any such relation employed within our analysis will always yield an exponent inequality 
that is less stringent than our bound \eqref{VI45}. For example, assuming that $\uit\sim\Omega_s \ell,$
the bound \eqref{VI35} implies that the linear phase-mixing contribution to entropy flux 
vanishes asymptotically for both $\ell,$ $\uit\to 0$ whenever the particle distributions have positive 
$p$th-order scaling exponents. In that case, only the nonlinear wave-particle interaction term 
can supply a non-vanishing entropy flux and the  bound \eqref{VI41} yields an exponent 
inequality $\sigma_p^F+2\sigma_p^{f_s}\leq 1$ for $p\geq 3.$ This inequality is clearly less restrictive
than \eqref{VI45}. The picture obtained by assuming $\uit\sim\Omega_s \ell,$ on the other 
hand, may be more physically relevant, even though the exponent inequalities that it yields 
are sub-optimal}}. As we discuss in section \ref{sec:VIIIa}, further analytical progress 
on gyrokinetic turbulence will require the control of delicate cancellations in \black{\eqref{VI15}}, 
our exact ``4/5th-law'' expressions for entropy flux. 

In summary, our analysis shows that the solutions $\bE,$ $\bB,$ $f_s,$ $s=1,...,S$ of the VML 
equations cannot remain smooth if there is persistent entropy production in the limit 
$Do\to\infty  .$ In fact, the solutions cannot have even a fractional smoothness which remains 
too high, or else entropy cascade is not possible. It is important to emphasize that the singularities  
that are required by our analysis need not develop in finite time from smooth Vlasov-Maxwell solutions 
with regular initial data. This is obvious for the collisionless limit of long-time steady-states as 
first considered by Krommes \& Hu \cite{krommes1994role,krommes1999thermostatted}, 
which corresponds to the limit first $t\to\infty$ and then $Do\to\infty.$ In this limit, phase-space 
mixing by ballistic streaming or other mechanisms has an infinite time to create fine 
structure down to collisional scales, and only subsequently are the collisional 
scales taken to zero. In freely-decaying turbulence without external forcing, 
singularities may be input as initial data, e.g. the solar wind originating 
in the superheated corona might have pre-existing turbulent fluctuations at all scales 
down to the Debye length. If smooth solutions of the collisionless Vlasov-Maxwell equations can indeed 
blow up in finite-time, then this would provide an additional source of singularities.  It is still unknown
whether initially smooth solutions of the (semi-relativistic) system \eqref{III1}-\eqref{III3} 
at vanishing collisionality will remain smooth, although it is known that any singularity 
formation requires particles moving with velocities near light-speed (see \cite{glassey1986singularity}, 
Proposition 9).  

\magc{More directly relevant for kinetic turbulence are theorems on the regularity of 
weak solutions of the Vlasov-Maxwell equations. The current best results seem to 
be those of \cite{besse2018regularity} for the DiPerna-Lions weak solutions of the (relativistic) 
Vlasov-Maxwell system, under an assumption that the particle energy densities $E_s(\bx,t)$ are 
square-integrable functions. By an application of averaging lemmas \cite{diperna1989global} and 
``non-resonant smoothing'' for particles with velocities bounded away from light-speed
\cite{bouchut2004nonresonant}, the latter paper proves that electromagnetic fields have regularity 
exponent $\sigma_2^F>6/(14+\sqrt{142})\doteq 0.2315.$ This value is remarkably close numerically to 
the critical value $\sigma_{cr}=\sqrt{5}-2\doteq 0.2361$ for non-vanishing entropy cascade, which we have shown 
to require $\sigma_p\leq \sigma_{cr}$ for $p\geq 3,$ under the additional assumption that all solution fields 
scale with the same exponent. Of course, there is no reason that such ``uni-scaling'' must hold and, 
even if it does, intermittency of the cascade could allow $\sigma_2^F>\sigma_{cr}.$   However,
the above numerical coincidence does show that monofractal (non-intermittent), uni-scaling 
solutions of the Vlasov-Maxwell equations with non-vanishing entropy production can exist 
in a narrow range only (if at all). Further conditional regularity results along the lines of 
\cite{bouchut2004nonresonant,besse2018regularity} would be very valuable,  for example, 
assuming some regularity exponents $\sigma_p^{f_s},\rho_p^{f_s}$ of particle distributions 
and deriving corresponding minimal regularity exponents $\sigma_p^F$ of the electromagnetic fields.  
Such results would cast considerable light on the range of scaling exponents allowed for the 
dissipative weak solutions of Vlasov-Maxwell equations hypothesized in this work. 
}

%Version of the argument with $Do$ very large but finite; Appendix. 

\section{Balances of Conserved Quantities in the Collisionless Limit}\lb{sec:VII} 

In this section we discuss the collisionless limit dynamics of quantities conserved 
for the total system (particles + fields) governed by the VML equations \eqref{III1}-\eqref{III3},
namely, the mass of each particle species, the total momentum, and the total energy. 
Since these quantities are absolutely conserved for any degree of collisionality, the weak solutions
of the VM equations \eqref{V7} obtained in the limit $Do\to\infty$ cannot develop any 
anomalies in the balances of these quantities of the same sort as the entropy-production anomaly 
\eqref{VI13}. On the other hand, there are collisional conversions of one form of these conserved 
quantities into other forms and these conversion terms may, in principle, remain non-zero 
and ``anomalous'' as $Do\to\infty.$ Such a situation occurs in the infinite Reynolds-number limit 
of compressible fluids, for example, where  total energy (kinetic + internal) is conserved 
but energy cascade leads to anomalous conversion of kinetic energy into internal energy 
\cite{aluie2013scale,eyink2017cascades1}. We show here that such anomalous conversion does 
{\it not} occur in kinetic turbulence of nearly collisionless plasmas and that all collisional conversion 
terms vanish in the limit $Do\to\infty,$ under reasonable assumptions. We establish this  both 
from the fine-grained point of view and in the coarse-grained description with finite resolutions 
$\ell,$ $\uit$ in position- and velocity-space. 

The results of the present section confirm naive expectations on the collisionless limit, 
while taking into account non-differentiability of limiting solutions. Results that are 
less expected can emerge, however, when one considers subsequent 
limits such as $\uprho_i/L_i\ll1$ (well-satisfied in the solar wind) and $\uprho_e/\uprho_i\ll 1$
(marginally satisfied in the solar wind), where $\uprho_i$ and $\uprho_e$ are ion and electron gyroradii,
respectively. In these secondary limits, anomalies by energy 
cascade through scales or anomalous conversion between different forms of energy may appear which
are described by the scale-resolved energy balance in phase-space that we derive below.
Likewise, the coarse-grained balance of electron momentum that we derive 
is the generalized Ohm's law valid in a turbulent plasma at a given length-scale,
which can lead to anomalous breakdown of magnetic flux-conservation and of the ``frozen-in'' property 
of field-lines \cite{eyink2006breakdown,eyink2015turbulent}. 

The limits $\uprho_i/L_i\ll1$ and $\uprho_e/\uprho_i\ll 1$ mentioned above 
have been discussed for a turbulent plasma generally within a gyrokinetic description, which becomes 
valid for gyrofrequencies much larger than rates of change of resolved scales 
\cite{schekochihin2008gyrokinetic,schekochihin2009astrophysical}. 
 In these gyrokinetic analyses,  energy and entropy balances are intertwined,  whereas in the full kinetic 
 description by VML equations their balance equations are completely separate in general. Nevertheless, 
 our coarse-graining in phase-space provides a regularization of short-distance divergences that 
can appear in these subsidiary limits and it thus provides a suitable non-perturbative tool for analysis 
of gyrokinetic turbulence. We shall discuss gyrokinetics briefly in the following section, after we derive 
the collisionless limit of the basic conservation laws here. 

\subsection{Mass Balances}\lb{sec:VIIa} 

Since we have assumed that collisions do not transform one particle species into another, 
there is no contribution from the collision integral to fine-grained mass balances \eqref{III10}.   
Assuming that strong limits of VML solutions exist as $Do\to\infty,$ the distributional 
mass balance equations $\partial_t\rho_{\star s}+\grad\bdot (\rho_{\star s}\bu_{\star s})=0$ hold
as a direct limit of \eqref{III10}. This same result may be obtained by 
integrating over $\bv$ the weak Vlasov equation \eqref{V7} for the limiting particle distribution 
$f_{\star s}$. 

The coarse-grained mass balance at length-scale $\ell$ for each particle species $s,$
\be \partial_t\orhos+\grad\bdot \overline{(\rho_s\bu_s)}=0, \lb{VII1} \ee
can be easily derived, either by coarse-graining the fine-grained balance \eqref{III10}  
or by integrating the coarse-grained Vlasov equation \eqref{V9} over $\obv$  
and using $\int d^3\ov \ \hbvs\, \ofs=\int d^3\ov \ \overline{\bv\, f_s}= \overline{\rho_s\bu_s}.$ 
In terms of spatial Favre averages, this can be written as: 
\be \partial_t\orhos+\grad\bdot(\orhos\tbus)=0 \lb{VII2} \ee
This is the same equation which holds for coarse-grained mass densities 
in compressible fluid theories \cite{aluie2013scale,eyink2017cascades1}. 

\subsection{Momentum Balances}\lb{sec:VIIb} 

We now derive the momentum balances that hold in the collisionless limit $Do\to\infty  .$ 
The total momentum density $\sum_s \rho_s\bu_s+ (1/4\pi c)\bE\btimes\bB$
of particles and fields satisfies a local conservation law for any degree of collisionality, 
so that it is not possible to have a ``dissipative anomaly'' of total momentum. 
However, it is possible, in principle, that collisional momentum transfers between 
different particle species might remain non-vanishing due to the divergence of 
velocity-gradients in the limit. We show that this does not happen under mild conditions.

\subsubsection{Fine-Grained Momentum Balances}\lb{sec:VIIb1} 

The drag force on species $s$ from collisions with species $s'$ can be estimated 
for the Landau collision integral \eqref{III9} by using integration by parts and the 
Cauchy-Schwartz inequality, in a similar fashion as for the estimation of $\overline{C}_{ss'}$
in eqs.\eqref{V2}-\eqref{V5}:
\bea 
&& \bR_{ss'}:=\int d^3v\ m_s \bv\, C_{ss'}\cr
&& = -\Gamma_{ss'}
\int d^3v \int d^3v' \frac{\bPi_{\bv-\bv'}}{|\bv-\bv'|}\bdot\left(\grad_\bp-\grad_{\bp'}\right)(f_s f_{s'})\cr
&& = -\Gamma_{ss'}\int d^3v  \int d^3v'  \left(\frac{f_sf_{s'}}{|\bv-\bv'|}\right)^{1/2}\cr
&& \hspace{25pt} \times \frac{\bPi_{\bv-\bv'}}{(f_sf_{s'}|\bv-\bv'|)^{1/2}}
\left(\grad_\bp-\grad_{\bp'}\right)(f_sf_{s'})
\lb{VII3} \eea
so that 
\bea 
&& |\bR_{ss'}(\bx,t)|\leq  \sqrt{\Gamma_{ss'}} \times\cr 
&& \sqrt{\int d^3v  \int d^3v'  \,\frac{f_sf_{s'}}{|\bv-\bv'|}}\times \cr
&& \sqrt{\Gamma_{ss'}\int d^3v  \int d^3v'  \,  \frac{|\bPi_{\bv-\bv'}\left(\grad_\bp-\grad_{\bp'}\right)(f_sf_{s'})|^2}{f_sf_{s'}|\bv-\bv'|}}\cr
&& \hspace{60pt} \leq C\sqrt{\Gamma_{ss'}\sigma(\bx,t)}
\lb{VII4} \eea
As shown in Appendix \ref{app:A2}, the integral under the first square-root factor remains finite as $Do\to\infty $
under very mild assumptions on the particle distribution functions.  
The integral under the second square-root is $\sigma(\bx,t)$ as defined in \eqref{III27}
and, invoking the hypothesis \eqref{III38} on the entropy production in 2-particle phase-space, 
this quantity remains finite pointwise in $(\bx,t)$ as $Do\to\infty  .$ Thus, the collisional drag force $\bR_{ss'}$
vanishes $\propto\sqrt{\Gamma_{ss'}}$ for all $s,s'$ in the collisionless limit. Assuming that a suitable strong limit exists 
$f_s,\,\bE,\, \bB\, \to \, f_{\star\,s},\bE_\star,\bB_\star$ as $Do\to \infty,$ which thus 
satisfies the Vlasov-Maxwell equations \eqref{V7}, then the fine-grained momentum balance for species $s$ in that 
limit solution becomes 
\be \partial_t (\rho_{\star s}\bu_{\star s}) +\grad_\bx\bdot(\rho_{\star s}\bu_{\star s}\bu_{\star s}
+\bP_{\star s}) = q_s n_{\star s} (\bE_\star)_{*s}. \lb{VII5} \ee
This is just the result that would be naively expected in the collisionless limit, with all
interspecies momentum transfer due to collisionless wave-particle interactions. 

\subsubsection{Coarse-Grained Momentum Balances}\lb{sec:VIIb2} 

A phase-space momentum balance at fixed resolutions $\ell,$ $\uit$ can be obtained 
by multiplying the coarse-grained kinetic equation \eqref{V1} with $\obv$ to obtain 
\bea 
&& \partial_t(m_s\obv\ofs)+\grad_\obx\bdot(m_s\hbvs\obv\ofs)+\grad_\obp\bdot(m_sq_s\hbE_{*s}\obv\ofs)\cr
&& \hspace{80pt} =q_s\hbE_{*s}\ofs +m_s\obv \overline{C}_s(f). \lb{VII6} \eea 
In the limit as $Do\to\infty$ recall from \eqref{V5} that $\overline{C}_s(f)\to 0$ pointwise in phase-space,
so that one may neglect the final term in the nearly collisionless limit for fixed $\ell,$ $\uit.$
By integrating \eqref{VII6} over velocities, it follows that 
\be \partial_t\overline{(\rho_s\bu_s)}+\grad\bdot\overline{(\rho\bu_s\bu_s+\bP_s)}=q_s\overline{(n_s\bE_{*s})}, 
\lb{VII7} \ee
for any fixed $\ell,$ $\uit$ and sufficiently large $Do.$ Here we have used the fact that the coarse-grained 
drag force $\bar{\bR}_s=\int d^3\ov\ \obv \,\overline{C}_s(f)\to 0$ 
in the limit as $Do\to \infty,$ assuming some uniform integrability in velocity of $\obv \,\overline{C}_s(f)$. 
In the idealized limit $Do\to\infty$ at fixed
$\ell$ one therefore obtains 
\be \partial_t \overline{(\rho_{\star s}\bu_{\star s})}
+\grad_\bx\bdot(\overline{\rho_{\star s} \bu_{\star s}\bu_{\star s}
+\bP_{\star s}}) = q_s \overline{n_{\star s} (\bE_\star)_{*s}}, \lb{VII8} \ee
a result consistent with \eqref{VII5} and which could also be obtained by coarse-graining 
that equation after first taking the collisionless limit.  The previous two equations can both be rewritten 
 in terms of spatial Favre averages, with \eqref{VII7}, for example, expressed equivalently as 
\be \partial_t(\orhos\tbus)+\grad\bdot(\orhos\tbus\tbus+\obP_s^*)=q_s\overline{n}_s\tbE_{*s} \lb{VII9} \ee
using the definitions \eqref{IV11} and \eqref{IV23}. These equations for $s=1,...,S$ fully 
specify the coarse-grained momentum balances of the particles in the collisionless limit. 

On the other hand, the momentum balance for the electromagnetic fields resolved to a spatial scale 
$\ell$ follows from the coarse-grained Maxwell equations \eqref{V6}: 
\bea
&& \partial_t\left(\frac{1}{4\pi c}\obE\btimes\obB\right)\cr
&&+\grad\bdot\left[\frac{1}{4\pi}\left(\obB\,\obB-\frac{1}{2}|\obB|^2\bI\right)
+\frac{1}{4\pi}\left(\obE\,\obE-\frac{1}{2}|\obE|^2\bI\right)\right] \cr
&& \hspace{60pt} = -\left(\overline{\varrho}\,\obE+\frac{1}{c}\obj\btimes\obB\right)
\lb{VII10} \eea
where the Lorentz reaction force on the righthand side acts as a source/sink of electromagnetic field
momentum.  It contains the coarse-grained charge and electric current densities, which are obtained from  
\be \overline{\varrho} = \sum_s q_s\overline{n}_s, \quad \obj= \sum_s q_s\overline{n}_s \tbus. 
\lb{VII11}\ee
An opposing Lorentz force is obtained by summing the righthand sides of \eqref{VII9} over 
$s=1,...,S,$ so that the coarse-grained balance of total momentum from \eqref{VII9},  
\eqref{VII10} becomes 
\bea
&& \partial_t\left(\sum_s \orhos\tbus+ \frac{1}{4\pi c}\obE\btimes\obB\right)\cr
&&+\grad\bdot\Big[ \ \left(\orhos\tbus\tbus+\bP_s^*\right) \cr
&& + \frac{1}{4\pi}\left(\obB\,\obB-\frac{1}{2}|\obB|^2\bI\right)
+\frac{1}{4\pi}\left(\obE\,\obE-\frac{1}{2}|\obE|^2\bI\right)\Big] \cr
&& \hspace{40pt} = \otau_\ell(\varrho,\bE)+(1/c)\otau_\ell(\bj \btimescom \bB), 
\lb{VII12} \eea
\black{where we use the rather obvious notation for the cross-product vector with $k$th
component $[\otau({\bf j}\btimescom {\bf B})]_k:=\epsilon_{klm}\otau({\rm j}_l,{\rm B}_m)$ and $\epsilon_{klm}$
the 3D completely antisymmetric Levi-Civita tensor.} 
Note, however, that the Lorentz force and its reaction force calculated from the coarse-grained 
Vlasov-Maxwell sytem \eqref{V6} do not exactly cancel and the total momentum at scales 
greater than $\ell$ is not exactly conserved! The righthand side of \eqref{VII12} represents 
a flux of momentum from unresolved scales $<\ell$ to resolved scales $>\ell.$ Since total 
momentum is exactly conserved for the Vlasov-Maxwell-Landau system \eqref{III1}-\eqref{III3} 
at any degree of collisionality, this ``momentum cascade'' must vanish as $\ell\to 0$ for a 
physical solution obtained in the limit $Do\to\infty.$ The estimates $\otau_\ell(\varrho,\bE)\sim (\delta_\ell \varrho)(\delta_\ell E),$ 
$\otau_\ell(\bj\btimescom\bB)\sim (\delta_\ell \dot{\j})(\delta_\ell B)$ following from \eqref{IV7} show that 
this flux of momentum will vanish as $\ell\to 0$ whenever limits $\varrho_\star,$ $\bj_\star,$ $\bE_\star,$ $\bB_\star$ 
remain spatially continuous or even when the limits satisfy weaker conditions that imply 
vanishing of the increments in a spatial-mean sense \footnote{For example, if the electromagnetic 
fields satisfy the finite-energy condition in \eqref{V8} and if also the particle distributions satisfy 
$\int d^3v \,(1+|\bv|)|\!|f_s(\cdot,\bv,t)|\!|_2<\infty$ for $s=1,...,S,$ where $|\!|\cdot |\!|_2$ is the spatial 
$L^2$-norm, then the Cauchy-Schwartz inequality $|\otau_\ell(\varrho,\bE)|\leq 2 \int d^3r\, G_\ell(\br) 
|\!|\delta_\br\varrho |\!|_2|\!|\delta_\br\bE |\!|_2$ and likewise
$|\otau_\ell(\bj\btimescom \bB)|\leq 2 \int d^3r\, G_\ell(\br) 
|\!|\delta_\br\bj |\!|_2|\!|\delta_\br\bB |\!|_2$ imply that these momentum fluxes vanish 
as $\ell\to 0.$ This follows from square-integrability of $\varrho,$ $\bj,$ which is a consequence 
of the bound $|\!|\varrho |\!|_2+|\!|\bj|\!|_2\leq \sum_s \int d^3v \,(1+|\bv|)|\!|f_s(\cdot,\bv,t)|\!|_2.$

$\quad$\magc{The previous considerations apply to any weak solution of the Vlasov-Maxwell equations. 
Of course, if the solution $f_{\star s},$ $\bE_\star,$ $\bB_\star$ arises as a suitable strong limit 
of a solution of the Vlasov-Maxwell-Landau equations, then $\sum_s \overline{\rho_{\star s}\bu_{\star s}} 
+ (1/4\pi c)\overline{\bE_\star\btimes\bB_\star}$ is exactly conserved}}

\subsection{Energy Balances}\lb{sec:VIIc} 

We finally derive the energy balances that hold in the collisionless limit $Do\to\infty.$
Since total energy density $\sum_s E_s + \frac{1}{8\pi}(|\bE|^2+|\bB|^2)$ of particles and fields 
is locally conserved by solutions of the VML system \eqref{III1}-\eqref{III3} for any degree of collisionality, 
there can be no anomaly in the conservation of total energy as $Do\to\infty.$ Just as for momentum
conservation, however, there are collisional conversions of energy from one type to another 
which might remain non-zero in the collisionless limit. We show here that such anomalous energy conversion 
does {\it not} occur in the limit $Do\to\infty,$ even if large velocity-gradients develop in the particle distribution
functions. We show this both in the fine-grained description and for the coarse-grained equations 
at fixed position and velocity resolutions $\ell,$ $\uit$ in the collisionless limit. Our energy balance 
equations will describe the transfers of energy simultaneously in phase-space and across scales $\ell,$ $\uit$
in phase-space. We thus recover and generalize previous work of Howes et al. 
\cite{howes2017prospectus,klein2017diagnosing} on fine-grained kinetic energy balance in phase space 
and of Yang et al. \cite{yang2017Aenergy,yang2017Benergy} on coarse-grained kinetic energy balance of bulk 
plasma flows in physical space and in length-scale $\ell.$  

\subsubsection{Fine-Grained Energy Balances}\lb{sec:VIIc1} 

A phase-space density of kinetic energy for particle-species $s$ was defined 
in \cite{howes2017prospectus,klein2017diagnosing} as $w_s(\bx,\bv,t)= (1/2)m_s|\bv|^2 f_s(\bx,\bv,t)$. 
The evolution of this density is easily obtained from the Vlasov-Landau 
kinetic equation \eqref{III1} to be 
\bea
&&  \partial_t w_s+
\grad_\bx\bdot\left(\bv w_s\right) +\grad_\bp\bdot\left(q_s \bE_* w_s \right) \cr 
&& \hspace{20pt} 
 =q_s\bv\bdot\bE f_s + (1/2)m_s|\bv|^2 C_s(f) 
\lb{VII13} \eea
The second term on the right arising from collision integral $C_s=\sum_{s'} C_{ss'}$ can be rewritten using 
the identity 
\bea
&& \frac{1}{2}m_s|\bv|^2\, C_{ss'} = \cr
&& \grad_\bv\bdot
\left[\frac{1}{2}\Gamma_{ss'}|\bv|^2\int d^3v' \frac{\bPi_{\bv-\bv'}}{|\bv-\bv'|}\bdot\left(\grad_\bp-\grad_{\bp'}\right)(f_s f_{s'})\right]\cr
&& \hspace{120pt} +{\mathcal R}_{ss'}(\bx,\bv,t),  \lb{VII14} \eea
with the divergence term representing a flux of kinetic energy in velocity space produced by collisions 
and with the second term representing the (signed) conversion of kinetic energy of species $s$ by collisions 
at phase-point $(\bx,\bv)$ into kinetic energy of species $s',$ given by  
\bea  && {\mathcal R}_{ss'}(\bx,\bv,t):= -\Gamma_{ss'}
\int d^3v' \frac{\bv\bdot\bPi_{\bv-\bv'}}{|\bv-\bv'|}\bdot\left(\grad_\bp-\grad_{\bp'}\right)(f_s f_{s'})\cr
&& =-\frac{\Gamma_{ss'}}{2}
\int d^3v' \frac{(\bv+\bv')\bdot\bPi_{\bv-\bv'}}{|\bv-\bv'|}\bdot\left(\grad_\bp-\grad_{\bp'}\right)(f_s f_{s'})\cr
&& \, 
\lb{VII15} \eea 
The expression in the second line is obtained by writing $\bv=\frac{1}{2}(\bv+\bv')+\frac{1}{2}(\bv-\bv')$ 
and using $\bw\bdot \bPi_\bw=\bzed.$ A simple estimate of this conversion term may be obtained 
by grouping the integrand into factors as 
\bea {\mathcal R}_{ss'}
&& = -\frac{1}{2}\Gamma_{ss'} \int d^3v' \ (\bv+\bv')\left(\frac{ f_sf_{s'}}{|\bv-\bv'|}\right)^{1/2}\cr
&& \hspace{1pt} \bdot \frac{\bPi_{\bv-\bv'}}{(f_sf_{s'}|\bv-\bv'|)^{1/2}}
\left(\grad_\bp-\grad_{\bp'}\right)(f_sf_{s'}), 
\lb{VII16} \eea
and applying the Cauchy-Schwartz inequality to obtain 
\bea 
&& \int d^3v\, |{\mathcal R}_{ss'}(\bx,\bv,t)|\leq  \Gamma_{ss'} \times\cr 
&& \sqrt{\frac{1}{4}\int d^3v  \int d^3v'  \,\frac{|\bv+\bv'|^2}{|\bv-\bv'|}f_sf_{s'}}\times \cr
&& \sqrt{\int d^3v  \int d^3v'  \,  \frac{|\bPi_{\bv-\bv'}\left(\grad_\bp-\grad_{\bp'}\right)(f_sf_{s'})|^2}{f_sf_{s'}|\bv-\bv'|}}\cr
&& \hspace{60pt} \leq C\sqrt{\Gamma_{ss'}\sigma(\bx,t)},
\lb{VII17} \eea
where the integral under the first square root is shown in Appendix \ref{app:A3} to be finite 
under mild assumptions. It follows that ${\mathcal R}_{ss'}\to 0$ in the sense of distributions 
as $Do\to\infty$. Note that the divergence term in \eqref{VII14} can also be show to vanish in 
the sense of distributions, by using an argument very similar to that for the term $\overline{C}_{ss'}$ in \eqref{V2}.  
We therefore conclude that in the limit $Do\to\infty$ the phase-space energy density satisfies
\be \partial_t w_{\star s}+
\grad_\bx\bdot\left(\bv w_{\star s}\right) +\grad_\bp\bdot\left(q_s (\bE_\star)_* w_{\star s} \right) 
=q_s\bv\bdot\bE_\star f_{\star s}.  
\lb{VII18} \ee
This is formally identical with the equation for $w_{\star s}$ argued to hold in the collisionless 
limit by \cite{howes2017prospectus}, eq.(2) or \cite{klein2017diagnosing}, eq.(2.6), 
but rewritten in a form that is meaningful and valid (in the distributional sense) even when, 
as expected, the particle distribution $f_{\star s}$ becomes non-differentiable in position and velocity. 

Since the physical-space energy density of particle species $s$ is given by $E_{s}=\int d^3v\, w_{s},$
we obtain from \eqref{VII18} by integrating over velocities and by using definitions \eqref{III12},\eqref{III17} 
that 
\be \partial_t E_{\star s} +\grad_\bx \bdot(E_{\star s}\bu_{\star s} +\bP_{\star s}\bdot\bu_{\star s}
+{\bf q}_{\star s}) =
\bj_{\star s}\bdot\bE_\star.  \lb{VII19} \ee
This same equation can be obtained from the $Do\to \infty$ limit of equation \eqref{III15} for $E_s,$ 
noting that its collisional contribution 
\be
Q_{ss'}+\bR_{ss'}\bdot\bu_s=\int d^3v\ \frac{1}{2}m_s|\bv|^2\, C_{ss'} \lb{VII20} \ee 
vanishes as $Do\to \infty$ by an estimate identical to \eqref{VII17}. Similarly, since 
$\bR_s\bdot\bu_s\to 0$ as $Do\to \infty,$ one obtains from \eqref{III20} the limiting 
equation for the bulk kinetic energy: 
\bea 
&& \partial_t (\frac{1}{2}\rho_{\star s}|\bu_{\star s}|^2) 
+\grad_\bx \bdot\left(\frac{1}{2}\rho_{\star s}|\bu_{\star s}|^2\bu_{\star s} 
+\bP_{\star s}\bdot\bu_{\star s}\right) \cr
&& \hspace{50pt} 
=\bP_{\star s}\bdots\grad_\bx\bu_{\star s}+\bj_{\star s}\bdot\bE_\star. \lb{VII21} \eea
From the vanishing of \eqref{VII20} we infer also that $Q_s\to 0$ as $Do\to 0$ 
and thus obtain from \eqref{III21} the limiting balance equation for the 
internal/fluctuational energy: 
\be \partial_t \epsilon_{\star s} +\grad_\bx \bdot(\epsilon_{\star s}\bu_{\star s} 
+{\bf q}_{\star s}) = -\bP_{\star s}\bdots\grad_\bx\bu_{\star s}. \lb{VII22} \ee
The results \eqref{VII19},\eqref{VII21}, \eqref{VII22} coincide, formally, with the 
results naively expected in the collisionless regime but are derived without 
assuming space-differentiability of solutions. 

Notice that the pressure-strain term
on the righthand sides of \eqref{VII21}, \eqref{VII22} must be carefully defined 
as a distributional limit $\bP_{\star s}\bdots\grad_\bx\bu_{\star s}
=\Dlim_{Do\to\infty}\bP_{s}\bdots\grad_\bx\bu_{s}.$ For the similar situation 
with compressible fluid turbulence, see \cite{eyink2017cascades1}. If the limiting 
fields $\bP_{\star s}$ and $\grad_\bx\bu_{\star s}$ exist as ordinary functions, 
then this distributional product will coincide with the ordinary pointwise product 
of functions. If $\bu_{\star s}$ is not classically differentiable, however, then this 
notion of product differs from the naive one. The degree of smoothness of 
$\bu_{\star s}$ is {\it a priori} not entirely obvious.  The inequality \eqref{VI45}
on scaling exponents of $\bE_\star,$ $\bB_\star,$ $f_{\star \green{s}}$ shows that 
these fields cannot be space-differentiable if there is a non-vanishing entropy 
production anomaly for species $s.$ The velocity field $\bu_{\star s},$ on the other hand,
is obtained from 0th and 1st velocity-moments of  $f_{\star s}$ by the formulas 
\eqref{III5}, \eqref{III6} and such moments are generally smoother than the particle 
distribution function appearing in the integrand (e.g. see section 3 of \cite{diperna1989global}). 
It is thus possible that  $\grad_\bx f_{\star \green{s}}$ exists only as a distribution/generalized 
function, while $\grad_\bx \bu_{\star \green{s}}$ exists as an ordinary function 
\footnote{The ``averaging lemmas'' of the type proved in section 3 of \cite{diperna1989global}
state, essentially, that the particle distribution functions $\overline{f}_{\star s,\uit}(\bx,\obv,t)$
coarse-grained in velocity space only, nevertheless have some fractional differentiability 
in space-time. Such results are obviously insufficient by themselves to guarantee 
that moments such as $\rho_{\star s},$ $\bu_{\star s}$ are classically differentiable}. 
Further detailed investigation,  both analytical and empirical, is required to settle this issue. 

% IS \bar{Q}_s NON_NEGLIGIBLE? 

%\vspace{20pt} 

% \section{Energy Cascade in Phase Space}

%!!!!!!!!!!!!!!!!!!!!!!!!!!!!

\subsubsection{Coarse-Grained Energy Balances}\lb{sec:VIIc2} 

We now consider the energy balances for solutions of the coarse-grained 
VM equations \eqref{V6} that are obtained in the nearly collisionless limit.  

\underline{Total Energy}: We may define a coarse-grained 
version of the phase-space kinetic energy density of particle species $s$
as $\overline{w}_s(\obx,\obv,t):=(1/2)m_s|\obv|^2 \ofs(\obx,\obv,t)$. It follows directly 
from the coarse-grained Vlasov-Landau equation \eqref{V1} that this energy density 
satisfies 
\bea
&&  
\partial_t \overline{w}_s +
\grad_\obx\bdot\left(\hbvs\overline{w}_s\right)
+\grad_\obp\bdot\left(q_s{\hbE}_{*s}\overline{w}_s \right) \cr
&& \hspace{20pt} 
 =q_s \obv\bdot\hbE_{*s}\ofs + (1/2)m_s|\obv|^2\overline{C}_s(f)
\lb{VII23} \eea
The ``renormalized'' quantities $\hbvs,$ $\hbE_{* s}$ are those given in \eqref{V10}-\eqref{V12}.
Because of the vanishing of the coarse-grained collision integral from estimate \eqref{V5}, 
we see that for fixed $\ell,$ $\uit$ and for sufficiently large (but finite) $Do$ 
the collisionless equation 
\be
\partial_t \overline{w}_s +
\grad_\obx\bdot\left(\hbvs\overline{w}_s\right)
+\grad_\obp\bdot\left(q_s{\hbE}_{*s}\overline{w}_s \right) 
 =q_s \obv\bdot\hbE_{*s}\ofs 
\lb{VII24}\ee
is satisfied \magc{to any specified accuracy}. In the idealized limit $Do\to\infty$ this becomes 
\bea
&&  
\partial_t \left(\frac{1}{2}m_s|\obv|^2 \overline{f}_{\star s}\right) +
\grad_\obx\bdot\left(\frac{1}{2}m_s|\obv|^2 \overline{\bv f_{\star s}}\right)\cr
&&  +\grad_\obp\bdot\left(\frac{1}{2}m_s|\obv|^2 \overline{q_s(\bE_{\star})_*f _{\star s}} \right) 
 \, = \, \obv \bdot \overline{q_s(\bE_\star)_{*}f_{\star s}} \cr
&& 
\lb{VII25}\eea
which further reduces to the equation \eqref{VII18} proposed in \cite{howes2017prospectus,klein2017diagnosing} 
in the limit as $\ell,$ $\uit\to 0.$ It must be stressed, however, that in dealing with real experimental data 
at fixed resolutions $\ell,$ $\uit,$ it is the equation \eqref{VII24} which will be satisfied by the 
measured energy density $\overline{w}_s$ and not the equation \eqref{VII18} 
suggested in \cite{howes2017prospectus,klein2017diagnosing}. The unresolved plasma turbulence at scales below 
$\ell,$ $\uit$ may lead to significant renormalization effects in the quantities 
$\hbvs,$ $\hbE_{* s}$ appearing in \eqref{VII24}.  

The spatial energy distribution of solutions to the coarse-grained Vlasov-Maxwell system 
\eqref{V6} is governed, for kinetic energy of particles, by the equation that comes from integrating 
\eqref{VII24} over $\obv$ and using definitions \eqref{III12},\eqref{III17}:
\be \partial_t \overline{E}_s +\grad_\obx \bdot(\overline{E_s\bu_s +\bP_s\bdot\bu_s+{\bf q}_s}) =
\overline{\bj_s\bdot\bE}. \lb{VII26}\ee
The same result is also obtained by coarse-graining \eqref{III15} and using $\overline{\bR_s\bdot\bu_s+Q_s}
=\int d^3\ov\ \frac{1}{2}|\obv|^2 \overline{C}_s(f)\to 0$ as $Do\to\infty.$  On the other hand, 
the evolution of the energy density of the resolved electromagnetic field is obtained 
from the coarse-grained Maxwell equations by the Poynting theorem: 
\bea 
&& \partial_t \left(\frac{|\obE|^2+|\obB|^2}{8\pi}\right)+
\grad_\obx\bdot\left(\frac{c\obE\btimes\obB}{4\pi}\right)=-\obj\bdot\obE. \cr
&& \,\!      
\lb{VII27}\eea 
Summing \eqref{VII26} over $s$ and adding \eqref{VII27} gives the balance equation 
for total energy density of coarse-grained solutions as  
\bea 
&& \partial_t \left(\sum_s \bar{E}_s+\frac{|\obE|^2+|\obB|^2}{8\pi}\right)\cr
&& +\grad_\obx\bdot\left(\sum_s \overline{E_s\bu_s+\bP_s\bdot\bu_s+{\bf q}_s}
      +\frac{c\obE\btimes\obB}{4\pi}\right)=\otau_\ell(\bj;\bE). \cr
&& \,\!      
\lb{VII28}\eea 
Just as for the coarse-grained momentum balance \eqref{VII12}, there is a source-term on the 
righthand side of \eqref{VII28} which represents a flux of energy from unresolved scales $<\ell$
to resolved scales $>\ell.$ Since total energy (particles $+$ fields) is conserved for the 
VML system \eqref{III1}-\eqref{III3}, this flux of energy must vanish for any collisionless limit 
of such solutions. Because of the estimate $\otau_\ell(\bj,\bE)\sim (\delta_\ell \bj)(\delta_\ell \bE)$ 
from \eqref{IV7},  the energy flux indeed vanishes as $\ell\to 0$ whenever limits 
$\varrho_\star,$ $\bj_\star,$ $\bE_\star,$ $\bB_\star$ are spatially continuous or 
satisfy even weaker regularity conditions \footnote{{\magc{See the discussion in  footnote [99].
In addition, as discussed there for momentum, the coarse-grained total energy $\sum_s \overline{E}_{\star s}
+\frac{1}{8\pi}\overline{|\bE_\star|^2+|\bB_\star|^2}$ will be conserved for any suitable 
strong limits $f_{\star s},$ $\bE_\star,$ $\bB_\star$ of VML solutions.}}}

As an aside, we note that current mathematical theory for global solutions of the 
Vlasov-Maxwell system does {\it not} provide weak solutions 
that conserve energy but instead guarantees only that total energy for solutions 
is non-increasing in time! Cf.\cite{diperna1989global}, p.740, remark 4.
The arguments for energy conservation which we made above 
may not apply, because the DiPerna-Lions theory guarantees only that 
$f_s,$ $\bE$ and $\bB$ are square-integrable and that 2nd-moments of $f_s$ with respect 
to $\bv$ exist. Such regularity properties are not enough to allow the 
equation \eqref{VII28} to be even written down, because they do not guarantee that heat fluxes 
${\bf q}_s$ (3rd moments) are finite. Even if energy density integrated over all space is 
considered, which eliminates the undefined ${\bf q}_s$ term, the DiPerna-Lions 
solutions are not guaranteed to satisfy the weak regularity conditions of the type discussed 
in footnote [99]  that imply that $\otau_\ell(\bj;\bE)\to 0$ as $\ell\to 0.$ While solutions 
with decreasing total energy are physically unrealistic as collisionless limits of VML solutions, 
one cannot rule out that weak Vlasov-Maxwell solutions with decreasing total energy might 
occur in other physical contexts (e.g. see discussion in section \ref{sec:IX}).

%While it is impossible to have a dissipative anomaly of total energy for any weak solution obtained 
%as a collisionless limit within the Vlasov-Maxwell-Landau theory, the latter model does not 
%describe all effects in physical plasmas. In particular, the VML theory does not directly describe the   
%bremsstrahlung radiation resulting from interspecies collisions with distinct charge-to-mass ratios.  
%The energy loss due to such ``free-free'' radiation can be calculated from the emissivity 
%for a plasma with given distribution functions $f_s(\bx,\bv,t)$ (Eidmann, 1975; Zheleznyakov, 1996),
%but the radiative reaction on the particles leads to a damping that must modify the kinetic 
%equations so that total energy is conserved (Kunze \& Rendall, 2001).  In principle, additional 
%radiation losses of energy by a supra-thermal tail of high-velocity particles might lead to a 
%dissipative anomaly of total energy?????

%However, there are turbulent conversions  
%between forms (electromagnetic, particle mean kinetic, particle fluctuation kinetic).   

\underline{Kinetic-Energy of Bulk Velocities}: 
The balance equation \eqref{VII26} describes the dynamics of the total kinetic energy 
of species $s$ calculated from the particle distribution resolved to scales $\ell,$ $\uit.$ 
However, one may furthermore divide the energy density $\overline{E}_s$ into separate 
contributions from the resolved bulk velocity $\tbus$ as defined in \eqref{IV20} and 
from the (intrinsic) resolved internal energy $\oeps$ defined in \eqref{IV24}. In particular,
the contributions from the bulk velocity $\tbus$ and from the coarse-grained fields $\obE,$ $\obB$ 
are often considered to be the only ``turbulent''  energy contributions at length-scale $\ell$, 
because these low-frequency fields are described by ``fluid-like'' equations and 
experience  a continual, reversible energy exchange due to Alfv\'enic wave oscillations 
(e.g. \cite{howes2015dynamical}, section 2(c)). In this view, $\oeps$ represents a quasi-thermal energy 
or energy of kinetic fluctuations not directly participating in the ``turbulence'' at scale $\ell$. 
We do not subscribe to this view, but it is nevertheless interesting to consider 
separately the kinetic energy dynamics of bulk flow and of the fluctuations. 

The balance equation for the bulk kinetic energy $(1/2)\orhos|\tbus|^2$ in the nearly 
collisionless limit is easily obtained from coarse-grained mass conservation \eqref{VII2}
and coarse-grained momentum conservation  \eqref{VII9}, yielding 
\bea
&&  \partial_t\left(\frac{1}{2}\orhos|\tbus|^2\right)+
\grad_\obx\bdot\left(\frac{1}{2}\orhos|\tbus|^2\tbus
+\obP_s^*\bdot\tbus\right)\cr 
&&\hspace{40pt} =\left(\orhos\ttau(\bu_s,\bu_s)
+\obP_s\right)\bdots\grad_\obx\tbus+q_s\overline{n}_s\tbE_{*s}\bdot\tbus \cr
&&\hspace{40pt} =\obP_s^*\bdots\grad_\obx\tbus+q_s\overline{n}_s\tbE_{*s}\bdot\tbus \
\lb{VII29} \eea
This same equation has been derived earlier in \cite{yang2017Benergy} for kinetic plasma turbulence
and it is very similar to the analogous equations for resolved kinetic energy in compressible
fluid turbulence \cite{aluie2013scale,eyink2017cascades1}. Obviously, the term 
$q_s\overline{n}_s\tbE_{*s}\bdot\tbus$ represents resolved wave-particle interactions. 
Based on the fluid turbulence 
analogy,  the term $-\orhos\ttau(\bu_s,\bu_s)\bdots\grad_\obx\tbus$ may be taken to represent 
energy flux arising from turbulent cascade, while $-\obP_s\bdots\grad_\obx\tbus$ represents 
resolved pressure-work. It should be remembered, however, that only the intrinsic resolved pressure 
tensor $\obP_s^*$ is calculable from the distribution function $\ofs$ resolved to scales 
$\ell,$ $\uit,$ and it is impossible from such coarse measurements of the particle distributions 
to compute the separate contributions of $\orhos\ttau(\bu_s,\bu_s)$ and $\obP_s.$ 

The limit in \eqref{VII29} with first $Do\to\infty$ and then $\ell\to 0$ must recover the equation 
\eqref{VII21} for $(1/2)\rho_{\star s}|\bu_{\star s}|^2,$ if the strong limits $\bE\to\bE_\star,$ $\bB\to\bB_\star,$ 
$f_s\to f_{\star s}$ exist as $Do\to 0.$ Indeed, since all of the other terms in \eqref{VII29}
then converge distributionally to the corresponding terms in \eqref{VII21}, one must have 
\be \stackrel[\ell\to 0]{}{\Dlim}  \obP_{\star s}^*\bdots\grad_\bx\tilde{\bu}_{\star s}
=\bP_{\star s}\bdots\grad_\bx\bu_{\star s}, \lb{VII30} \ee 
where the product on the righthand side is the same quantity that appears in \eqref{VII21}. 
The result \eqref{VII30}, if correct, means that there is no ``pressure-work defect'' 
of the type that appears in compressible fluid shocks \cite{eyink2017cascades1}. 
 This result would be expected, in particular,
if the gradient $\grad_\bx\bu_{\star s}$ exists as an ordinary function. In that case, 
\be  
\stackrel[\ell\to 0]{}{\Dlim}  \overline{\rho}_{\star s}\ttau(\bu_{\star s},\bu_{\star s})
\bdots\grad_\obx\widetilde{\bu}_{\star s} = 0 
\lb{VII31} \ee 
as well. This last relation can be interpreted as the statement that there is a vanishing energy 
flux in the order of limits first $Do\to\infty$ and then $\ell\to 0.$ This is a quite reasonable 
conclusion, since the collisional transfer of energy from species $s$ to other species, $\bR_s\bdot\bu_s,$ 
vanishes as $Do\to\infty$ according to \eqref{VII4}. Thus, there is physically no ``sink'' 
for an energy cascade to small scales. 

This tentative conclusion, that there is ``no energy cascade to small scales in a collisionless plasma'', 
must be carefully interpreted. The solar wind is a nearly collisionless plasma with Kolmogorov-type 
spectra observed at scales above the (thermal) ion gyroradius $\uprho_i,$ that are generally interpreted 
as an energy-cascade ``inertial range'' \magc{of, primarily, incompressible shear-Alfv\'en waves}. In fact, there is direct 
evidence of non-zero energy flux in this range from empirical studies of third-order structure functions (e.g. \cite{marino2008heating,coburn2015third}). \magc{This cascade is described by the balance equation 
of the resolved mechanical energy in the bulk velocities of the particles (mostly from protons, or $H^+$ ions) and 
electromagnetic fields, obtained by combining the eqs.\eqref{VII27},\eqref{VII29}:
\bea
&&  \partial_t\left(\sum_s \frac{1}{2}\orhos|\tbus|^2+ \frac{|\obE|^2+|\obB|^2}{8\pi} \right)\cr 
&& \hspace{10pt} + \grad_\obx\bdot\left[ \sum_s \left(\frac{1}{2}\orhos|\tbus|^2\tbus+\obP_s^*\bdot\tbus\right) 
+ \frac{c\obE\btimes\obB}{4\pi}\right]\cr 
&&\hspace{60pt} =\sum_s \left(\obP_s^*\bdots\grad_\obx\tbus+\obj_s\bdot \widetilde{\boeps}_s\right)
\lb{VII3b}  \eea 
with $\widetilde{\boeps}_s$ an ``electromotive force'' generated by unresolved turbulent fluctuations 
of bulk velocity and density for particles of species $s$: 
\bea \widetilde{\boeps}_s &:=&  \frac{1}{c}\widetilde{\tau}(\bu_s\btimescom\bB) 
+ \frac{1}{\overline{n}_s}[\otau(n_s,\bE)+\frac{1}{c}\tbus\btimes\otau(n_s,\bB)] \cr 
&&
%&\,=& \boeps_{u_s} + \boeps_{n_s}
\lb{VII31c} \eea
so that $Q_{\ell,F}:=\sum_s \obj_s\bdot \widetilde{\boeps}_s$ represents a flux of electromagnetic energy to the unresolved scales.}
Thus for length-scales 
$\ell$ in the range $L_i\gg \ell\gg \rho_i,$ one would expect non-vanishing values of the ion kinetic-energy flux 
$\calQ_{\ell,i}:=-\overline{\rho}_i\ttau(\bu_i,\bu_i)\bdots\grad_\obx\tilde{\bu}_i$
\magc{and of $Q_{\ell,F}$}. This does not contradict 
the conclusion \eqref{VII31}, which involves the limit $\ell\to 0$ with $\uprho_i$ fixed or, equivalently, length-scales 
$\ell\ll \uprho_i.$ In order to develop an Onsager-type theoretical description of the energy-cascade ``inertial-range'' 
of the solar wind at scales $\ell\gg \uprho_i,$ one would need to consider after the limit $Do\to\infty$ a subsequent 
limit $\uprho_i/L_i\to 0,$ corresponding to a long energy inertial-range of scales. It is quite plausible that limits 
exist $\bE_\star\to\bE_\bullet,$ $\bB_\star\to\bB_\bullet,$ $f_{\star s}\to f_{\bullet s},$ $s=i,e$ 
as $\uprho_i/L_i\to 0,$ leading to a kinetic description with a turbulent cascade of ion kinetic nergy:
\be  \calQ_{\bullet i}:=\stackrel[\ell\to 0]{}{\Dlim}  \overline{\rho}_{\bullet i}\ttau(\bu_{\bullet i},\bu_{\bullet i})
\bdots\grad_\obx\widetilde{\bu}_{\bullet i} \neq  0 \lb{VII31a} \ee 
More precisely,
one expects that this limit lies within the regime of validity 
\footnote{We emphasize that $\uprho_i\ll \ell$ is a sufficient 
condition for validity of a gyrokinetic description at length-scale $\ell,$ but not a necessary condition. When 
there is scale-anisotropy in directions perpendicular and parallel to the local magnetic field, then 
$\ell\sim\ell_\perp\ll \ell_{\|}.$ In that case, $\uprho_i/\ell\gg (\uprho_i/v_A)\omega=\sqrt{\beta_i}(\omega/\Omega_i)$
for $\omega\sim v_A/\ell_{\|}$, and the fundamental condition $\omega\ll \Omega_i$ required for validity of 
gyrokinetics is guaranteed by $\rho_i\ll \ell$ and $\beta_i\sim 1$. See section \ref{sec:VIIIa} for further discussion.
On the other hand, it is likely that gyrokinetics remains valid for $\ell\lesssim \rho_i$ 
\cite{schekochihin2008gyrokinetic,schekochihin2009astrophysical}} of a gyrokinetic description 
\cite{schekochihin2008gyrokinetic,schekochihin2009astrophysical}. 
A full treatment of the $\uprho_i/L_i\to 0$ limit is beyond the scope of the current paper,
but we shall discuss briefly the relationship of our analysis with gyrokinetic 
theory in section \ref{sec:VIIIa}. We likewise do not consider in detail the limit $\uprho_e/\uprho_i\to 0$
(heavy ion limit) which idealizes the ``ion dissipation range''  of the solar wind over the interval of 
length-scales $\ell$ satisfying $\uprho_i\gg \ell\gg \uprho_e$ \footnote{Of course, in the solar wind where 
the dominant ionic species by mass consists of protons, $\uprho_i/\uprho_e\doteq 43,$ assuming that 
$T_i\doteq T_e.$ Thus, the mathematical limit $\uprho_e/\uprho_i\to 0$ is only marginally applicable to the solar wind.},
where a gyrokinetic description is expected to be valid at least for the electrons. See section \ref{sec:VIIIa} 
for brief remarks. 

%Above division and Aluie division may not be appropriate.
%
%On the other hand, 
%\bea 
%&& \partial_t (\frac{1}{2}\overline{\rho_s|\bu_s|^2}) 
%+\grad_\bx \bdot\left(\frac{1}{2}\overline{\rho_s|\bu_s|^2\bu_s} +\overline{\bP_s\bdot\bu_s}\right) \cr
%&& \hspace{50pt} 
%=\overline{\bP_s\bdots\grad_\bx\bu_s}+\overline{\bj_s\bdot\bE} +\overline{\bR_s\bdot\bu_s} \eea
%These two must agree in the limit as $\ell\to 0$!

%\newpage 

%$\,\!$

%\newpage

\underline{Kinetic-Energy of Fluctuations}: 
The balance equation for $\overline{\epsilon}_s^*=\overline{\epsilon}_s+\frac{1}{2}\orhos\ttau(\bu_s;\bu_s)$
can be obtained by subtracting equation \eqref{VII26} for $\overline{E}_s$ and
equation \eqref{VII29} for $(1/2)\orhos|\tbus|^2,$ giving: 
\bea
&&  \partial_t \overline{\epsilon}_s^* +
\grad_\obx\bdot\bigg( \overline{\epsilon_s\bu_s}+\overline{{\bf q}}_s+\overline{\tau}(\bP_s;\bu_s) \cr
&& \hspace{42pt} -\obP_s\cdot\ttau(\rho_s,\bu_s)/\orhos+\frac{1}{2}\orhos\ttau(\bu_s;\bu_s,\bu_s)\bigg)\cr 
&& \hspace{30pt} = -\obP_s^*\bdots\grad_\obx\tbus 
+q_s \overline{n}_s \ttau(\bE_{*s};\bu_s).  \cr
&& \, 
\lb{VII32} \eea
Note that the term $-\obP_s^*\bdots\grad_\obx\tbus$ on the righthand side differs only in sign 
from the corresponding term on the righthand side of \eqref{VII29}, so that this quantity acts to 
exchange kinetic energy between bulk flow and fluctuations. 
Even after taking the limit $Do\to\infty,$ \eqref{VII32} is quite distinct from the equation obtained by coarse-graining 
\eqref{VII22} for the fine-grained limit field $\epsilon_{\star s},$  or: 
\be \partial_t \overline{\epsilon}_{\star s} +\grad_\bx \bdot(\overline{\epsilon_{\star s}\bu_{\star s}} +\overline{{\bf q}}_{\star s}) = 
-\overline{\bP_{\star s}\bdots\grad_\bx\bu_{\star s}}. \lb{VII33}\ee
In particular, note that \eqref{VII32} contains a non-vanishing wave-particle interaction term 
$q_s \overline{n}_s \ttau(\bE_{*s};\bu_s)$ which is entirely absent from \eqref{VII33}. These two equations 
must agree in the limit $\ell\to 0,$ on the other hand, and in that limit
the term $q_s \overline{n}_{\star s} \ttau(\bE_{\star *s};\bu_{\star s})\to 0$
under plausible regularity assumptions, as in footnote [99]. 

It is interesting to refine the spatial-balance equation \eqref{VII32} for kinetic energy of fluctuations 
in order to follow the transfer through phase-space. For that purpose, we 
define a {\it phase-space density of fluctuation energy} at scales $\ell,$ $\uit$ by 
\be \overline{z}_s(\obx,\obv,t):= \frac{1}{2}m_s|\obv-\tbus|^2\ofs(\obx,\obv,t) \lb{VII34}\ee
so that $\overline{\epsilon}_s^*=\int d^3\ov\  \overline{z}_s.$ 
A tedious calculation (see Appendix \ref{app:O}) 
yields the following balance equation for $\overline{z}_s$: 
\bea
&&  \partial_t\overline{z}_s +
\grad_\obx\bdot\left(\hbvs\overline{z}_s
+\obP_s^*\bdot(\tbus-\obv)\ofs/\overline{n}_s\right)+\grad_\obp\bdot\left(q_s {\hbE}_{*s}\overline{z}_s\right) \cr 
&& \hspace{20pt} =\orhos\ttau(\bu_s,\bu_s)\bdots \grad_\obx((\tbus-\obv)\ofs/\overline{n}_s)\cr
&& \hspace{100pt} -m_s(\hbvs\obv\ofs-\obv\,\obv\ofs_{,\ell})\bdots\grad_\obx\tbus\cr 
&& \hspace{40pt} \mbox{(turbulent redistribution of energy)}\cr
&& \hspace{25pt}+\obP_s\bdots \grad_\obx((\tbus-\obv)\ofs/\overline{n}_s) \cr 
&& \hspace{40pt} \mbox{(energy redistribution by resolved pressure)}\cr
&& \hspace{25pt}-m_s\Big(\obv\,\obv\ofs_{,\ell} -\tbus\obv\ofs-\hbvs\tbus\ofs \cr
&& \hspace{100pt} +\tbus\tbus\ofs-\ttau(\bu_s,\bu_s)\ofs\Big)\bdots \grad_\obx\tbus \cr
&& \hspace{40pt} \mbox{(work by mean-velocity gradient)}\cr
&& \hspace{25pt} -m_s\ttau(\bu_s,\bu_s)\bdots\grad_\obx\tbus \ofs\cr
&& \hspace{40pt} \mbox{(energy input from turbulent cascade)}\cr
&& \hspace{25pt} +q_s(\obv-\tbus)\bdot(\hbE_{*s}-\tbE_{*s})\ofs \cr
&& \hspace{40pt} \mbox{(energy input \& redistribution by EM field)} \cr
&& \, 
 \lb{VII40}\eea
The above equation \eqref{VII40} for $\overline{z}_s$ gives more insight into the flow of kinetic energy 
through phase-space than does the corresponding equation  \eqref{VII24} for $\overline{w}_s,$
because it describes locally in phase-space the turbulent interactions of the kinetic velocity fluctuations 
with the bulk velocity for particle species $s.$  
The five terms on the righthand side have been arranged so that the first two vanish 
after integration over $\obv$  and the last three terms yield after integration the expressions
\be  -\obP_s\bdots\grad_\obx\tbus, \quad -\orhos\ttau(\bu_s,\bu_s)\bdots\grad_\obx\tbus, \quad  
q_s n_s \ttau(\bE_{*s};\bu_s), 
 \lb{VII41} \ee
which appear as sources of $\bar{\epsilon}_s^*$ in \eqref{VII32}. The physical meaning 
of these five terms are briefly indicated in parentheses beneath each. As discussed 
below \eqref{VII29}, it might be argued to be more appropriate 
to combine the first two energy redistribution terms. This would yield an 
expression proportional to the intrinsic stress tensor $\obP_s^*,$ rather than 
separate contributions proportional to $\orhos\ttau(\bu_s,\bu_s)$ and $\obP_s$. 
Likewise, it might be more appropriate to combine the third and fourth terms, since 
both represent work performed by the resolved strain, acting against the stress of 
fluctuating velocities, on the one hand, and against the mean stress, on the other hand. 

%This is important, for instance, in the discussion of Landau damping in a turbulent 
%plasma \cite{howes2017prospectus,klein2017diagnosing}.  

Nothing very exciting emerges from the equation  \eqref{VII40} in the limit $Do\to\infty$ alone 
\footnote{
Taking the limit $\ell,\uit\rightarrow 0$ after first taking the collisonless limit $Do\to\infty$ 
simplifies the equation \eqref{VII40} to 
%\bea
%&&  \partial_t z_{\star s}\! + \!
%\grad_\bx\bdot\left(\bv z_{\star s}
%\!+\! \bP_{\star s}  \bdot (\bu_{\star s}-\bv)\frac{f_{\star s}}{n_{\star s}}\right)
%\!+\!\grad_\bp\bdot\left(q_s (\bE_\star)_* z_{\star s}\right) \cr
%&& \hspace{75pt}=\bP_{\star s} \bdots \grad_\bx\left((\bu_{\star s}-\bv)\frac{f_{\star s}}{n_{\star s}}\right) \cr
%&& \hspace{80pt} -m_s(\bv-\bu_{\star s})(\bv-\bu_{\star s})\bdots \grad_\bx\bu_{\star s}, \cr
%&&
%\lb{VII42} \eea
$\partial_t z_{\star s}\! + \!
\grad_\bx\bdot\left(\bv z_{\star s}
\!+\! \bP_{\star s}  \bdot (\bu_{\star s}-\bv)f_{\star s}/n_{\star s}\right)
\!+\!\grad_\bp\bdot\left(q_s (\bE_\star)_* z_{\star s}\right) 
=\bP_{\star s} \bdots \grad_\bx\left((\bu_{\star s}-\bv)f_{\star s}/n_{\star s}\right) 
-m_s(\bv-\bu_{\star s})(\bv-\bu_{\star s})\bdots \grad_\bx\bu_{\star s}, $
which yields eq.\eqref{VII22} for $\epsilon_{\star s}$ after integration over velocities. This same equation
could also be obtained formally from Howes' equation \eqref{VII18} for $w_{\star *},$ 
the collisionless kinetic equation \eqref{V7} for $f_{\star s},$ and the momentum balance \eqref{VII5}
for $\bu_{\star s}.$ Note, however, that $f_{\star s}$ will not be classically differentiable 
when there is a non-vanishing entropy cascade for species $s$ and $n_{\star s},$ $\bu_{\star s}$
may not be differentiable as well. Thus, the two terms on the righthand side must be 
carefully interpreted as distributional limits of the corresponding terms on the righthand side of 
\eqref{VII40} (i.e. the second and third)}, but more interesting possibilities emerge if one considers 
the secondary limit $\uprho_i/L_i\to 0,$ which permits an asymptotic energy cascade to small scales. 
If one assumes that strong limits exist $\bE_\star\to\bE_\bullet,$ $\bB_\star\to\bB_\bullet,$ 
$f_{\star s}\to f_{\bullet s},$ $s=i,e$ as $\uprho_i/L_i\to 0,$ then taking this limit in 
\eqref{VII40} (after first taking $Do\to\infty$) and only then taking $\ell,\uit\rightarrow 0$ gives 
\bea
&&  \partial_t z_{\bullet s}\! + \!
\grad_\bx\bdot\left(\bv z_{\bullet s}
\!+\! \bP_{\bullet s}  \bcirc (\bu_{\bullet s}-\bv)\frac{f_{\bullet s}}{n_{\bullet s}}\right)
\!+\!\grad_\bp\bdot\left(q_s (\bE_\bullet)_* z_{\bullet s}\right) \cr
&& \hspace{10pt} = {\mathcal R}_{\bullet s}(\bx,\bv,t) \cr
&& \hspace{30pt} \mbox{(turbulent redistribution of energy)}\cr
&& \hspace{15pt}+\bP_{\bullet s} \bodots \grad_\bx((\bu_{\bullet s}-\bv)f_{\bullet s}/n_{\bullet s}) \cr 
&& \hspace{30pt} \mbox{(energy redistribution by resolved pressure)}\cr
&& \hspace{15pt}-m_s(\bv-\bu_{\bullet s})(\bv-\bu_{\bullet s})\bodots \grad_\bx\bu_{\bullet s} \cr
&& \hspace{30pt} \mbox{(work by mean-velocity gradient)}\cr
&& \hspace{15pt} +\calQ_{\bullet s}(\bx,t)\circ f_{\bullet s} \cr
&& \hspace{30pt} \mbox{(energy input from turbulent cascade)}. \cr
&&
\lb{VII43} \eea
The four terms on the righthand side of \eqref{VII43} are taken to be distributional limits of the 
corresponding first four terms on the righthand side of \eqref{VII40}. As one can see, 
there is a possible anomalous redistribution of energy ${\mathcal R}_{\bullet s},$ 
which vanishes upon integration over velocities, and a possible anomalous input of energy 
$\calQ_{\bullet s}$ from turbulent cascade. These conclusions must be considered tentative,
since they require a rigorous study of the limit $\uprho_i/L_i\to 0,$ which we do not attempt here.
In the next section we discuss the problem very briefly. 

%\newpage

\section{Relation to Prior Works}\lb{sec:VIII}

\subsection{Gyrokinetic Turbulence}\lb{sec:VIIIa}

All prior work on entropy cascade in plasma turbulence has been developed, essentially, within the 
framework of gyrokinetics. We therefore must briefly review gyrokinetic theory and its physical 
basis, in order to make comparisons with our own work. 

\subsubsection{Concise Review of Gyrokinetic Theory}

Nonlinear gyrokinetic equations capable 
of describing turbulent cascades were first derived in the seminal paper of Frieman \& Chen 
\cite{frieman1982nonlinear} 
and subsequently extensively investigated theoretically and numerically by the plasma fusion community.  
Modern approaches to nonlinear gyrokinetics exploit powerful Hamiltonian and geometric methods 
\cite{brizard2007foundations,krommes2012gyrokinetic}. The application of gyrokinetics to astrophysical 
and space plasmas was pioneered in papers of Howes et al. \cite{howes2006astrophysical}, 
Schekochihin et al. \cite{schekochihin2008gyrokinetic,schekochihin2009astrophysical}, 
which also first proposed and developed the theory of entropy cascades in plasma turbulence. 
Our review of gyrokinetic theory  and especially the role of entropy in gyrokinetic turbulence
shall follow closely the discussions in \cite{howes2006astrophysical,schekochihin2008gyrokinetic,schekochihin2009astrophysical}.
More general gyrokinetic theories of entropy cascade are possible (e.g. \cite{kunz2015inertial}), but the scaling 
predictions have been less developed in those generalizations and the original theoretical work therefore 
provides a more adequate basis of comparison with our results. 

Although not necessary to achieve a gyrokinetic reduction \cite{brizard2007foundations,krommes2012gyrokinetic},
many treatments, including that of \cite{howes2006astrophysical,schekochihin2008gyrokinetic,schekochihin2009astrophysical}  
start from a decomposition of fields into ``background'' and ``fluctuation'' contributions
\be f_s=F_{s}+\delta f_s, \quad \bB=\bB_0+\delta \bB, \quad \bE =\delta \bE \quad (\bE_0=\bzed) 
\lb{VIII1}\ee
with the further assumption of {\it (i) fluctuation amplitudes small relative to backgrounds}: 
\be \delta f_s/F_s \sim \delta B_\perp/B_0\sim \delta B_\|/B_0\sim c\delta E_\perp/v_{th,s} B_0\sim \epsilon, 
\lb{VIII2}\ee
where $\epsilon\ll 1$ is a dimensionless parameter that quantifies this smallness and the subscripts 
$\|$ and $\perp$ denote vector components parallel and perpendicular to $\bB_0,$ respectively. 
With $u_\perp\sim c\delta E_\perp/B_0$ giving the $\bE\btimes\bB$ drift velocity, the fourth condition 
in \eqref{VIII1} can be restated as $u_\perp/v_{th,s}\sim \epsilon.$ For applications to astrophysical and 
space plasmas (e.g. the solar wind), the condition (i) is perhaps the most dubious of the various assumptions 
discussed here. A second assumption, very essential for the validity of gyrokinetics, is 
{\it (ii) frequency of fluctuations small relative to the gyrofrequency}:
\be  \omega/\Omega_s \sim \epsilon. \lb{VIII3} \ee 
This condition is often found to be satisfied over very broad ranges 
of scales in a turbulent plasma.  It imposes no direct restriction on the perpendicular 
length-scale $\ell_\perp$ or perpendicular wavenumber $k_\perp\sim 1/\ell_\perp$ relative to the thermal 
gyroradius $\uprho_s$, which may be taken to satisfy $k_\perp \uprho_s\sim 1.$ However, 
if one takes $\omega\sim v_{th,s}k_\| $ in order to admit Landau resonances, then {\it (iii) scale-anisotropy
of fluctuations} is required: 
\be  k_\|/k_\perp \sim \epsilon. \lb{VIII4}\ee
%Similarly, assuming $\omega\sim v_A k_\|,$ the Alfv\'en-wave frequency, implies that $k_\|/k_\perp \sim 
%\epsilon/\sqrt{\beta_i}.$ Here $\beta_i$ is the ion beta parameter $\beta_i\sim 1.$
This condition is also often observed to be satisfied over wide ranges of scales. 
If electric fields are assumed electrostatic, $\delta \bE=-\grad\varphi,$ to leading order, 
then scale-anisotropy implies 
$\delta E_\|/\delta E_\perp\sim \epsilon.$
Whenever the above conditions hold initially, then gyrokinetic theory implies that they are dynamically 
maintained, with a slow evolution of the background fields on $\sim 1/\epsilon^3\Omega_s$ time scales. 

Gyrokinetic theory obtains closed evolutionary equations by seeking approximations to solutions of 
the Vlasov-Maxwell-Landau equations as asymptotic series $\delta a\sim \sum_{i\geq 1} \delta a^{(i)} \epsilon^i $
for all fluctuation fields $\delta a,$ $\,$ as $\,\epsilon\to 0.$ \\ Following 
\cite{howes2006astrophysical,schekochihin2008gyrokinetic,schekochihin2009astrophysical}, 
we consider here the simple case where all background 
distributions are isotropic Maxwellian, $F_s=n_{s} (m_s/2\pi T_s)^{3/2}\exp(-m_s v^2/2T_s),$ with 
temperature $T_s$ of species $s$ (in energy units) and where the background magnetic field is uniform, $\bB_0=B_0\hat{{\bf z}}.$ 
Then it is found in \cite{howes2006astrophysical,schekochihin2008gyrokinetic,schekochihin2009astrophysical} 
that 
\be \delta f_s^{(1)}=-\frac{q_s\varphi(\bx,t)}{T_{s}}F_{s}(v,t)+h_s(\bX_s,v,v_\perp,t) \lb{VIII5}\ee 
where the first term in \eqref{VIII5} gives the adiabatic, Boltzmann response and the second term $h_s$ 
is the ring distribution function which describes for each species $s$ the distribution of the gyrocenters 
\be  \bX_s = \bx+\bv_\perp\btimes\hat{{\bf z}}/\Omega_s. \lb{VIII6}\ee
The time-evolution of the ring distribution functions $h_s$ is obtained from the {\it gyrokinetic equations}:  
\bea 
&&\frac{\partial h_s}{\partial t} +v_\| \frac{\partial h_s}{\partial z} +
\frac{c}{B_0}\{\langle\chi\rangle_{\bX_s},h_s\} \cr
&& \hspace{50pt} = \frac{q_s F_s}{T_s}\frac{\partial \langle\chi\rangle_{\bX_s}}{\partial t} 
+ \left(\frac{\partial h_s}{\partial t}\right)_c 
\lb{VIII7} \eea 
where the gyrokinetic electromagnetic potential is defined by $\chi:=\varphi-\bv\bdot{\bf A}/c$ in terms of the 
usual scalar $\varphi$ and vector ${\bf A}$ potentials, where 
\be \langle a\rangle_{\bX_s} = \frac{1}{2\pi} \int_0^{2\pi} 
d\theta\, a(\bX_s-\bv_\perp(\theta)\btimes\hat{{\bf z}}/\Omega_s,v_\|,\bv_\perp(\theta),t) \lb{VIII8} \ee 
is the {\it ring-average} over cyclotron motions with velocities $\bv_\perp(\theta)=
v_\perp[(\sin\theta) \hat{\bf x}+(\cos\theta)\hat{\bf y}]$, the spatial 
{\it Poisson bracket} is defined by $\{a,b\}:=\hat{{\bf z}}\bdot (\grad_{\bX_s}a\btimes\grad_{\bX_s}b),$ and 
$(\partial h_s/\partial t)_c$ is the collisional contribution from the linearized and gyro-averaged Landau operator.  
The evolution of the electromagnetic fields $\varphi,$ $A_\|,$ $\delta B_\|$ is likewise obtained from the 
Maxwell equations in a reduced, gyro-averaged form. See \cite{howes2006astrophysical}, eqs.(26)-(28). Together 
with the kinetic equations for the ring distribution functions $h_s,$ these completely specify the dynamics. 
One has the freedom in these equations to take $\int d^3x\, \varphi =\int d^3x\, h_s=0,$ and,   in fact, 
to any order in the expansion in $\epsilon$, one can impose $\int d^3x\, \delta f_s=0.$ The nonlinear 
Poisson bracket term arises, of course,  from the wave-particle interaction term $(q_s/m_s)(\bE_*\bdot\grad_\bv) f_s$ 
in the Vlasov-Landau equation \eqref{III1}. 
The $\bv$-gradient of $\delta f_s^{(1)}$ contributes an $\bX_s$-gradient of $h_s$ because of the 
$\bv_\perp$-dependence of the gyrocenter $\bX_s$ in \eqref{VIII6}. Although there is no direct ``advection'' 
in velocity-space for the gyrokinetic equation, the Poisson bracket term represents this effect, 
which creates fine-scale velocity structure. 

\subsubsection{Gyrokinetic H-Theorems}

The gyrokinetic $H$-theorem for entropy has been discussed in
\cite{howes2006astrophysical,schekochihin2008gyrokinetic,schekochihin2009astrophysical}, 
whose results we briefly summarize. Assuming the smallness
of fluctuations \eqref{VIII1} the phase-space entropy density \eqref{III22} can be 
Taylor-expanded as 
\bea \sit(f_s) &\doteq &\sit(F_s) -(1+\ln F_s)\delta f_s - \frac{(\delta f_s)^2}{2 F_{s}}. \cr
&& \, \lb{VIII9}  \eea
With $\int d^3x\, \delta f_s=0,$ the entropy of species $s$ becomes 
 \be S(f_s) = S(F_s) - \int d^3x\int d^3v\  \frac{(\delta f_s)^2}{2 F_{s}}, \lb{VIII10} \ee 
The second law of \eqref{III25}-\eqref{III27} can be written as 
\be \frac{d}{dt}\sum_s S(f_s) = -\sum_{ss'} \int d^3x\int d^3v\, \ln f_s\, C_{ss'}(f_s,f_{s'})\geq 0, \lb{VIII11} \ee
with the logarithm on the RHS expanded as 
\bea  \ln f_s %&=& \ln F_s + \ln\left(1+ \frac{\delta f_s}{F_s}\right) \cr 
&\doteq & \ln F_s +  \frac{\delta f_s}{F_s}. \lb{VIII12} \eea
For a Maxwellian $F_{s}$ with temperature $T_s$ 
\be \ln F_{s}= -\frac{m_s v^2}{2T_s} + \log(c n_s/T_s^{3/2}), \lb{VIII13} \ee  
for a constant $c,$ and thus the contribution from $\ln F_{s}$ on the RHS of \eqref{VIII11} vanishes 
because of the equations
$\int d^3v\, C_{ss'}=0$ and $\sum_{ss'}\int d^3v\, (1/2)m_s|\bv|^2 C_{ss'}=0.$
The contribution from $\delta f_s/F_s$ in \eqref{VIII11} then gives the final quadratic-order $H$-theorem 
\bea && \frac{d}{dt}\sum_s \left[S(F_s) - \int d^3x\int d^3v\  \frac{(\delta f_s)^2}{2 F_{s}} \right] \cr
&& \hspace{20pt} 
= -\sum_s  \int d^3x\int d^3v\,  \frac{\delta f_s}{F_s} \left( \frac{\partial \delta f_s}{\partial t}\right)_c\ \geq 0\cr
&&
\lb{VIII14}  \eea
where 
\be \left( \frac{\partial \delta f_s}{\partial t}\right)_c=\sum_{s'} \left[C_{ss'}(F_s,\delta f_{s'})+
C_{ss'}(\delta f_s,F_{s'})\right] \lb{VIII15} \ee 
is the linearized collision integral and the condition $\int d^3x\, \delta f_s=0$ has been used again 
to eliminate the contribution from $C_{ss'}(F_s,F_{s'})$ on the RHS of \eqref{VIII14}. 

In the work \cite{howes2006astrophysical,schekochihin2008gyrokinetic,schekochihin2009astrophysical}, 
this $H$-theorem has been further reformulated as an equation for the dissipation of a ``generalized energy'' or 
``free energy''.  Noting that the entropy per volume for the Maxwellian $F_s$ is 
\be S(F_s)/V = n_s \ln(T_s^{3/2}/c n_s) + \frac{3}{2}n_s, \lb{VIII16} \ee 
then \cite{howes2006astrophysical}, Appendix B1, shows that to leading order $dn_s/dt=0$ (their Eq.(B3)). 
Thus, the entropy balance for a single species $s$ reduces, per volume, to 
\bea &&  \frac{1}{T_s}\frac{dE_{0s}}{dt}-\frac{d}{dt}\left[\int \frac{d^3x}{V}\int d^3v\  \frac{(\delta f_s)^2}{2 F_{s}} \right] \cr
&& \hspace{20pt} 
= - \int \frac{d^3x}{V} \int d^3v\,  \frac{\delta f_s}{F_s} \left( \frac{\partial \delta f_s}{\partial t}\right)_c\ 
+ \frac{1}{T_s}Q_s, \cr
&&
\lb{VIII17} \eea
with $E_{0s}=(3/2)n_s T_s$ the kinetic energy density for the Maxwellian $F_s,$ and $Q_s$ 
the collisional heat exchange defined in \eqref{III18}. The above eq.\eqref{VIII17} is the ``heating equation'' 
derived as eq.(B15) of \cite{howes2006astrophysical}, Appendix B2. As already discussed there, this 
``heating equation'' implies that the temperatures $T_s$ evolve on a time-scale $\sim 1/\epsilon^3\Omega_s,$
an order $O(\epsilon^{-2})$ longer than the evolution time-scale $1/\epsilon\Omega_s$ of the ring distribution functions $h_s.$ 
Because of the condition 
$\int d^3x\, \delta f_s=0$, one has also 
\be E_{0s} = \int \frac{d^3x}{V}\int d^3v\ \frac{1}{2}m_s |\bv|^2 f_s, \lb{VIII18} \ee
which shows that $E_{0s}$ is just the volume-average of the particle energy density $E_s$ 
defined in \eqref{III16}. 
Using this equation and the slow time evolution of $T_s,$ 
eq.\eqref{VIII17} is rewritten to leading order as
\cite{howes2006astrophysical,schekochihin2008gyrokinetic,schekochihin2009astrophysical}  
 \bea &&  \frac{d}{dt}\int \frac{d^3x}{V}\int d^3v\  \left[ \frac{1}{2}m_s |\bv|^2 f_s- \frac{T_s (\delta f_s)^2}{2 F_{s}} \right] \cr
&& \hspace{20pt} 
= -\int \frac{d^3x}{V} \int d^3v\,  \frac{T_s\delta f_s}{F_s} \left( \frac{\partial \delta f_s}{\partial t}\right)_c\ 
+ Q_s, \cr
&&
\lb{VIII19} \eea
which is equivalent to eq.(B11) in \cite{howes2006astrophysical}, Appendix B1, or eq.(9) in \cite{schekochihin2008gyrokinetic}. 
Summing over $s$ gives a valid formulation of the $H$-theorem for gyrokinetics, but the quantity 
in the square brackets is sign-indefinite. Using conservation of total energy with space density $E=\sum_s E_s + 
\frac{1}{8\pi}(|\bE|^2+|\bB|^2)$,  one can instead introduce a {\it free energy} or {\it generalized energy} 
with volume-average density 
\bea W %&=& E-\sum_{s} T_{s}[s(f_s)-s(F_s)] \cr 
           &=& \int \frac{d^3x}{V}\left[ \sum_s \int d^3v\ \frac{T_s(\delta f_s)^2}{2F_s} +  \frac{|\bE|^2+|\bB|^2}{8\pi}\right], \cr
           && \, 
           \lb{VIII20} \eea
which is non-negative and also dissipated, according to the balance equation 
\be \frac{dW}{dt}= \sum_s \int \frac{d^3x}{V} \int d^3v\,  \frac{T_s\delta f_s}{F_s} \left( \frac{\partial \delta f_s}{\partial t}\right)_c\ \leq 0. 
\lb{VIII21}\ee     
This coincides with the  eq.(B19) derived in \cite{howes2006astrophysical}, Appendix B3, or eq.(11) in 
\cite{schekochihin2008gyrokinetic} 
for the case of no external forcing.  Here we have emphasized how \eqref{VIII21} arises from the 
more general Vlasov-Maxwell-Landau model, but it can also be derived directly within the gyrokinetic 
description for the first-order fluctuations $\delta f_s^{(1)}$ in \eqref{VIII5}. See eqs.(73),(74) in 
\cite{schekochihin2009astrophysical}. 

Unfortunately, there is no obvious analogue of this ``free energy'' for the full VLM model in general. 
The analogous quantity would seem to be
\be  W= \sum_s T_s H(f_s|F_s) + 
\int \frac{d^3x}{V} \frac{|\bE|^2+|\bB|^2}{8\pi} \lb{VIII21a} \ee
where $F_s$ is a global Maxwellian with density $n_{0s}$ and temperature $T_s$ 
and the {\it relative entropy} is 
\be
H(f_s|F_s) =\int \frac{d^3x}{V}\int d^3v\, \left[ f_s\log(f_s/F_s) - f_s + F_s\right] 
                 \geq 0. 
\lb{VIII21b} \ee
Indeed, for a single-species plasma and for time-independent equilibrium parameters 
$n_{0},$ $T,$ the quantity $W/T$ is well-known to be both non-negative, convex and dissipated 
\cite{arsenio2013solutions}. A simple calculation gives
\begin{eqnarray} 
&& H[f_s|F_s] = -\left(S[f_s]-\frac{N_s}{N_{0s}}S[F_s]\right)/V \cr 
&& +\frac{1}{T_s} \int \frac{d^3x}{V}\ \left(E_s-\frac{3}{2}n_s T_s\right) 
    -\int \frac{d^3 x}{V}\ (n_s-n_{0s}) \cr
&&\,     
\lb{VIII21c} \end{eqnarray} 
If the parameters $n_s,$ $T_s$ of the reference Maxwellian are chosen so that 
\be \int d^3x \int d^3v\ \delta f_s=\int d^3x \int d^3v\ \frac{1}{2}|\bv|^2 \delta f_s=0, 
\lb{VIII21d} \ee
for $f_s=F_s+\delta f_s,$ then the last two terms in \eqref{VIII21c} vanish, $N_s=N_{0s}$ 
and $H[f_s|F_s]=(S[F_s]-S[f_s])/V\geq 0.$ The densities $n_{0s}$ are time-independent 
but the temperatures $T_s$ specified by \eqref{VIII21d} generally vary in time. By means 
of \eqref{VIII21c} one can write
\begin{eqnarray} 
&& W= \sum_s T_s \left(S[F_s]-S[f_s]\right)/V \cr 
&&+ \int \frac{d^3x}{V}\ \left(\sum_s E_s +\frac{|\bE|^2+|\bB|^2}{8\pi} -\sum_s \frac{3}{2}n_s T_s\right).  \cr
&&\, 
\lb{VIII21e} \end{eqnarray} 
Using the conservation of total energy and $d S[F_s]/dt=\frac{3}{2}(N_{0s}/T_s)dT_s/dt,$
it then follows that 
\be 
\frac{dW}{dt}= \sum_s \frac{dT_s}{dt} \left(S[F_s]-S[f_s]\right)/V 
-\sum_s T_s\frac{d}{dt} S[f_s]/V. \lb{VIII21f} \ee
The first term on the right is positive if, as seems plausible, $dT_s/dt>0.$ The second 
term on the left also cannot be shown to be negative, because the symmetrization 
argument using $s\leftrightarrow s'$ and $\bp\leftrightarrow \bp'$ with the 
Landau collision integral giving \eqref{III27} also takes $T_s\leftrightarrow T_{s'}.$ Only 
for $T_s=T$ and $dT/dt=0$ does one obtain $\frac{dW}{dt}= -T \int \frac{d^3x}{V}\ \sigma\leq 0$ exactly. 
The gyrokinetic result \eqref{VII21} holds to leading order because in that case similarly $T_s-T_{s'}=O(\epsilon^2)$
for $s\neq s'$ and $dT_s/dt=O(\epsilon^3\Omega_s)$ (see \cite{howes2006astrophysical}, footnote 8, p.595). 

\subsubsection{Scaling Exponent Predictions}

Gyrokinetics is expected to provide an asymptotic description as $\epsilon\to 0$ of a class of 
exact solutions of the Vlasov-Maxwell-Landau (VML) equations, including solutions that describe 
turbulent cascades of energy and entropy. 
A theory of these cascades may therefore be developed either within the reduced gyrokinetic description 
or within the more comprehensive VML model.  
Although energy and entropy are separate quantities with their own distinct balances for 
VML solutions, these quantities are intertwined into the single invariant $W$
in the works \cite{schekochihin2008gyrokinetic,schekochihin2009astrophysical} on astrophysical gyrokinetic turbulence. 
The cascades of $W$ discussed 
in those works are partially associated to energy cascade in the full VML description, and partially to entropy cascade. 
However, the flux of $W$ at the smallest collisionless scales which matches onto the anomalous entropy 
production by collisions (see section 2.5 in \cite{howes2006astrophysical}, section 5 in \cite{schekochihin2008gyrokinetic},  
sections 7.9.3, 7.12 in \cite{schekochihin2009astrophysical}) must be entirely due to entropy cascade in the VML 
description, since no energy dissipation anomalies are possible in the $Do\to \infty$ limit.  

Anomalous entropy production, both in the gyrokinetic and in the full VML description, requires short-distance 
divergences of solutions in phase-space, which must be regularized to allow for a dynamical description
in the collisionless limit. One may study this limit $Do\to \infty$ either before or after the limit $\epsilon\to 0.$ 
%However, the simple regularization using phase-space coarse-graining that we have applied to VML does not 
%apply straightforwardly to the gyrokinetic equations. One difficulty is that the Maxwell equations for the 
%fields $\varphi,$ $A_\|,$ $B_\|$ are written in physical space, whereas the ring distribution functions $h_s$
%are written in gyrocenter phase-space. The ring-averages required to relate these two spaces make
%the gyrocentric system a non-local set of integro-partial-differential equations rather than a local set 
%of partial differential equations. More importantly, the gyrokinetic equations for the ring-distributions 
%$h_s$ are not local conservations laws, like the kinetic equations \eqref{III2} for the distribution functions $f_s,$
%with all derivatives appearing 
Taking the limit first $\epsilon\to 0,$ then $Do\to\infty $ can be achieved with a suitable distributional or ``weak'' formulation 
of the gyrokinetic model equations, which we shall not attempt to develop here \footnote{This should be 
possible, in principle. The weak formulation of the reduced Ampere's law for $A_{|\!|},$ $\delta B_{|\!|}$ is 
straightforward. One can define $h_s$ to be a weak solution of the collisionless version of \eqref{VIII7} if 
$\int \left[\frac{\partial \psi}{\partial t} +v_{|\!|} \frac{\partial \psi}{\partial z} +
\frac{c}{B_0}\{\langle\chi\rangle_{\bX_s},\psi\} \right] h_s$
$ = -\int \frac{q_s F_s}{T_s}\frac{\partial \langle\chi\rangle_{\bX_s}}{\partial t} \psi$
for all smooth test functions $\psi=\psi(\bX_s,z,v_\perp,v_{|\!|},t),$ where integration is with respect 
to the measure $d^2\bX_s\, dz\,2\pi v_\perp dv_\perp\,dv_{|\!|}\,dt.$ This formulation suffices if the 
electromagnetic scalar and vector potentials $\varphi,$ ${\bf A}$ are at least once-differentiable 
in space and time. This is probably a reasonable assumption generally, but seems not to be true 
for the joint KAW/ion entropy cascade predicted by \cite{schekochihin2009astrophysical} 
in the limit $m_e/m_i\to 0$, 
when the scalar potential is expected to become extremely rough. See further discussion below}.  
Alternatively, one take first the limit $Do\to \infty$ of regularized VML solutions and then take 
$\epsilon\to 0$ as a subsidiary limit. This second order of limits is required if the collisional 
phase-space cutoff scales $\ell_c,$ $\uit_c$ (see section 2.5 in \cite{howes2006astrophysical}, 
section 5 in \cite{schekochihin2008gyrokinetic},  sections 7.9.3, 7.12 in \cite{schekochihin2009astrophysical},
and Appendix \ref{app:C}) 
are too small for the gyrokinetic approximation to be valid at those scales. In this order of limits, the 
coarse-graining regularization of VML solutions employed in the present work applies and all of our rigorous 
estimates of entropy flux carry over to gyrokinetics. Note that $\ell$ in our estimates should be 
understood to represent $\ell_\perp,$  when the scale-anisotropy $\ell_\|\gg\ell_\perp$ implied by 
\eqref{VIII4} holds in the limit $\epsilon\to 0.$  All fields are then smoother along the $\bB_0$-direction and, 
for fixed displacement length $r,$ increments are smaller for $\br\| \bB_0$ than for  $\br\perp \bB_0.$ 
Thus, averages over an isotropic kernel $G=G(r)$ are dominated by the increments with displacements 
$\br\perp \bB_0$ \cite{aluie2010scale}. Similar statements apply to $\uit,$ as $\delta_\bw$-increments 
are likewise dominated by the most singular direction in velocity space.

Based on these remarks, we may directly compare our exact inequalities \eqref{VI45} on the 
scaling exponents $\sigma_p^F:=\min\{\sigma_p^E,\sigma_p^B\},$ $\sigma_p^{f_s},$ and 
$\rho_p^{f_s}$ of orders $p\geq 3,$ required for entropy cascade, with the scaling predictions 
for gyrokinetic turbulence in \cite{schekochihin2008gyrokinetic,schekochihin2009astrophysical}. 
Those papers derive predictions 
for spectral exponents, or orders $p=2,$  but their results may be assumed to apply to all orders $p$ if 
intermittency effects can be ignored. Since the scaling exponents in question are non-increasing in $p,$ 
this ``mean-field'' approximation necessarily overestimates the true exponent values for $p\geq 3.$ 
Gyrokinetic theory assumes that background fields are smoother than fluctuations, so that 
$\sigma_p^B=\sigma_p^{\delta B},$ $\sigma_p^{f_s}=\sigma_p^{\delta f_s},$ and 
$\rho_p^{f_s=}\rho_p^{\delta f_s}.$ The first-order gyrokinetic result \eqref{VIII5} for $\delta f_s$ also implies 
that $\sigma_p^{\delta f_s}:=\min\{\sigma_p^\varphi,\sigma_p^{h_s}\}.$ \magc{Another 
general prediction of gyrokinetics is the relation $\uit/v_{th,s}\sim\ell/\rho_s$ that connects scaling 
in position and velocity space.  This relation is a consequence of the nonlinear perpendicular 
phase-mixing mechanism for entropy cascade in gyrokinetics, in which velocity-space structure 
arises from position-space structure due to the dependence of ring gyroradii on perpendicular velocity 
(Figure 1 in \cite{schekochihin2008gyrokinetic}; Figure 10 in \cite{schekochihin2009astrophysical}).} 
An immediate consequence is that velocity-space and position-space 
exponents are equal, or $\rho_p^{f_s}=\sigma_p^{f_s},$ in gyrokinetic trbulence. 

Specific predictions for scaling exponents in possible entropy cascade ranges of 
gyrokinetic turbulence have been developed phenomenologically in 
\cite{schekochihin2008gyrokinetic,schekochihin2009astrophysical} , 
for the particular case of a Maxwellian, two-species (electron-ion) plasma. The work 
\cite{schekochihin2009astrophysical} 
considered three different situations, which we briefly summarize here: 

\underline{(a) KAW/ion entropy cascade ($\uprho_e\ll \ell\ll\uprho_i$)}:  Section 7.9 of 
\cite{schekochihin2009astrophysical} 
considered an entropy cascade passively driven by a kinetic Alfv\'en wave (KAW) cascade, 
assuming $\uprho_i/L_i\lesssim 1,$ $m_e/m_i\ll 1$.  
Their predictions, expressed in terms of scaling of increments, are: 
\be ``\delta_\ell E\mbox{{\,}''} \sim \ell^{-1/3}, \quad \delta_\ell B\sim \ell^{2/3} \lb{VIII22}\ee
\be  \delta_\ell f_i\sim \ell^{1/6}, \quad \delta_\uit f_i \sim\uit^{1/6}  \lb{VIII23}\ee  
so that 
\be \sigma_p^F= -\frac{1}{3}, \quad \sigma^f_p=\frac{1}{6}, \quad \rho^f_p=\frac{1}{6}. 
\lb{VIII24}\ee
Quotation marks ``$\,$'' appear around the electric-field term in \eqref{VIII22}
because increments no longer suffice to define scaling exponents, in the same manner 
as in \eqref{VI33},  when the exponents become negative. Instead, one must use 
some sort of smooth low-pass or band-pass filter, e.g. wavelet coefficients as in 
\cite{eyink1995besov,jaffard1997multifractal} 
\footnote{If the electric field has the extreme irregularity
predicted in \eqref{VIII22} as $m_e/m_i\to 0,$ then the coarse-graining analysis of the 
present paper cannot be used to analyze that limit. For example, the phase-space 
Favre-average electric field $\hat{\bE}_s$ that appears in the coarse-grained Vlasov equation 
\eqref{V9} is no longer well-defined by \eqref{IV10}, because the fields $\bE$ and $f_s$ are too 
irregular in the limit $m_e/m_i\to 0$ for their pointwise product $\bE f_s$  to be 
{\it a priori} meaningful. This is largely a technical issue, but it means that the 
coarse-graining regularization as employed in this paper probably does not 
suffice to study the KAW/ion entropy cascade in the $m_e/m_i\to 0$ limit.}.  

\underline{(b) Pure ion entropy cascade ($\uprho_e\ll \ell\ll\uprho_i$)}: 
Section 7.10 of \cite{schekochihin2009astrophysical} , under the same limit conditions 
$\uprho_i/L_i\lesssim 1,$ $m_e/m_i\ll 1$ but assuming now no KAW cascade 
and assuming also $h_e=0$, predicted: 
\be \delta_\ell E\sim \ell^{1/6}, \quad \delta_\ell B \mbox{ (or  $\delta^3_\ell B$)}\sim \ell^{13/6} \lb{VIII25}\ee
\be  \delta_\ell f_i\sim \ell^{1/6}, \quad \delta_\uit f_i \sim\uit^{1/6}   \lb{VIII26}\ee  
so that 
\be \sigma_p^F= \frac{1}{6}, \quad \sigma^f_p=\frac{1}{6}, \quad \rho^f_p=\frac{1}{6}. \lb{VIII27}\ee
In this case magnetic fluctuations are very small over the range considered, so that the 
entropy cascade is self-driven by the electrostatic fields arising from fluctuations in the ion distribution. 
Note that the high smoothness of the magnetic field (scaling exponent $>2$) implies that its 
1st-order increments scale as $\delta B_\ell\sim \ell.$ Thus, the scaling exponent 
as defined in \eqref{VI33} is $\sigma_p^B=1$. To obtain instead $\sigma_p^B=13/6,$ one must replace 
the 1st-order increments in \eqref{VI33} with 3rd-order increments, so that the $O(\ell)$,
$O(\ell^2)$ terms in the Taylor-expansion are cancelled. See \cite{eyink1995besov,jaffard1997multifractal} 
for a general discussion. 

\underline{(c) Electron entropy cascade ($\ell\ll \uprho_e$)}:  
Section 7.12 of \cite{schekochihin2009astrophysical}, assuming $\uprho_e<\uprho_i\lesssim L_i,$ considered 
a pure electron entropy cascade, with contributions of ion distribution $h_i$ neglected
(e.g. because the gyroaveraging makes its contributions subdominant in powers of $m_e/m_i$) 
\be \delta_\ell E\sim \ell^{1/6}, \quad \delta_\ell B \mbox{ (or  $\delta^3_\ell B$)}\sim \ell^{13/6} \lb{VIII28}\ee
\be  \delta_\ell f_e\sim \ell^{1/6}, \quad \delta_\uit f_e \sim\uit^{1/6}   \lb{VIII29}\ee  
so that 
\be \sigma_p^F= \frac{1}{6}, \quad \sigma^f_p=\frac{1}{6}, \quad \rho^f_p=\frac{1}{6}. \lb{VIII30}\ee
The scaling exponents are identical to those for the pure ion entropy cascade and, indeed, 
the physics is very similar, with electrostatic fields created by fluctuations in the electron distribution 
driving cascade of electron entropy. 

Comparing these various predictions with our inequalities \eqref{VI45}, the first observation 
is that our exact constraints required for an entropy cascade to exist are well-satisfied by the predictions 
of \cite{schekochihin2009astrophysical} for all three cases. Secondly, the inequalities are not satisfied
as near-equalities, but instead with the predicted exponents yielding a value considerably below 
the upper bound in \eqref{VI45}. A somewhat similar situation occurs also in incompressible fluid 
turbulence, where the corresponding inequality $\sigma^u_p<1/3$ is satisfied with values of $\sigma^u_p$ 
much smaller than $1/3$ for $p\geq 3.$ For incompressible turbulence, this is a consequence 
of space-time intermittency (see e.g. \cite{eyink1995local}), but, \magc{as there,  a ``mean-field'' approximation
which neglects effects of intermittency should be approximately valid for exponents of order $p\simeq 3.$}
We believe that the large gap is due instead to the strong depletion of nonlinearity in gyrokinetics, 
arising from substantial cancellations in the ring-averages \eqref{VIII8}, and which is not taken into account 
in our upper bounds \eqref{VI34}-\eqref{VI37} on entropy flux. In order to compensate for the reduced 
nonlinearity, more singular scaling behavior than what follows from \eqref{VI45} 
is thus required in gyrokinetic turbulence in order to sustain the cascade of entropy. 

\subsection{Empirical Studies}\lb{sec:VIIIb}

We here briefly review the available evidence for kinetic entropy cascades from  empirical studies 
and discuss also some promising situations in space plasmas where they are likely to exist. 

Numerical simulations of gyrokinetic turbulence have provided, so far, the best direct evidence 
for nonlinear entropy cascades in turbulent plasmas \footnote{\black{A recent 
numerical simulation of a hybrid Vlasov-Maxwell system \cite{cerri2018dual} provides the 
best evidence so far of an ion entropy cascade beyond the gyrokinetic description. This 
simulation employs a fully kinetic Vlasov-Mawell equation for ions, but a generalized Ohm's law 
for an electron fluid. A full discussion of this simulation would be too lengthy for the current 
paper, but we note here briefly that the spectra observed in \cite{cerri2018dual} over the 
range $1\lesssim \uprho_i k_\perp\lesssim 10$ of perpendicular wavenumbers and 
$1\lesssim m_\|\lesssim 10$ of Hermite velocity modes are equivalent to $\sigma_2^E=-1/6,$ 
$\sigma_2^B=5/6,$ and $\sigma_2^{f_i}=\rho_2^{f_i}=-1/2$ in our language \cite{mhaskar2017local}. The authors
of \cite{cerri2018dual} interpreted this range as an ion entropy cascade driven by linear 
phase-mixing along field-lines, similar to that discussed in 
\cite{zocco2011reduced,kanekar2015fluctuation}. The observed scaling is consistent with an ion entropy 
cascade according to our analysis, but we find that all three terms \eqref{VI35}-\eqref{VI37} 
of the entropy flux could contribute, with the nonlinear wave-particle interaction terms perhaps even dominant.
The range $\uprho_i k_\perp\gtrsim 10,$ $m_\|\gtrsim 10$ observed in the simulation of 
\cite{cerri2018dual} is very unlikely to correspond to an asymptotic ion entropy cascade 
according to our analysis, unless due to an extreme phase-space intermittency}}
The studies \cite{tatsuno2009nonlinear,tatsuno2010gyrokinetic}
has considered decaying, electrostatic turbulence in a spatially 2D setting, with no 
variations parallel to $\bB_0,$ in order to eliminate damping by the Landau resonance. 
The spatial domain-size was $2\pi\uprho\times 2\pi\uprho,$ with $\uprho$ the gyroradius. A smooth, 
unstable initial condition was chosen for $\delta f,$ perturbed by small-amplitude white noise, 
together with the corresponding electrostatic potential $\varphi.$ This initial configuration was evolved 
under the gyrokinetic dynamics for three cases with decreasing collisionality ($Do=48,$ 118, 440)
and correspondingly increased numerical resolution. The collisional entropy production was found to be 
only weakly dependent on $Do$ and spectrally-local, nonlinear fluxes of entropy were observed to small scales 
in position-space and velocity-space.  The scaling behavior found in this study was quite close to that 
predicted in cases (b),(c) above, with Fourier spectra $E_h(k_\perp), E_{E_\perp}(k_\perp)\sim k_\perp^{-4/3}$ 
and identical scaling in the Hankel-transform velocity spectrum of $h$. A similar study 
\cite{navarro2011free} has also considered electrostatic gyrokinetic turbulence, but now in 3D 
and ion temperature gradient-driven. A statistical steady-state was reached with artificial 
hyperdiffusion added in position and velocity space. Despite the fact that such dissipation 
acted effectively at all scales, this study observed scale-local, nonlinear entropy cascade and obtained 
spectra similar to those in the study \cite{tatsuno2009nonlinear,tatsuno2010gyrokinetic}. 

In a different direction, the paper \cite{howes2011gyrokinetic}
performed a 3D, fully electromagnetic, gyrokinetic simulation of an ion-electron plasma 
designed to reproduce the turbulent KAW/ion entropy cascade of 
\cite{schekochihin2009astrophysical} 
(case (a) above). The size of the spatial domain was $L_\perp=2\pi\uprho_i$ and $L_\|\gg L_\perp,$ with a $128^3$
spatial grid able to resolve the electron gyroradius $\uprho_e\doteq \uprho_i/42.8.$
The simulation was driven by an ``antenna current''  set up to mimic energy input from a 
critically-balanced cascade of Alfv\'en waves and collisions were incorporated by a fully conservative, 
linearized collision operator. The field spectra observed were close to $E_{E_\perp}(k_\perp)\sim 
k_\perp^{-1/3}$ and $E_{B_\perp}(k_\perp), E_{B_\|}(k_\perp)\sim k_\perp^{-2.8},$ with the 
latter somewhat steeper than the $k_\perp^{\black{-7/3}}$ spectrum predicted in 
\cite{schekochihin2009astrophysical}. This steepening was plausibly explained by the 
finiteness of the mass ratio $m_e/m_i$ and the damping of KAW modes by Landau 
resonance with electrons, which peaks in the simulation at $k_\perp\uprho_e\sim 1$ but 
is increasing roughly as a power-law over the entire $k_\perp$-range.  The important point 
here is that the collisionless input into $h_i$ by the Landau resonance with ions
peaked at $k_\perp\uprho_i\sim 1$ but the collisional ion heating peaked at higher wave number 
$k_\perp\uprho_i\sim 20$. This is consistent with the presence of an ion entropy cascade. 
\magc{See also \cite{told2015multiscale,navarro2016structure}.}\\

\vspace{-16pt} 
Entropy cascade should occur not only within \black{numerical simulations} but 
quite ubiquitously at small scales in turbulent plasmas of very weak 
collisionality, with the solar wind and the terrestrial magnetosheath as likely examples. 
We know of no direct evidence of non-vanishing entropy flux in such environments, although 
high-resolution measurements of ion distribution functions in the magnetosheath do reveal 
complex velocity-space structure \cite{servidio2017magnetospheric}. Furthermore, {\it in situ} observations 
of magnetic field spectra broadly agree with gyro-simulations exhibiting entropy cascade. 
As recently reviewed \cite{sahraoui2013scaling}, solar wind spectra are well fit as power-laws  
$E_{B_\perp}(k_\perp)\sim k_\perp^{-x}$ for $1/\uprho_i\lesssim k_\perp\lesssim 1/\uprho_e$ 
and $E_{B_\perp}(k_\perp)\sim k_\perp^{-y}$ for $1/\uprho_e\lesssim k_\perp,$ with a distribution 
of exponents $x\in [2.5,3.1]$ 
peaked at $x=2.8,$ and $y\in [3.5,-5.5]$ peaked at $y=4.$ In the terrestrial \black{magneto}sheath, 
paper \cite{huang2014kinetic}
reports similar scaling but with $x\in [2.4,3.5]$ peaked at $x=2.9,$ and
$y\in [4,7.5]$ peaked at $y=5.2.$  Clearly, the magnetic spectra observed in the range $1/\uprho_i\lesssim 
k_\perp\lesssim 1/\uprho_e$ for both the solar wind and heliosheath agree reasonably well with 
the simulation of the KAW/ion entropy cascade in \cite{howes2011gyrokinetic}. Another paper \cite{sahraoui2009evidence} 
reported in the solar wind an electric spectrum $E_{E_\perp}(k_\perp)\sim k_\perp^{-0.3}$ fitted over the decade $k_\perp
\uprho_i\in [0.43,4.3]$,  roughly consistent with the prediction $E_{E_\perp}(k_\perp)
\sim k_\perp^{-1/3}$ of \cite{schekochihin2009astrophysical} for the KAW/ion entropy cascade. See as well 
\cite{salem2012identification}. 
At sub-electron scales $1/\uprho_e\lesssim k_\perp$ the magnetic spectra 
reported for both the solar wind and \black{magneto}sheath in these references appear also to be roughly in agreement 
with the prediction $E_{B_\perp}(k_\perp)\sim k_\perp^{-16/3}$ of \cite{schekochihin2009astrophysical} for the electron 
entropy cascade. Agreement is clearly best for the \black{magneto}sheath where, as pointed out in 
\cite{sahraoui2013scaling,huang2014kinetic}, 
the signal-to-noise ratio of measurements is higher than for the solar wind and 
where, therefore, the spectral slopes are more reliable. 

Although reasonably identified as entropy cascades, these turbulent space plasmas are likely not accurately described 
by gyrokinetics all the way down to collisional scales. 
The gyrokinetic approximation is estimated break down in the solar wind at a length-scale between $\uprho_i$ and $\uprho_e$
\cite{howes2006astrophysical}, %$\omega\sim\Omega_i$ and ,  
but the collisional cutoffs for both ion and electron entropy cascades %in such weakly collisional environments 
should lie at much smaller scales.  
The cutoff scale for the ion entropy cascade is $\ell_c\sim\uprho_i  Do_i^{-3/5}$ 
within gyrokinetic theory \cite{schekochihin2009astrophysical}, where 
ion-scale Dorland number is given by $Do_i=1/\nu_{ii}\tau_{\rho_i}$ for ion-ion Coulomb collision rate
$\nu_{ii}$ and eddy-turnover rate $\tau_{\rho_i}$ at the ion gyroradius. In the solar wind at 1 AU 
$\nu_{ii}\sim 3\times 10^{-7}$ Hz and $\uprho_i\sim 100$ km. From $\tau_{\rho_i}\sim \varepsilon^{-1/3}\uprho_i^{2/3}$ and 
using $\varepsilon \sim 10^4\, {\rm m}^2/{\rm sec}^3$ from 3rd-moment measurements \cite{coburn2015third}, 
one can estimate $\tau_{\rho_i}\sim 10$ sec. Thus, $Do_i\sim 10^5$ and the collisional cutoff scale for ion 
entropy cascade calculated within gyrokinetics is $\ell_c\sim 10^{-3}\uprho_i$ or smaller.  
Similar estimates apply to the cutoff $\ell_c\sim \uprho_e Do_e^{-3/5}$ for the electron entropy cascade, 
with electron-scale Dorland number $Do_e=1/\nu_{ei}\tau_{\rho_e}.$ Note that the electron-ion collision rate $\nu_{ei}$ 
is larger than $\nu_{ii}$ by a factor of $(m_i/m_e)^{1/2}$ but the electron-scale turnover rate $\tau_{\rho_e}$ is smaller than $\tau_{\rho_i}$ 
by a comparable factor. If these various estimates are accurate, entropy cascades in the solar wind and terrestrial 
magnetosheath must extend down to scales well below those where gyrokinetics is valid. 

The description of such kinetic cascades is one of the principal motivations for the theory developed in the present work.
Measured magnetic and electric spectra in the solar wind 
\cite{sahraoui2013scaling,salem2012identification}
and in the magnetosheath \cite{mangeney2006cluster,matteini2017electric}
indicate that the turbulence at sub-electron scales in those environments is probably ``electrostatic,''   
with electric fluctuations much larger than magnetic fluctuations. Therefore, the dominant contribution to the entropy flux 
is presumably the electric-field contribution \black{\eqref{VI24}} from the wave-particle interaction. 
Future work will exploit this formalism to elucidate further the physics of this phase-space cascade. 

%$\,$
%\newpage

\subsection{Turbulent Magnetic Reconnection}\lb{sec:VIIIc}

The results on coarse-grained momentum balance in section \ref{sec:VIIb} of this paper 
also make connection with prior work on turbulent magnetic reconnection and provide it with 
a deeper theoretical foundation. As is well-known, the momentum balance equations  
for an electron-ion plasma yield a ``generalized Ohm's law'' for the electric field 
\cite{vasyliunas1975theoretical,bhattacharjee1999impulsive,priest2000magnetic}. 
For a turbulent plasma, the coarse-grained momentum balance equations, \eqref{VII7} or \eqref{VII9}, for the 
two species $s=i,e$ can be combined, using the formula $\bj=e(n_i\bu_i-n_e\bu_e)$ for the 
electric current and assuming quasi-neutrality ($n_e=n_i=n$), to give: 
\bea 
&& \tbE+\frac{1}{c}\widetilde{\bu_i\btimes \bB}=\frac{1}{\overline{n}e}\overline{\bR} 
+\frac{m_i}{m_i+m_e}\frac{\overline{\bj\btimes\bB}}{\bar{n}ec}\cr
&& \hspace{30pt} -\frac{1}{\overline{n}e}\grad\bdot\left( \frac{m_i\overline{\bP}_e-m_e \overline{\bP}_i }{m_i+m_e} \right)\cr 
&& \hspace{30pt} +\frac{m_em_i}{\bar{n}e^2(m_i+m_e)}\left[\partial_t\obj+\grad\bdot(\overline{
\bj\bu_i+\bu_i\bj-\bj\bj/ne})\right], \cr
&&\, \lb{VIII31} \eea
Here we have retained the collisional drag forces $\pm \overline{\bR}$ on the electrons/ions, respectively. 
Unresolved turbulent eddies can be considered to contribute two new terms to this 
coarse-grained Ohm's law. One is the {\it velocity-fluctuation induced electric field} defined by 
\be \magc{\widetilde{\boeps}_{u_i}} =  \frac{1}{c}\widetilde{\tau}(\bu_i \btimescom\bB) := 
\frac{1}{c}[\widetilde{\bu_i\btimes\bB}-\widetilde{\bu_i}\btimes\tbB]. \lb{VIII32} \ee 
This effect was already considered in the theory of turbulent reconnection for an incompressible fluid 
by Matthaeus \& Lamkin  (\cite{matthaeus1986turbulent}, section X.D) 
and, in the density-weighted Favre-formulation, by \cite{eyink2015turbulent}, section 6. 
For a compressible flow, however, there is another turbulence effect. Because it is $\obE,$ $\obB$ 
that appear in the coarse-grained Maxwell equations \eqref{V6} and not $\tbE,$ $\tbB,$ one should write 
\be   \tbE+\frac{1}{c}\tilde{\bu}_i\btimes\tbB= \obE+\frac{1}{c}\tilde{\bu}_i\btimes\obB + \magc{\widetilde{\boeps}_n}  
\lb{VIII33}\ee 
with {\it density-fluctuation induced electric field} 
\be  \magc{\widetilde{\boeps}_{n}} :=\frac{1}{\bar{n}}(\otau(n,\bE)+\tilde{\bu}_i\btimes\otau(n,\bB)/c). \lb{VIII34} \ee 
Here we have used the general relation $\widetilde{b}=\overline{b}+\otau(n,b)/\overline{n}$ between unweighted 
and Favre-weighted spatial coarse-graining, analogous to \eqref{IV15}. This second electric-field 
contribution from turbulent density fluctuations was pointed out in \cite{eyink2015turbulent}, eq.(6.11). 
\magc{The sum of these two electric fields $\widetilde{\boeps}_i=\widetilde{\boeps}_{u_i}+\widetilde{\boeps}_n$ 
coincides with the ``turbulent electromotive force'' defined in eq.\eqref{VII31c} for $s=i.$} 

Magnetic reconnection at length-scale $\ell$ in a turbulent plasma is thus governed by the 
generalized Ohm's law 
\bea 
&& \obE+\frac{1}{c}\tilde{\bu}_i\btimes\obB =-\magc{\widetilde{\boeps}_i}
+\frac{1}{\overline{n}e}\overline{\bR}+\frac{1}{\bar{n}ec}
\overline{\bj\btimes\bB}-\frac{1}{\overline{n}e}\grad\bdot\overline{\bP}_e\cr 
&& \hspace{30pt} +\frac{m_e}{\bar{n}e^2}\left[\partial_t\obj+\grad\bdot(\overline{
\bj\bu_i+\bu_i\bj-\bj\bj/ne})\right], \cr
&&\, \lb{VIII35} \eea
assuming for simplicity a small mass ratio $m_e/m_i\ll 1,$ which recovers eqs.(6.2),(6.10) 
of \cite{eyink2015turbulent}.  
In  \cite{eyink2015turbulent}, eq.(6.2) the collisional drag force was represented by an Ohmic field 
$\bR/e n=\eta\bj$ with Spitzer resistivity $\eta$ and it was argued from this representation that the drag 
term is negligible in a weakly collisional plasma such as the solar wind. Strictly speaking, such an 
argument is only valid for coarse-graining length $\ell\gg \lambda_{mfp,e},$ the mean-free path 
of the electrons, since it is only at such scales that that the drag force is correctly represented 
by Ohmic resistivity \cite{braginskii1965transport}. On the other hand, the estimate \eqref{VII4} in the present 
work shows more generally that the collisional drag term vanishes as $Do\to 0$ at any fixed length-scale $\ell$ 
in the coarse-grained momentum balance equations \eqref{VII7} or \eqref{VII9}. 

It was further shown in \cite{eyink2015turbulent}
that all of the microscopic non-ideal electric fields terms on the righthand side 
of the generalized Ohm's law \eqref{VIII31} are negligible in the inertial-range of the solar wind. Assuming 
the scaling of increments that are observed at length scales $\uprho_i\ll\ell\ll L_i$ in the solar wind and that are 
expected generally for MHD-like turbulence, the analysis showed that the non-ideal 
terms are all suppressed by powers of $\delta_i/\ell$ or $(\delta_i/\ell)^2$ at length scale $\ell$, with 
$\delta_i$ the ion skin-depth. The non-ideal terms are thus like (infrared) irrelevant variables in the 
technical RG sense.  Here we may note that the plasma dynamics in the ``inertial-range'' of the solar wind, 
for $\uprho_i\ll\ell\ll L_i$, has been previously argued to be governed by ``kinetic RMHD'' in the works 
\cite{schekochihin2009astrophysical} , section 5, and \cite{kunz2015inertial}, by means of gyrokinetic theory. 
In particular, the dominant component of incompressible, shear-Alfv\'en waves in that range 
was argued to be described by ``reduced MHD'' (RMHD) and the magnetic field to be governed by the ideal 
induction equation.  Our analysis here and in  \cite{eyink2015turbulent} agrees with the latter conclusion. However, papers
\cite{schekochihin2009astrophysical,kunz2015inertial} both go on to argue that, as a consequence, the magnetic 
field at inertial-range scales is ``frozen in'' to the ion flow, e.g. ``At $k_\perp\uprho_i\ll1$, ions (as well as the electrons) 
are magnetized and the magnetic field is frozen into the ion flow'' \cite{schekochihin2009astrophysical}. 
This statement is incorrect. Insofar as the ideal induction equation holds in the inertial-range 
of the solar wind, it does not imply magnetic flux-freezing at those scales, and insofar as the 
ideal induction equation implies magnetic flux-freezing, it is not valid in the inertial-range 
of the solar wind. 

As pointed out in \cite{eyink2015turbulent}, an ``ideal Ohm's law'' holds in the inertial range of the solar wind only in the sense 
that the equality 
\be \tbE+\frac{1}{c}\widetilde{\bu_i\btimes \bB}=\bzed \lb{VIII36} \ee
is well-satisfied for length scales $\ell\gg\uprho_i.$ Validity of the ideal 
Ohm's law in this ``weak'' or ``coarse-grained'' sense, however, does {\it not} imply that the 
magnetic field at those scales is ``frozen-in'' to the velocity $\tilde{\bu}_i.$ This becomes 
obvious if one rewrites the ``ideal Ohm's law"  \eqref{VIII36} equivalently as 
\be \obE+\frac{1}{c}\tilde{\bu}_i\btimes\obB =-\magc{\widetilde{\boeps}_i} , \lb{VIII37} \ee 
which makes apparent that the turbulent 
electromotive force \magc{$\widetilde{\boeps}_i$} breaks flux-freezing at those scales. 
\magc{Keeping the contribution $\obj_i\bdot\widetilde{\boeps}_i$ to energy cascade 
in eq.\eqref{VII3b} while discarding $\widetilde{\boeps}_i$ spuriously from the Ohm's law eq.\eqref{VIII37}  in order to infer 
``flux-freezing'' at scales $\ell\gg\uprho_i$ is a fundamental inconsistency.} As recognized in the work of Lazarian \& Vishniac 
\cite{lazarian1999reconnection}, reconnection must occur for eddies at {\it all} scales $\ell$ 
in a turbulent plasma. In fact, due to the turbulent contributions, 
magnetic-flux conservation may be anomalous and violated in the limit first $\max\{\uprho_i,\delta_i\}/L_i\to 0,$ then $\ell/L_i\to 0$ 
\cite{vishniac1999fast,eyink2006breakdown}.  Magnetic flux-structures with dimensions much larger 
than $\uprho_i$ or $\delta_i$ which are embedded in a turbulent inertial-range may therefore undergo reconnection at rates 
which are independent of microscopic physics and determined solely by the inertial-range turbulence. 
A concrete example of this type has been studied numerically in \cite{lalescu2015inertial} using a database of 
incompressible MHD turbulence, where it was shown the electric field \magc{$\widetilde{\boeps}_{u_i}$} induced by 
turbulent velocity fluctuations accounts for the reconnection at inertial-range scales. An empirical study 
in  \cite{eyink2015turbulent}
using spacecraft data suggests that in the solar wind the compressible contribution \magc{$\widetilde{\boeps}_n$} 
plays a relatively small role and that inertial-range reconnection there is also due primarily to the ``ideal" electric field
\magc{$\widetilde{\boeps}_{u_i}$} induced by velocity fluctuations of unresolved eddies. 

Similar remarks hold for reconnection of magnetic structures at sub-ion scales, which is 
generally treated by rewriting the generalized Ohm's law to refer to the electron fluid. 
Turbulent reconnection at sub-ion scales $\ell<\uprho_i$ may likewise be treated by 
by rewriting the coarse-grained Ohm's law \eqref{VIII31} in terms of the electron bulk velocity, 
yielding 
\bea 
&& \tbE+\frac{1}{c}\widetilde{\bu_e\btimes \bB}=\frac{1}{\overline{n}e}\overline{\bR} 
-\frac{m_e}{m_i+m_e}\frac{\overline{\bj\btimes\bB}}{\bar{n}ec}\cr
&& \hspace{30pt} -\frac{1}{\overline{n}e}\grad\bdot\left( \frac{m_i\overline{\bP}_e-m_e \overline{\bP}_i }{m_i+m_e} \right)\cr 
&& \hspace{30pt} +\frac{m_em_i}{\bar{n}e^2(m_i+m_e)}\left[\partial_t\obj+\grad\bdot(\overline{
\bj\bu_e+\bu_e\bj+\bj\bj/ne})\right]. \cr
&&\, \lb{VIII38} \eea
For weak collisionality and $m_e/m_i\ll 1$
\bea 
&& \obE+\frac{1}{c}\tilde{\bu}_e\btimes\obB =-\magc{\widetilde{\boeps}_e} 
-\frac{1}{\overline{n}e}\grad\bdot\overline{\bP}_e\cr 
&& \hspace{30pt} +\frac{m_e}{\bar{n}e^2}\left[\partial_t\obj+\grad\bdot(\overline{
\bj\bu_e+\bu_e\bj+\bj\bj/ne})\right], \cr
&&\, \lb{VIII39} \eea
with $\widetilde{\boeps}_e$ \magc{given by eq.\eqref{VII31c} for $s=e$.} 
The estimates in  \cite{eyink2015turbulent} show that the contributions from the 
electron pressure tensor and electron inertia are suppressed by powers of $\delta_e/\ell$ \footnote{The 
estimate in \cite{eyink2015turbulent}, section 4.2.4 can be rewritten as $(1/\overline{n}e)\grad\bdot \overline{{\bf P}}_e\sim
(\delta_e/\ell)\sqrt{\beta_e} v_{th,e} B/c$}. Therefore, when $\ell\gg \delta_e,$ then the Ohm's 
law referred to the electron fluid is ``ideal''  but magnetic fields are nevertheless not frozen-in into 
the velocity $\tilde{\bu}_e$ because of the turbulent contribution \magc{$\widetilde{\boeps}_e
=\widetilde{\boeps}_{u_e}+\widetilde{\boeps}_n$}. 
When $\ell\sim \rho_i\sim \delta_i$ (assuming $\beta_i\sim 1$), then 
the non-ideal electric fields are suppressed by a factor of only $\sim 1/43$ relative to the turbulent 
contributions and need not be entirely negligible. When $\ell\sim\delta_e\sim \rho_e$ then the non-ideal 
contributions will begin to dominate.  Price et al. \cite{price2016effects,price2017turbulence}
have suggested based upon 
3D PIC simulations that the $\otau(n,\bE)/\overline{n}$ contribution in \magc{$\widetilde{\boeps}_n$} 
plays an important (but not dominant) role in dayside magnetopause reconnection observed by MMS,
with turbulence self-driven by the reconnection itself.  Magnetic reconnection of ion and electron-scale 
structures is also observed in the terrestrial magnetosheath \cite{retino2007situ,yordanova2016electron}.
There is strong pre-existing turbulence in this environment which should contribute significantly 
to reconnection of magnetic structures at length-scales $\ell\sim \rho_i\sim \delta_i.$
 
\section{Conclusions and Outlook}\lb{sec:IX} 

This paper has systematically explored the hypothesis \cite{krommes1994role,krommes1999thermostatted}
that entropy production in a weakly coupled, multi-species plasma may remain non-zero in the 
limit of vanishing collisionality. This hypothesis implies that there will be thermalization of the plasma 
or a tendency of velocity distribution functions to evolve toward Maxwellian, even as the dimensionless 
collision rate tends to zero. This tendency is consistent with particle distribution functions for 
driven systems remaining very far from Maxwellian and with large mean entropy-production 
in long-time steady states. The earlier conjecture of \cite{schekochihin2008gyrokinetic,schekochihin2009astrophysical} 
that such non-vanishing dissipation may occur by a turbulent cascade of entropy through phase-space,
based on gyrokinetic theory, has been shown here to be the necessary consequence of an 
entropy production anomaly. In close analogy with Onsager's ``ideal turbulence" theory
for incompressible fluids, we have shown that the dynamics of the plasma at fixed 
length and velocity scales in the collisionless limit is governed by a ``weak'' or ``coarse-grained'' 
solution of the Vlasov-Maxwell equations. Although smooth solutions of the Vlasov-Maxwell system 
conserve entropy, the solutions \magc{suggested by our analysis} violate that conservation law by a 
nonlinear cascade of entropy. We obtain an explicit formula for the entropy flux through 
phase-space, \magc{which we use to predict specific correlations (down-gradient transport)   
and specific types of} singularities/scaling exponents required to sustain a non-vanishing entropy cascade. 
Our results are consistent with gyrokinetics, but are more general, 
because they do not require any of the specific conditions assumed for validity 
of gyrokinetic theory (evolution rates small compared with gyrofrequencies,  scale-anisotropy, etc.).
Our sole assumption is weak collisionality. Our conclusions are thus widely applicable, 
holding, for example, at all scales in the solar wind smaller than the Coulomb mean-free-path 
length and larger than the Debye screening length. The collisionless entropy cascade discussed in this work 
should occur and be observable at sub-ion and sub-electron scales in the solar wind and the 
terrestrial magnetosheath.

We have also considered in this paper the balances of the standard collisional invariants: 
mass, momentum, and energy. Although conserved overall, these quantities can be 
converted from one form to another by Coulomb collisions of the particles 
(e.g. momentum may be transferred 
from one particle species to another). We show that such collisional transfers cannot be 
anomalous, but instead must vanish in the collisionless limit. Anomalies may appear in 
subsidiary limits, however, such as gyroradii small compared with turbulence injection scales 
($\uprho_s/L_s\ll 1$). For example, the electron momentum equation reduces in that limit 
to an ideal Ohm's law, but only in a ```weak'' or ``coarse-grained'' sense that does not 
imply the frozen-in property of magnetic flux \magc{and that predicts instead reconnection 
of ``magnetic eddies'' at all inertial-range scales}.  Likewise, energy transfers through length-scales 
and velocity-space may be anomalous in such a small gyroradius limit, including a novel 
phase-space redistribution effect. The energy balance 
equations that we derive in this work,  resolved simultaneously in phase-space and in scale,   
generalize and unify previous results in the literature 
\cite{howes2017prospectus,klein2017diagnosing,yang2017Aenergy,yang2017Benergy}. 
They provide a basis for the study both of turbulent energy cascade and of nonlinear Landau 
damping in a turbulent setting. 

Because energy is not dissipated by collisions in the Vlasov-Maxwell-Landau theory,
it is useful to address briefly the question of the ultimate sink of energy cascaded to small 
\magc{length-}scales. 
The answer to this question is clearly situation-dependent. In some cases, there may be 
no sink at all, with energy simply accumulating \magc{in kinetic velocity fluctuations after cascading 
to small length-}scales. This seems to be the case in 
the solar wind,  where turbulent cascade appears to provide the energy required to offset the 
``cooling'' due to adiabatic expansion \cite{kiyani2015dissipation}. Of course, this energy input does not  
necessarily correspond to a temperature increase of a Maxwellian velocity distribution, but 
may correspond instead to non-Maxwellian tails and supra-thermal particle production. 
In other cases, e.g. the solar corona, the particle kinetic energy cascaded to small scales may be carried 
off by electromagnetic radiation. This process is not described within Maxwell-Vlasov-Landau theory, 
which assumes elastic Coulomb collisions that conserve the total kinetic energy of charged particles. 
Radiative processes such as bremmstrahlung involve inelastic particle collisions with emission of photons 
and their treatment requires separate consideration of plasma emissivity \cite{zheleznyakov1996radiation}. 
\magc{Likewise, thermal radiation which carries off both energy and entropy requires a kinetic model of 
the photon gas that is coupled with the kinetic equations for the charged particles \cite{dreicer1964kinetic,oxenius2012kinetic}}. 
A theory of plasma turbulence \magc{based upon the Vlasov-Maxwell-Landau equations alone} cannot 
directly answer the question of the ultimate fate of cascaded energy but it should provide the inputs 
(e.g. particle distribution functions at small scales) necessary to address that question.     

The present paper is intended to provide an exact, systematic framework for describing 
plasma turbulence at collisionless scales and should serve as a useful starting point for further  
investigations, not only theoretical but also numerical and experimental.  
Our analysis provides the foundation for numerical modelling of kinetic 
plasma turbulence by a ``large-eddy simulation'' (LES) methodology in phase-space
\cite{meneveau2000scale,schmidt2015large}. For experimentalists, our results provide 
a concrete model of ``resolution effects''. Our results show that finite-resolution measurements 
in a turbulent plasma can lead to substantial ``renormalizations'' of bare quantities 
that must be taken into account in interpreting observational data.  These are all important directions
to pursue in future work.  

\acknowledgements{I am grateful to the Princeton Plasma Physics Laboratory for its 
support during my visit in the fall of 2017, when most of the work on this paper was done. 
I also thank the PPPL theory group, and especially Amitava Bhattacharjee, Gregory Hammett,
John Krommes \black{and Hantao Ji}, for discussions during that visit. Finally, I wish to acknowledge 
Nicholas Besse, \black{Silvio Cerri}, Herbert Spohn, \black{Bogdan Teaca}, 
Ethan Vishniac, and Minping Wan for very useful conversations.}

\appendix

\vspace{.3in} 
\section{Derivation of Eq.(\ref{VII40}) in the Main Text}\lb{app:O}

\noindent 

We define a {\it phase-space density of fluctuation energy} at scales $\ell,$ $\uit$ as in 
eq.(\ref{VII34}) of the main text by 
\be \overline{z}_s(\obx,\obv,t):= \frac{1}{2}m_s|\obv-\tbus|^2\ofs(\obx,\obv,t) \lb{VII34a}\ee
so that $\overline{\epsilon}_s^*=\int d^3\ov\  \overline{z}_s.$ A phase-space balance 
equation for this quantity can be obtained by decomposing it as 
\bea 
&& \frac{1}{2}m_s|\obv-\tbus|^2\ofs\cr
&& =\left(\frac{1}{2}m_s|\obv|^2\ofs\right)-(m_s\obv\ofs)\bdot\tbus
+\left(\frac{1}{2}m_s|\tbus|^2\right)\ofs \cr
&&  \lb{VII35}\eea 
and then a lengthy but straightforward calculation using eq.(\ref{VII24}), eq.(\ref{VII6}), 
the equations 
\be \tilde{D}_{t,s} \left(m_s\tbus\right) +(1/\overline{n}_s)\grad_\obx\bdot\obP_s^*=q_s\tbE_{s*}  \lb{VII36}\ee 
\be \tilde{D}_{t,s}\left(\frac{1}{2}m_s|\tbus|^2\right) +(\tbus/\overline{n}_s)\bdot
\grad_\obx\bdot\obP_s^*=q_s \tbus\bdot \tbE_{s*}  \lb{VII37}\ee 
following from (\ref{VII2}), (\ref{VII9}) with $\tilde{D}_{t,s}:=\partial_t+\tbus\bdot\grad_\obx,$
and finally eq.(\ref{V9}) gives in the nearly collisionless limit 
\bea
&&  \partial_t\overline{z}_s +
\grad_\obx\bdot\left(\hbvs\overline{z}_s
+\obP_s^*\bdot(\tbus-\obv)\ofs/\overline{n}_s\right)+\grad_\obp\bdot\left(q_s {\hbE}_{*s}\overline{z}_s\right) \cr 
&& \hspace{20pt} =\obP_s^*\bdots \grad_\obx((\tbus-\obv)\ofs/\overline{n}_s) \cr 
&& \hspace{25pt} -(\bu_s-\hbvs)\bdot\grad_\obx\left(\frac{1}{2}m_s|\obv-\tbus|^2\right) \cdot \ofs\cr
&& \hspace{25pt} + q_s (\obv-\tbus)\bdot(\hbE_{*s}-\tbE_{*s})\ofs. \cr
&& \, 
 \lb{VII38}\eea
Noting that $(1/2)\grad_\obx|\obv-\tbus|^2=-\grad_\obx\tbus\bdot(\obv-\tbus),$ we may then 
rewrite the second term on the right by adding and subtracting a term proportional to 
$\obv\,\obv\ofs_{,\ell}$, giving  
\bea
&& (\bu_s-\hbvs)\bdot\grad_\obx\left(\frac{1}{2}|\obv-\tbus|^2\right) \cdot \ofs \cr 
&& \hspace{20pt} =  (\hbvs\obv\ofs-\obv\,\obv\ofs_{,\ell})\bdots\grad_\obx\tbus \cr
&& \hspace{25pt} +\Big(\obv\,\obv\ofs_{,\ell} -\tbus\obv\ofs-\hbvs\tbus\ofs +\tbus\tbus\ofs\Big)\bdots \grad_\obx\tbus. \cr
&& 
 \lb{VII39}\eea
Here the first contribution has a vanishing $\obv$-integral and thus represents a redistribution 
of fluctuational energy in velocity space, whereas the $\obv$-integral of the second term (and of the sum of the terms) 
is easily checked to give $\obP_s^*\bdots\grad_\obx\tbus.$ Substituting (\ref{VII39}) into 
(\ref{VII38}) we finally obtain the desired balance equation for $\overline{z}_s$: 
\bea
&&  \partial_t\overline{z}_s +
\grad_\obx\bdot\left(\hbvs\overline{z}_s
+\obP_s^*\bdot(\tbus-\obv)\ofs/\overline{n}_s\right)+\grad_\obp\bdot\left(q_s {\hbE}_{*s}\overline{z}_s\right) \cr 
&& \hspace{20pt} =\orhos\ttau(\bu_s,\bu_s)\bdots \grad_\obx((\tbus-\obv)\ofs/\overline{n}_s)\cr
&& \hspace{100pt} -m_s(\hbvs\obv\ofs-\obv\,\obv\ofs_{,\ell})\bdots\grad_\obx\tbus\cr 
&& \hspace{40pt} \mbox{(turbulent redistribution of energy)}\cr
&& \hspace{25pt}+\obP_s\bdots \grad_\obx((\tbus-\obv)\ofs/\overline{n}_s) \cr 
&& \hspace{40pt} \mbox{(energy redistribution by resolved pressure)}\cr
&& \hspace{25pt}-m_s\Big(\obv\,\obv\ofs_{,\ell} -\tbus\obv\ofs-\hbvs\tbus\ofs \cr
&& \hspace{100pt} +\tbus\tbus\ofs-\ttau(\bu_s,\bu_s)\ofs\Big)\bdots \grad_\obx\tbus \cr
&& \hspace{40pt} \mbox{(work by mean-velocity gradient)}\cr
&& \hspace{25pt} -m_s\ttau(\bu_s,\bu_s)\bdots\grad_\obx\tbus \ofs\cr
&& \hspace{40pt} \mbox{(energy input from turbulent cascade)}\cr
&& \hspace{25pt} +q_s(\obv-\tbus)\bdot(\hbE_{*s}-\tbE_{*s})\ofs \cr
&& \hspace{40pt} \mbox{(energy input \& redistribution by EM field)} \cr
&& \, 
 \lb{VII40a}\eea
 
\section{Bounds on Phase-Space Integrals}\lb{app:A} 

\subsection{The Integral in Estimate (\ref{V4})}\lb{app:A1} 

In the upper bound (\ref{V4}) on the coarse-grained collision integral, there appears the 
following integral over 2-particle phase-space: 
\be I=\int d^3r \int d^3v  \int d^3v'  \,
G_\ell(\br)|(\grad H)_\uit(\bv-\obv)|^2 \frac{f_sf_{s'}}{|\bv-\bv'|}. \lb{A1} \ee 
We shall show that this integral remains finite as $\Gamma\to 0$ under reasonable 
assumptions. First, we assume that 
\be n_s(\bx,t):=\int d^3v\, f_s(\bx,\bv,t)  <\infty \lb{A2} \ee  
for all $s,$ uniformly in $\Gamma,$ so that no infinite spatial densities appear in the collisionless
limit. Second, we assume that the distributions $f_s$ are locally square-integrable
for all species, so that  
\be \int_B d^3v\, f_s^2(\bx,\bv,t)  <\infty \lb{A3} \ee  
for all bounded open sets of velocities $B$ and for all $s,$ uniformly in $\Gamma.$ Note that 
the square-integrability of the distribution functions is generally assumed in theories 
of gyrokinetic turbulence, so that second-order structure functions and spectra are well-defined [23,24]. 
Square-integrability is also a natural assumption guaranteeing 
that the wave-particle term $\bE_* f_s$ in the Vlasov-Maxwell equation is pointwise 
well-defined [66]. 

Divide the integral $I$ into two contributions as $I=I_>+I_<,$ corresponding to the 
conditions $|\bv-\bv'|\geq 1$ and $|\bv-\bv'|\leq 1,$ respectively. Then
\bea
I_>:&=&\int d^3r  
 \stackrel[|\bv-\bv'|\geq 1]{}{ \int d^3v\int d^3v'}
G_\ell(\br)|(\grad H)_\uit(\bv-\obv)|^2 \frac{f_sf_{s'}}{|\bv-\bv'|} \cr
&\leq& \int d^3r  \int d^3v\int d^3v' \
G_\ell(\br)|(\grad H)_\uit(\bv-\obv)|^2 f_sf_{s'}\cr
&\leq & \max|(\grad H)_\uit|^2 \,\cdot\, \overline{n_s n_{s'}}(\obx,t)
\lb{A4} \eea
and is bounded uniformly in $\Gamma.$ On the other hand, applying Cauchy-Schwartz to $I_<$ 
gives
\bea
I_<:&=&\int d^3r  
 \stackrel[|\bv-\bv'|\leq 1]{}{ \int d^3v\int d^3v'}
G_\ell(\br)|(\grad H)_\uit(\bv-\obv)|^2 \frac{f_sf_{s'}}{|\bv-\bv'|} \cr
&\leq& \sqrt{
\int d^3r  
 \stackrel[|\bv-\bv'|\leq 1]{}{ \int d^3v\int d^3v'}
\frac{G_\ell(\br)|(\grad H)_\uit(\bv-\obv)|^2}{|\bv-\bv'|^2} }\cr 
&&\times
\sqrt{
\int d^3r  
 \stackrel[|\bv-\bv'|\leq 1]{}{ \int d^3v\int d^3v'}
G_\ell(\br)|(\grad H)_\uit(\bv-\obv)|^2 f_s^2 f_{s'}^2 }\cr
&& 
\lb{A5} \eea
The integral inside the first square-root is finite in 3D and defines a 
constant depending only upon $\ell,$ $\uit.$ The integral inside the second square 
root has the upper bound 
\bea
&& \int d^3r  
 \stackrel[|\bv-\bv'|\leq 1]{}{ \int d^3v\int d^3v'}
G_\ell(\br)|(\grad H)_\uit(\bv-\obv)|^2 f_s^2 f_{s'}^2 \cr
&&\leq \max|(\grad H)_\uit|^2 \int d^3r\, G_\ell(\br) \cr
&& \hspace{30pt} \times \left(\int_{B_\uit(\obv)} d^3v\ f_s^2\right)
\left(\int_{B_{\uit+1}(\obv)} d^3v'\ f_{s'}^2 \right)
\lb{A6} \eea
since the support of $|(\grad H)_\uit(\bv-\obv)|^2$ is contained inside the ball
$B_\uit(\obv)$ of radius $\uit$ around $\obv$ in velocity-space, 
with our assumptions on $H.$ We thus conclude that $I_<$ is also bounded 
uniformly in $\Gamma.$

\subsection{The Integral in Estimate (\ref{VII4})}\lb{app:A2} 

In the upper bound (\ref{VII4}) on the drag force $\bR_{ss'}$ there appears the 
following integral: 
\be J=\int d^3v  \int d^3v'  \, \frac{f_sf_{s'}}{|\bv-\bv'|}. \lb{A7} \ee 
We shall show that this integral remains finite as $\Gamma\to 0$ under reasonable 
assumptions, which include \eqref{A2} and a strengthening of \eqref{A3}, according to which
\be \int d^3v\, f_s(\bx,\bv,t) \int_{B_1(\bv)} d^3v'\, f_{s'}^2(\bx,\bv',t)  <\infty \lb{A8} \ee  
The proof again proceeds by dividing the integral $J$ into two contributions $J_<,$ $J_>$ 
corresponding to the conditions $|\bv-\bv'|\geq 1$ and $|\bv-\bv'|\leq 1.$ Clearly, 
\be
J_>:=
 \stackrel[|\bv-\bv'|\geq 1]{}{ \int d^3v\int d^3v'}
\frac{f_sf_{s'}}{|\bv-\bv'|} 
\leq n_s(\bx,t) n_{s'}(\bx,t). 
\lb{A9} \ee
On the other hand, 
\bea
J_<:&=& 
 \stackrel[|\bv-\bv'|\leq 1]{}{ \int d^3v\int d^3v'}\frac{f_sf_{s'}}{|\bv-\bv'|} \cr
&& \hspace{10pt} =\int d^3v\, f_s \int_{B_1(\bv)} d^3v'\, \frac{f_{s'}}{|\bv-\bv'|} \cr 
&& \hspace{10pt} \leq \sqrt{4\pi} 
\int d^3v \, f_s \sqrt{\int_{B_1(\bv)} d^3v' \, f_{s'}^2} \
\lb{A10} \eea
by applying Cauchy-Schwartz to the inner integral and by using $\int_{B_1(\bv)} \frac{d^3v'}{|\bv-\bv'|^2}=4\pi$
in 3D. Now apply Cauchy-Schwartz to the outer integral, giving 
\be J_<\leq \sqrt{4\pi\int d^3v\, f_s\times \int d^3v\, f_s \int_{B_1(\bv)} d^3v'\, f_{s'}^2}<\infty \lb{A11} \ee 
together with \eqref{A2} and \eqref{A8}.

\subsection{The Integral in Estimate (\ref{VII17})}\lb{app:A3}  

In the upper bound (\ref{VII17}) on the conversion term ${\mathcal R}_{ss'}$ there appears the 
following integral:
\be K=\frac{1}{4}\int d^3v  \int d^3v'  \, \frac{|\bv+\bv'|^2}{|\bv-\bv'|} f_sf_{s'}.  \lb{A12}\ee 
We show that this integral remains finite as $Do\to\infty$ under the conditions 
\eqref{A2},\eqref{A8}, and with also the further reasonable conditions 
\be K_s(\bx,t):=E_s(\bx,t)/m_s=\frac{1}{2}\int d^3v\, |\bv|^2 f_s(\bx,\bv,t)<\infty \lb{A13}\ee 
and
\be \int d^3v\, |\bv|^2 f_s(\bx,\bv,t) \int_{B_1(\bv)} d^3v'\, f_{s'}^2(\bx,\bv',t)  <\infty  \lb{A14}\ee  
As with the preceding integrals, we divide the integral $K$ into two contributions $K_<,$ $K_>$ 
corresponding to the conditions $|\bv-\bv'|\geq 1$ and $|\bv-\bv'|\leq 1$ and bound these 
two integrals separately. 

Using $|\bv+\bv'|^2\leq 2(|\bv|^2+|\bv'|^2),$ we get 
\bea
K_>:&=& \frac{1}{4}
 \stackrel[|\bv-\bv'|\geq 1]{}{ \int d^3v\int d^3v'}
\frac{|\bv+\bv'|^2}{|\bv-\bv'|} f_sf_{s'}\cr 
&\leq& \frac{1}{2}\int d^3v\int d^3v' \, (|\bv|^2+|\bv'|^2)\, f_sf_{s'}\cr 
&=& K_s n_{s'} + K_{s'} n_s \, <\, \infty. 
\lb{A15}\eea

On the other hand, for $|\bv-\bv'|\leq 1,$
\be |\bv+\bv'|^2=|2\bv+(\bv'-\bv)|^2\leq (2|\bv|+1)^2 \leq  2(4|\bv|^2+1) \lb{A16}\ee 
so that 
\bea
K_<:&=&\frac{1}{4}
 \stackrel[|\bv-\bv'|\leq 1]{}{ \int d^3v\int d^3v'}\frac{|\bv+\bv'|^2}{|\bv-\bv'|}f_sf_{s'} \cr
&=& \frac{1}{4}\int d^3v\, f_s \int_{B_1(\bv)} d^3v'\, \frac{|\bv+\bv'|^2}{|\bv-\bv'|}f_{s'} \cr 
&\leq & \frac{1}{2} \int d^3v\, (1+4|\bv|^2) f_s \int_{B_1(\bv)} d^3v'\, \frac{f_{s'}}{|\bv-\bv'|}\cr
&&
\lb{A17}\eea
In the same manner as for $J_<$ in \eqref{A10} and \eqref{A11},  
we apply Cauchy-Schwartz to the inner integral and then apply 
Cauchy-Schwartz to the outer integral, giving 
\bea
&& K_< \leq \sqrt{\pi} 
\int d^3v \, (1+4|\bv|^2) f_s \sqrt{\int_{B_1(\bv)} d^3v' \, f_{s'}^2} \cr
&& \leq \sqrt{\pi\int d^3v\, (1+4|\bv|^2) f_s} \cr
&& \times \sqrt{\int d^3v\, (1+4|\bv|^2) f_s \int_{B_1(\bv)} d^3v'\, f_{s'}^2}<\infty
\lb{A18}\eea
using \eqref{A2},\eqref{A8}, \eqref{A13},\eqref{A14}. 

\section{Estimating the Collisional-Cutoff or Dissipation Scales}\lb{app:C} 

The estimate (\ref{V5}) on the coarse-grained collision integral derived in the main text 
provides a means to estimate the ``cut-off scales" $\ell_c,$ $\uit_c$ in phase-space where 
particle collisions begin to compete with the turbulent renormalization effects due to 
ideal Vlasov-Maxwell dynamics. We here follow this approach to make more explicit determinations 
of such collisional-cutoff or dissipation scales. First, however, we shall review the derivation 
of similar viscous cut-offs in incompressible fluid turbulence, which suggests the approach to be 
followed also within kinetic turbulence. As we shall see, an improvement of the estimate (\ref{V5})   
is required and also an additional phenomenological assumption analogous to Kolmogorov's 
``refined similarity hypothesis'' in incompressible fluid turbulence. 

\subsection{Viscous-Cutoff Scale in Incompressible Fluid Turbulence}\lb{app:C1} 

In incompressible fluid turbulence, the role of the coarse-grained collision integral is played 
by the viscous diffusion term $\nu \triangle \overline{\bu}$ in the coarse-grained Navier-Stokes equation. 
See \cite{uriel1995turbulence}, eqs.(III.1,2) or \cite{eyink2010notes}, Chapter II(D). The Cauchy-Schwartz estimate analogous 
to (\ref{V4}) for the collision integral is eq.(III.3) in \cite{eyink2018review} or, in detail,  
\bea && \left|\nu\triangle\overline{\bu}(\obx,t)\right|\leq \frac{1}{\ell}
\sqrt{\nu\ {\rm vol(supp(}G_\ell))} \cr
&& \hspace{20pt} \times \sqrt{ \int d^3r\  |(\grad G)_\ell(\br)|^2\varepsilon(\obx+\br,t)}, \lb{C1} \eea 
where ${\rm vol(supp(}G_\ell))$ is the volume of the compact support of the scaled kernel $G_\ell.$ This 
volume is $C \ell^3$ for some $\ell$-independent constant $C,$ so that we may rewrite \eqref{C1} instead as 
\be \left|\nu\triangle\overline{\bu}(\obx,t)\right|\leq \frac{1}{\ell}
\sqrt{\nu\, C' \int d^3r\  \Phi_\ell(\br) \varepsilon(\obx+\br,t)}, \lb{C2} \ee 
with $\Phi:=|\grad G|^2/\int |\grad G|^2$ another $C^\infty,$ compactly-supported, unit-normalized test function, 
$\Phi_\ell(\br)=(1/\ell^3)\Phi(\br/\ell),$ and $C'=C \int |\grad G|^2$ is a new $\ell$-independent constant. The integral inside the square root in 
\eqref{C2} thus represents viscous dissipation (smoothly) averaged over a volume of order $\sim \ell^3$
and therefore can be estimated by an appeal to the ``Kolmogorov refined similarity hypothesis'' as  of order 
\be \overline{\varepsilon}_\ell(\obx,t):=\int d^3r\  \Phi_\ell(\br) \varepsilon(\obx+\br,t) \sim (\delta u(\ell))^3/\ell.  \lb{C3} \ee
with $\delta u(\ell):=\sup_{|\br|<\ell}|\delta_\br \bu(\bx,t)|.$ This hypothesis is unproved but has enjoyed considerable 
empirical succcess; see \cite{uriel1995turbulence}, section 8.6.2. We therefore obtain
\be \left|\nu\triangle\overline{\bu}(\obx,t)\right|\leq C'' 
\sqrt{\frac{\nu (\delta u(\ell))^3}{\ell^3}} \lb{C4} \ee 
with $C''$ a constant of order unity. The bound \eqref{C4} is, in general, a large over-estimate 
of $\nu \triangle \overline{\bu}$. A better estimate is provided by 
\be \left|\nu\triangle\overline{\bu}(\obx,t)\right|\sim  \nu \frac{\delta u(\ell)}{\ell^2} \lb{C5}, \ee
which is established as a rigorous upper bound in \cite{eyink2018review}, footnote [16] or 
\cite{eyink2010notes}, Chapter II(D), but which should also be a good order-of-magnitude estimate. 

Interestingly, however, the estimates 
\eqref{C4} and \eqref{C5} coincide when local Reynolds-number $Re_\ell:=\ell \delta u(\ell)/\nu\simeq 1,$ 
which is also the standard criterion used to identify the local viscous cut-off scale $\ell_\nu$ in incompressible fluid turbulence 
(see e.g. \cite{uriel1995turbulence}, section 8.5.5).  
This criterion can be rationalized by estimating the ``Reynolds-stress'' term $\grad\bdot \otau_\ell(\bu,\bu)$ that arises 
in the coarse-grained Navier-Stokes equation as a turbulent renormalization effect of unresolved eddies (see eq.(III.6) 
in \cite{eyink2018review}). 
A rigorous bound can be derived of the form 
\be |\grad \bdot \otau(\bu,\bu)| \leq C \frac{(\delta u(\ell))^2}{\ell}, \lb{C6} \ee  
using cumulant methods (e.g. see \cite{eyink2015turbulent}, Appendix B or the detailed derivation in 
\cite{eyink2010notes}, Chapter II(D)). 
The exact upper bound \eqref{C6} should also be a good order-of-magnitude estimate of $|\grad \bdot \otau(\bu,\bu)|$, unless 
there is a substantial depletion of nonlinearity. Equating $|\grad \bdot \otau(\bu,\bu)|\sim |\nu\triangle\overline{\bu}|$
to determine $\ell\sim\ell_\nu$ and using \eqref{C6} for $|\grad \bdot \otau(\bu,\bu)|$ and either \eqref{C4} or \eqref{C5} 
for $|\nu\triangle\overline{\bu}|,$ one finds that the condition $Re_\ell:=\ell \delta u(\ell)/\nu\simeq 1$ indeed provides the criterion 
for appearance of viscous effects locally in the coarse-grained equations. 

If there is a substantial depletion of nonlinearity, one 
may instead proceed by {\it defining} an ``eddy-turnover rate'' $\omega_\ell^{eddy}$ and a 
coarse-grained ``dissipation rate''  $\omega_\ell^{diss},$  at each length-scale $\ell,$  by the equations 
\be \omega^{eddy}_\ell \delta u(\ell) := |\grad \bdot \otau(\bu,\bu)|, 
\qquad \omega^{diss}_\ell(\delta u(\ell))^2 :=\overline{\varepsilon}_\ell. 
 \lb{C7} \ee
Depletion of nonlinearity means that $ \omega^{eddy}_\ell,$ $\omega^{diss}_\ell\ll \delta u(\ell)/\ell$.
Balancing $|\grad \bdot \otau(\bu,\bu)|$ with the sharp estimate \eqref{C5} of $|\nu\triangle\overline{\bu}|$ then yields 
\be \omega^{eddy}_\ell \simeq \nu/\ell^2, \lb{C8} \ee
i.e. turnover-rate $\sim$ viscous diffusion rate, as the criterion to determine $\ell\sim \ell_\nu.$ On the other hand, balancing 
$|\grad \bdot \otau(\bu,\bu)|$ with the looser estimate \eqref{C2} of $|\nu\triangle\overline{\bu}|$ gives 
\be \frac{(\omega^{eddy}_\ell)^2}{\omega^{diss}_\ell} \simeq \nu/\ell^2. \lb{C9} \ee 
As long as $\omega^{eddy}_\ell\sim \omega^{diss}_\ell,$ the two criteria \eqref{C8},\eqref{C9} will select the same $\ell\sim \ell_\nu.$ Empirical
evidence suggests that there is not a strong depletion of nonlinearity in incompressible fluid turbulence, so that 
$\omega^{eddy}_\ell\sim \omega^{diss}_\ell\sim \delta u(\ell)/\ell$ and the conditions \eqref{C8},\eqref{C9} then coincide 
with the naive criterion $Re_\ell:=\ell \delta u(\ell)/\nu\simeq 1.$

\subsection{Improved Estimation of Coarse-Grained Collision Integral}\lb{app:C2} 

In kinetic theory, on the other hand, there are well-known effects that may lead to depletion of nonlinearity, 
such as rapid wave oscillations, fast gyromotion of particles, dynamical alignment of vectors, etc. It would 
be desirable have a sharp estimate of the coarse-grained collision integral  analogous to \eqref{C5}, in order 
to obtain a criterion like \eqref{C8}, involving only the ideal small-eddy turnover time and consistent with 
depletion of nonlinearity. Unfortunately, the Landau collision integral has much greater complexity than the 
viscous diffusion term in hydrodynamic turbulence, so that it is not at all obvious how to derive an estimate 
of the coarse-grained collision integral similar to \eqref{C5}. An exact analogue of the hydrodynamic 
estimate \eqref{C4} can be derived, however, by a modest improvement of the estimate (\ref{V5}) in the main 
text and this result can be employed to derive a criterion analogous to \eqref{C9} for collisional cut-off scales 
$\ell_c$, $\uit_c$ in kinetic turbulence. 

To obtain the desired improvement of  (\ref{V5}), we make a slightly different  factorization of the integrand in (\ref{V3}), 
now moving the $(\grad H)_\uit$ into the second factor: 
\bea
&& \overline{C}_{ss'}(\obx,\obv,t) = -\frac{\Gamma_{ss'}}{m_s\uit} \int d^3r \int_{|\bv-\obv|<C\uit} d^3v  \int d^3v'  \cr 
&& \hspace{40pt} G_\ell^{1/2}(\br)\left(\frac{f_sf_{s'}}{|\bv-\bv'|}\right)^{1/2} \cr
&& \hspace{35pt} \bdot \ \frac{G_\ell^{1/2}(\br)(\grad H)_\uit(\bv-\obv) \bPi_{\bv-\bv'}}{(f_sf_{s'}|\bv-\bv'|)^{1/2}}
\left(\grad_\bp-\grad_{\bp'}\right)(f_sf_{s'})\cr
&& 
\lb{C10} \eea
We have assumed here that $H_\uit$ has compact support, contained inside a ball of radius $C\uit,$
so that the $\bv$-integration can be restricted to that ball centered around $\obv.$
We then apply the Cauchy-Schwartz inequality to obtain
\bea 
&& |\overline{C}_{ss'}(\obx,\obv,t)|\leq  \frac{\sqrt{\Gamma_{ss'}}}{m_s\uit} \cr 
&& \times \sqrt{\int d^3r \int_{|\bv-\obv|<C\uit} d^3v  \int d^3v'  \
G_\ell(\br) \frac{f_sf_{s'}}{|\bv-\bv'|}} \cr
&& \times \sqrt{ \begin{array}{l} 2 \int d^3r \int d^3v  \int d^3v'  \,
G_\ell(\br)|(\grad H)_\uit(\bv-\obv)|^2 \cr 
\hspace{60pt} \times \varsigma_{ss'}(\obx+\br,\bv,\bv',t) \cr
\end{array}}
\cr
&&
\lb{C11} \eea
where $\varsigma_{ss'}(\bx,\bv,\bv',t)$ is the entropy production rate in the 2-particle phase-space due 
to collisions of particle species $s,s',$ which is given by the corresponding term in the sum over $s,$ $s'$ 
in Eq.(\ref{III37}) in the main text that defines $\varsigma(\bx,\bv,\bv',t)$. Since the estimates 
\eqref{A4}-\eqref{A6} in section B.1 of this Supplement in fact depended only upon the compact support property 
of $H_\uit,$ they show for the first square-root factor essentially that 
\bea 
&& \int d^3r \int_{|\bv-\obv|<C\uit} d^3v  \int d^3v'  \
G_\ell(\br) \frac{f_sf_{s'}}{|\bv-\bv'|} \cr
&& \hspace{50pt} \leq C' \frac{\ofs(\obx,\obv,t) \overline{n}_{s'}(\obx,t)}{v_{th,ss'}}\uit^3 \lb{C12} \eea 
with $v_{th,ss'}:=\max\{v_{th,s},v_{th,s'}\}$ and with $C'$ a constant depending upon $(\obx,\obv,t)$ and 
$f_s,f_s',$ but not upon $\ell,$ $\uit.$ We leave details to the reader and note here only that $\ofs(\obx,\obv,t)$ and 
$\overline{n}_{s'}(\obx,t)$ represent in fact local r.m.s. values in the averages over $\br,$ $\bv,$ $\bv',$ which we 
for simplicity replaced with the usual coarse-grained values, assuming that they are of similar orders of magnitude. 
For the second square-root factor in \eqref{C11}, we write  
\be |(\grad H)_\uit(\bv-\obv)|^2 = \frac{1}{\uit^3} \Psi_\uit(\bv-\obv) \times \int |\grad H|^2 \lb{C13}\ee 
with $\Psi=|\grad H|^2\big/ \int |\grad H|^2$ another $C^\infty,$ compactly-supported, unit-normalized test function. Putting 
all of these estimates together, we obtain our final improvement of  (\ref{V5}), for some $\ell,$ $\uit$-independent constant $C''$:  
\be 
\overline{C}_{ss'}(\obx,\obv,t) \leq C'' \sqrt{\nu_{ss'}\,\overline{\varsigma}_{s,\ell,\uit}(\obx,\obv,t)\, \ofs(\obx,\obv,t)}\times \frac{v_{th,ss'}}{\uit}
\lb{C14} \ee
where $\nu_{ss'}:=\Gamma_{ss'} \bar{n}_{s'}/m_s^2 v_{th,ss'}^3$ is essentially the Spitzer-Harm Coulomb collision rate of particle 
species $s$ with particle species $s',$ and we have defined 
\bea 
&& \overline{\varsigma}_{s,\ell,\uit}(\obx,\obv,t):= \sum_{s'}\int d^3r \, G_\ell(\br) \int d^3w \, \Psi_\uit(\bw) \int d^3v'  \cr
&& \hspace{80pt} \times \varsigma_{ss'}(\obx+\br,\obv+\bw,\bv',t) \lb{C15}\eea
representing total collisional entropy production of species $s$ per phase-space volume, coarse-grained at scales 
$\ell,$ $\uit$ and evaluated at point $(\obx,\obv,t).$ 

The latter quantity may be used to define a {\it (coarse-grained) dissipation rate} 
$\omega^{diss}_{s,\ell,\uit}(\obx,\obv,t)$ for particle species $s$ by setting 
\be \omega^{diss}_{s,\ell,\uit}(\obx,\obv,t) \frac{(\delta f_s(\ell,\uit))^2}{\ofs}:=\overline{\varsigma}_{s,\ell,\uit}(\obx,\obv,t) 
\lb{C16}\ee
where 
\be (\Delta\, \sit)_{s,\ell,\uit}:=\overline{\sit(f_s)}-\sit(\overline{f}_s)\sim  \frac{(\delta f_s(\ell,\uit))^2}{\ofs} 
\lb{C17} \ee 
is a measure of the kinetic entropy of species $s$ residing at scales $\ell,$ $\uit$ in phase-space, with 
$\delta f_s(\ell,\uit)(\obx,\obv,t)=\sup_{|\br|<\ell,|\bw|<\uit}|\delta_{\br,\bw}f_s(\obx,\obv,t)|.$ For the estimate 
on the right side of \eqref{C17}, see \cite{eyink2017cascades1}, footnote [132]. In these terms, the bound \eqref{C14} may be rewritten as 
\be 
\overline{C}_{ss'}(\obx,\obv,t) \leq C'' \sqrt{\nu_{ss'}\,\omega^{diss}_{s,\ell,\uit}}\
\frac{v_{th,ss'}}{\uit}\ \delta f_s(\ell,\uit). 
\lb{C18}\ee
Another way to represent the estimate \eqref{C14} follows from a natural kinetic analogue of the Kolmogorov ``refined similarity 
hypothesis (RHS)'', according to which the coarse-grained entropy production rate $\overline{\varsigma}_{s,\ell,\uit}(\obx,\obv,t)$ 
should scale in the same manner as the phase-space entropy flux $\varsigma^{*flux,s}_{\ell,\uit}(\obx,\obv,t)$ given by (\ref{VI15}) 
in the main text, so that 
\bea 
\hspace{-20pt} 
\overline{\varsigma}_{s,\ell,\uit}(\obx,\obv,t)\sim \max\bigg\{&& \frac{\uit\,(\delta_\uit f_s) (\delta_\ell\! f_s)}{\ell f_s},  \cr
&& \frac{q_s(\delta_\ell E)(\delta_\ell f_{s})(\delta_\uit f_s)}{m_s\uit f_s}, \cr 
&& \frac{\ov\,q_s(\delta_\ell B)\,(\delta_\uit f_s)(\delta_\ell f_{s})}{c\,m_s \uit f_s} 
\bigg\}. \lb{C19}\eea
The three terms on the right side of \eqref{C19} arise from the estimates (\ref{VI23})-(\ref{VI25}) of entropy-flux 
contributions in the main text and we assume, naturally, that the largest contribution to flux dominates the scaling
(neglecting the fourth flux term from (\ref{VI26}) as always smaller). Assuming this kinetic RHS implies a corresponding 
estimate of the entropy cascade rate,  as an upper bound: 
\be \omega^{diss}_{s,\ell,\uit}(\obx,\obv,t) =O\left(\max\left\{\frac{\uit}{\ell}, \frac{q_s(\delta_\ell E)}{m_s\uit }, 
\frac{\ov\,q_s(\delta_\ell B) }{c\,m_s \uit} 
\right\}\right), \lb{C20} \ee
The true entropy flux rate (and, assuming the kinetic RHS, the coarse-grained dissipation rate) 
can be much smaller than this upper limit, if there is substantial depletion of nonlinearity. We therefore prefer to employ 
the general bound \eqref{C18}, without making use of the more specific estimate in \eqref{C20}. 

\subsection{Estimation of Turbulence-Generated Terms in Coarse-Grained Equations}\lb{app:C3} 

As emphasized in the main text, the Vlasov-Maxwell equations ``in the coarse-grained sense" (\ref{V9}) differ from the 
naive Vlasov-Maxwell equations, because  turbulent renormalization effects from unresolved eddies produce correction 
terms to the naive equations at each set of scales $\ell,$ $\uit$ in phase-space. The coarse-grained collision integral 
can be neglected at those scales $\ell,$ $\uit$ where it is much smaller than (the largest of) these turbulence-induced 
correction terms, and this is the defining characteristic of the ``collisionless range'' of scales. One can easily see from 
eqs.(\ref{V12})-(\ref{V15}) in the main text that the small-eddy contributions to the time-evolution of $\ofs$ have the form: 
 \bea  
 && (\partial_t\ofs)^{eddy}=\grad_\obx\bdot \left(\hbw_s\, \of_s\right)+  
q_s\grad_\obp\bdot\bigg(\otau(\bE,f_s) \cr 
&& \hspace{20pt} + \frac{1}{c}\obv\,\btimes\,\otau(\bB,f_s) + 
 \frac{1}{c}\widehat{(\bw\btimes\bB)}_s\, \of_s
 \bigg)\lb{C21} \eea
Simple expressions can be readily obtained for each of the four contributions, which permit their 
magnitudes to be estimated as follows:  
\bea
\grad_\obx\bdot(\hbw_s\,\of_s)  &=& -\frac{1}{\ell} \int d^3r\, (\grad G)_\ell(\br)\bdot 
\langle \bw(\delta_\br\delta_\bw f_s) \rangle_{\uit} \cr 
&=& O\left(\frac{\uit\,(\delta_\uit \delta_\ell\! f_s)}{\ell}\right)
\lb{C22} \eea
\be
q_s\grad_\obp\bdot \otau(\bE,f_s) = \frac{q_s}{m_s}
\otau_\ell(\bE;\grad_\obv\overline{f}_{s,\uit}) = O\left(\frac{q_s\delta_\ell E\,(\delta_\uit\delta_\ell \! f_{s})}{m_s \uit}\right)
\lb{C23} \ee 
\bea
q_s\grad_\obp\bdot\left(\frac{1}{c}\obv\,\btimes\,\otau(\bB,f_s)\right)
&=& \frac{q_s}{m_s c}\, \epsilon_{ijk}\, \overline{v}_i \, \otau_\ell(B_j, \partial_{\bar{v}_k}\overline{f}_{s,\uit}) \cr
&=& O\left(\frac{\ov}{c} \cdot \frac{q_s\delta_\ell B\,(\delta_\uit\delta_\ell \! f_{s})}{m_s \uit}\right) \cr 
&&\, 
\lb{C24} \eea

\vspace{-20pt} 
\noindent and 
\bea
q_s\grad_\obp\bdot\left( \frac{1}{c}\widehat{(\bw\btimes\bB)}_s\, \of_s\right)
&=& -\frac{q_s}{m_s \uit} \int d^3w \ (\grad H)_\uit(\bw) \cr 
&& \hspace{50pt} \bdot \frac{\bw}{c}\btimes \langle \bB\,\delta_\bw f_s \rangle_{\ell} \cr 
&= & O\left( \frac{q_s}{m_sc} \overline{B}\,\delta_\uit \! f_s\right)
\lb{C25} \eea
with the rightmost terms providing rigorous upper bounds. Note, as usual, that the fourth term
is negligible compared with the others (in fact, vanishing exactly when $H$ is radially symmetric)
and can be dropped. In these bounds we have introduced the following notation for the maximum 
double-increment (both in $\br$ and in $\bw$):  
\be \delta_\ell\delta_\uit f_s:=\sup_{|\br|<\ell,|\bw|<\uit}|\delta_\br\delta_\bw f_s| \sim 
\min\{\delta_\ell f_s,\delta_\uit f_s\}. \lb{C26} \ee 
The estimate in \eqref{C26} for the double-increment is also seen to be an exact upper bound, using the identities 
$\delta_\br\delta_\bw f_s(\bx,\bv)=\delta_\br f_s(\bx,\bv+\bw)-\delta_\br f_s(\bx,\bv)$
and likewise $\delta_\br\delta_\bw f_s=\delta_\bw f_s(\bx+\br,\bv)-\delta_\bw f_s(\bx,\bv),$
where the first identity is used if $f_s$ is smoother in $\bx$ and the second identity if $f_s$ is smoother in $\bv.$

An ``eddy-turnover rate'' $\omega^{eddy}_{s,\ell,\uit}(\obx,\obv,t)$ in phase-space is naturally defined by the equality 
\be \omega^{eddy}_{s,\ell,\uit}(\obx,\obv,t) \, \delta f_s(\ell,\uit) :=  (\partial_t\ofs)^{eddy}(\obx,\obv,t). \lb{C27} \ee
One can readily see from the estimates \eqref{C22}-\eqref{C25} that an upper bound follows 
\be \omega^{eddy}_{s,\ell,\uit}(\obx,\obv,t) =O\left(\max\left\{\frac{\uit}{\ell}, \frac{q_s(\delta_\ell E)}{m_s\uit }, 
\frac{\ov\,q_s(\delta_\ell B) }{c\,m_s \uit} 
\right\}\right), \lb{C28} \ee
of the same form as \eqref{C20} for $\omega^{diss}_{s,\ell,\uit}(\obx,\obv,t).$ When there is large depletion of nonlinearity, 
however, one can expect that $\omega^{eddy}_{s,\ell,\uit}(\obx,\obv,t)$ is much smaller in magnitude 
than the bound \eqref{C28}. We shall therefore not use the latter bound in our determination of collisional cut-off scales. 
Even when there is strong nonlinearity depletion, however, it is plausible to expect that $\omega^{eddy}_{s,\ell,\uit}(\obx,\obv,t) 
\sim \omega^{diss}_{s,\ell,\uit}(\obx,\obv,t),$ with similar magnitudes and identical scaling in $\ell,$ $\uit$. 
Despite the physical plausibility of these expectations, it is far from clear how to prove their validity.

\subsection{Criterion for Collisional-Cutoff Scales}\lb{app:C4} 

The collisionless range of scales for particle species $s$ is characterized by the condition that 
$|(\partial_t\ofs)^{eddy}|\gg |\overline{C}_{s}|$ and, likewise, cut-off scales $\ell_c,$ $\uit_c$ where collisions with particles 
of species $s'$ become important for species $s$ are specified by \\

\vspace{-20pt} 
\be |(\partial_t\ofs)^{eddy}(\obx,\obv,t)|\simeq |\overline{C}_{ss'}(\obx,\obv,t)|.  \lb{C29}\ee 
Note that \eqref{C29} is a pointwise condition in phase-space and thus the cut-off scales $\ell_c(\obx,\obv,t),$
$\uit_c(\obx,\obv,t)$ are local quantities, with fluctuating values reflecting phase-space intermittency of the entropy cascade. 
Employing the upper bound (\ref{C18}) as an estimate of $|\overline{C}_{ss'}(\obx,\obv,t)|$ and recalling 
the definition \eqref{C27} of $\omega^{eddy}_{s,\ell,\uit}(\obx,\obv,t)$ in terms of $(\partial_t\ofs)^{eddy}(\obx,\obv,t),$
the condition \eqref{C29} can be approximately rewritten as
\be \frac{\left(\omega^{eddy}_{s,\ell,\uit}\right)^2}{\omega^{diss}_{s,\ell,\uit}} \simeq \nu_{ss'} 
\left(\frac{v_{th,ss'}}{\uit}\right)^2. \lb{C30}\ee 
Because \eqref{C18} is only an upper bound on the coarse-grained collision integral, the true values of 
$\ell_c(\obx,\obv,t),$ $\uit_c(\obx,\obv,t)$ defined by \eqref{C29} could be smaller than those 
specified by \eqref{C30}. On the other hand, under the reasonable scaling hypothesis 
$ \omega^{eddy}_{s,\ell,\uit}(\obx,\obv,t) \sim \omega^{diss}_{s,\ell,\uit}(\obx,\obv,t),$
the condition \eqref{C30} reduces to 
\be \omega^{eddy}_{s,\ell,\uit}  \simeq \nu_{ss'} \left(\frac{v_{th,ss'}}{\uit}\right)^2 \lb{C31}\ee 
and thus essentially coincides with the heuristic criterion employed by Schekochihin et al. (2008,2009): 
see \cite{schekochihin2009astrophysical}, eq.(251) and \cite{schekochihin2008gyrokinetic}, section 2. 

Clearly the conditions \eqref{C29} or \eqref{C30} are satisfied for any fixed $\ell,$ $\uit$ in the formal limit 
$\nu_{ss'}\to 0$ (or $Do_{ss'}\to\infty$). To determine how large $Do_{ss'}$ must be in order 
to neglect collisions at specific values of $\ell,$ $\uit$ requires concrete scaling laws for 
$\omega^{eddy}_{s,\ell,\uit},$ $\omega^{diss}_{s,\ell,\uit}$ in terms of $\ell,$ $\uit,$ which 
will depend upon the circumstances (specific plasma parameters) and also, presumably, will 
fluctuate from point to point in phase space.  One important, general point is already clear, 
however, from the fact that \eqref{C29} or \eqref{C30} provide a single condition to determine {\it two}
free parameters $\ell,$ $\uit.$ There is obviously an undetermined degree of freedom, which may be 
taken to be the slope $\beta$ in the $\log(1/\ell)-\log(1/\uit)$ plane along which $\log(1/\ell),\log(1/\uit)\to \infty.$
In other words, one can impose as an {\it a priori} relation, with {\it any} choice of $\beta>0,$
\be \ell \sim \delta_s (\uit/v_{th,s,s'})^\beta,  \lb{C32} \ee 
where, for example, $\delta_s$ is the skin-depth for particle species $s,$ so that $\ell,$ $\uit$ vanish together.  
Substituting the relation \eqref{C32} into \eqref{C29} or \eqref{C30} then uniquely determines $\ell_c^{(\beta)},$ $\uit_c^{(\beta)}$
for that choice of $\beta>0.$ One should expect there to be a non-trivial $\beta$-dependence, since 
different choices of that free parameter will weight differently the linear advection contribution and the 
wave-particle interaction contribution to the rates $\omega^{eddy}_{s,\ell,\uit},$ $\omega^{diss}_{s,\ell,\uit}.$  

\section{The 4/5th-Law for Entropy Cascade in Kinetic Turbulence}\lb{app:D} 

In the main text we derived an explicit expression Eq.(\ref{VI15}) for entropy flux rate in phase space, or
\bea
&& \varsigma^{*flux,s}_{\ell,\uit}= - \hbw_s\bdot\grad_\obx\of_s 
-\frac{q_s}{m_s} \otau_\ell(\bE,\overline{f}_{s,\uit})\bdot \frac{\grad_\obv\of_s}{\of_s} \cr 
&& +\frac{q_s}{m_s c} \otau_\ell(\bB,\overline{f}_{s,\uit})\bdot \frac{(\obv\btimes\grad_\obv)\of_s}{\of_s}
-\frac{q_s}{m_s c} \widehat{(\bw\btimes\bB)}_s \bdot\grad_\obv\of_s. \cr 
&& \lb{D1} \eea 
As we remarked there, the four quantities that appear in this entropy flux can all be expressed in terms 
of phase-space increments of the VML solutions $f_s,$ $s=1,...,S$ and $\bE,$ $\bB$ and these 
formulae provide the kinetic theory analogues of ``4/5th laws" for entropy cascade. The 
concrete connection with turbulent ``4/5th-laws" is not needed to derive the scaling exponent constraints 
Eq.(\ref{VI45}) in the main text, but we discuss such relations here for their general interest. 

We have explained in the main text how to write the entropy flux in terms of increments, by means of the 
general relation eq.(\ref{IV7}) for the correlation terms $\otau_\ell(\bE,\overline{f}_{s,\uit}),$ $\otau_\ell(\bB,\overline{f}_{s,\uit})$,  
the identities eq.(\ref{V11}), eq.(\ref{V15}) for $\hbw_s,$ $\widehat{(\bw\btimes\bB)}_s,$ and the equations  
eq.(\ref{IV8})-eq.(\ref{IV9}) for the gradients $\grad_\obx\ofs,$ $\grad_\obv\ofs.$ We shall now  
write down the explicit expressions in terms of increments, adopting notations that are purposely chosen to make 
the connection with traditional ``4/5th laws" more obvious. Taking $\bz=(\br,\bw)$ to denote a displacement 
in  six-dimensional phase-space, we use the notation 
\be \langle a_\bz \rangle_\bz := \int d^6z\,  G_\ell(\br) H_\uit(\bw) \, a_{\bz} \lb{D2} \ee
to indicate the average over $\bz$ with respect to the kernels $G_\ell(\br) H_\uit(\bw).$ Here $a_\bz$ is any 
quantity depending upon $\bz,$ possibly through $\br$ or $\bw$ alone. One can then easily check using the 
aforementioned equations in the main text that 
\be
\hbw_s\bdot\grad_\obx\of_s  = \left\langle \grad_{\br''}\bdot\left[\bw\,\frac{(\delta_\bw f_s)\, (\delta_{\br''}f_s)}{\ofs}\right]\right\rangle_{\bz,\bz''}
\lb{D3} \ee
\bea 
&& \otau(\bE,\overline{f}_{s,\uit})\bdot \frac{\grad_\obv\ofs}{\ofs} = 
\left\langle \grad_{\bw''}\bdot\left[\frac{(\delta_\br\bE)\,(\delta_\br f_s)\, (\delta_{\bw''}f_s)}{\ofs}\right]\right\rangle_{\bz,\bz''} \cr
&& \hspace{40pt} - \, \left\langle \grad_{\bw''}\bdot\left[\frac{(\delta_\br\bE)\,(\delta_{\br'} f_s)\, (\delta_{\bw''}f_s)}{\ofs}\right]\right\rangle_{\bz,\bz',\bz''}
\lb{D4} \eea
%\newpage 
\bea
&& -\otau_\ell(\bB,\overline{f}_{s,\uit})\bdot \frac{(\obv\btimes\grad_\obv)\of_s}{\of_s} \cr 
&&  \hspace{15pt} = \left\langle \grad_{\bw''}\bdot\left[ \obv\btimes\frac{(\delta_\br\bB)\,(\delta_\br f_s)\, (\delta_{\bw''}f_s)}{\ofs}\right]\right\rangle_{\bz,\bz''}\cr
&&  \hspace{15pt} - \,  \left\langle \grad_{\bw''}\bdot\left[\obv\btimes\frac{(\delta_\br\bB)\,(\delta_{\br'} f_s)\, (\delta_{\bw''}f_s)}{\ofs}\right]\right\rangle_{\bz,\bz',\bz''}\lb{D5} \eea 
\bea
&& \widehat{(\bw\btimes\bB)}_s \bdot\grad_\obv\of_s \cr 
&& = \left\langle \grad_{\bw''}\bdot\left[\bw\btimes\bB(\obx+\br)\,\frac{(\delta_\bw f_s)\, (\delta_{\bw''}f_s)}{\ofs}\right]\right\rangle_{\bz,\bz''} 
\lb{D6} \eea
with multiple averages over $\bz=(\br,\bw),$ $\bz'=(\br',\bw'),$ etc. indicated by corresponding multiple subscripts. These formulas  
may be compared with standard expressions for (anisotropic) 4/5th-laws both in incompressible fluid turbulence, such as 
\cite{uriel1995turbulence}, Eq.(6.8), and in gyrokinetic turbulence, such as \cite{plunk2009theory}, Eq.(4.52) or 
\cite{plunk2010two}, Eq.(6.9).  The resemblance is quite clear for the 
two middle contributions \eqref{D4}-\eqref{D5} from nonlinear wave-particle interactions, which are cubic in terms of solutions fields.
The other two, \eqref{D3} from linear advection and the last term \eqref{D6}, have a similar form but are only quadratic in solution fields.   
In the main text we in fact sketched the derivation of two different versions of the ``kinetic 4/5th law'' based upon the above formulas: 
Eq.(\ref{VI10}), which is an ``ensemble version'' (or globally spatially-averaged version) analogous to that of Kolmogorov, and 
Eq.(\ref{VI14}), which is a ``deterministic, local version'' like that of Duchon-Robert \cite{duchon2000inertial}. We shall further elaborate on both of these here. 

The derivation of the ``ensemble version'' Eq.(\ref{VI10}) mostly follows standard arguments for the fluid case,
except for the one important difference that there is no ``statistical homogeneity'' in velocity-space for kinetic turbulence.
On the other hand, the total integrals over velocity-space can be presumed to exist, so that one can instead  
{\it integrate} over $\obv$ rather than average.  Integrating the phase-space entropy balance Eq.(\ref{VI18}) 
over velocity then gives a physical-space entropy balance 
\be \partial_t s[\ofs]+ \grad\bdot \bJ_{S}^{*res,s} = \sigma^{*flux,s}_{\ell,\uit} \lb{D7} \ee 
with $\sigma^{*flux,s}_{\ell,\uit}$ the term corresponding to species $s$ in the sum of Eq.(\ref{VI21}). Whereas 
the four terms in Eq.\eqref{D1} all represent entropy production rate per unit phase-space volume and per unit time,
the corresponding terms in $\sigma^{*flux,s}_{\ell,\uit}$ give entropy production rates per unit physical-space volume 
and per unit time. We may now average over space or, assuming statistical homogeneity, average over an ensemble 
of solutions and obtain Eq.(\ref{V10}) by the arguments in the main text.  The resulting ``4/5th-law'' for kinetic entropy cascade
written out in full detail is:
\bea 
&& \langle\sigma_\star\rangle =\left\langle \grad_{\br''}\bdot\left[\bw\,\frac{(\delta_\bw f_s)\, (\delta_{\br''}f_s)}{\ofs}\right]\right\rangle_{\ell,\uit} \cr 
&& \hspace{10pt} + \, \frac{q_s}{m_s}\left\langle \grad_{\bw''}\bdot\left[\frac{(\delta_\br\bE)\,(\delta_\br f_s)\, (\delta_{\bw''}f_s)}{\ofs}\right]\right\rangle_{\ell,\uit} \cr 
&& \hspace{10pt}- \, \frac{q_s}{m_s}\left\langle \grad_{\bw''}\bdot\left[\frac{(\delta_\br\bE)\,(\delta_{\br'} f_s)\, (\delta_{\bw''}f_s)}{\ofs}\right]\right\rangle_{\ell,\uit} \cr 
&& \hspace{10pt}+\, \frac{q_s}{m_sc}\left\langle \grad_{\bw''}\bdot\left[ \obv\btimes\frac{(\delta_\br\bB)\,(\delta_\br f_s)\, (\delta_{\bw''}f_s)}{\ofs}\right]\right\rangle_{\ell,\uit}\cr 
&&  \hspace{10pt}- \,  \frac{q_s}{m_sc}\left\langle \grad_{\bw''}\bdot\left[\obv\btimes\frac{(\delta_\br\bB)\,(\delta_{\br'} f_s)\, (\delta_{\bw''}f_s)}{\ofs}\right]\right\rangle_{\ell,\uit}\cr
&& \hspace{10pt}+\,\frac{q_s}{m_sc} \left\langle \grad_{\bw''}\bdot\left[\bw\btimes\bB(\obx+\br)\,\frac{(\delta_\bw f_s)\, (\delta_{\bw''}f_s)}{\ofs}\right]\right\rangle_{\ell,\uit} \cr
&&
\lb{D8} \eea
valid in the ``collisionless range'' $L\gg\ell\gg \ell_c,$ $U\gg \uit\gg \uit_c,$ where $\left\langle\cdot^{\!\,}\right\rangle_{\ell,\uit}$ means 
that all increments $\bz,$ $\bz',$ etc. which appear inside the bracket have been independently averaged with respect to $G_\ell H_\uit$, 
that $\obv$ has been integrated over all of velocity-space, and that $\obx$ has been averaged over all of physical-space. 

The local, deterministic form of this 4/5th Law in Eq.(\ref{VI14}) can be likewise derived following the arguments 
of \cite{duchon2000inertial}, which are briefly sketched in the main text. The result has exactly the same form as \eqref{D8} except 
that, on both sides of the equation, averages of $\obx$ over all of space are replaced with averages of $(\obx,t)$ over 
$\varphi(\obx,t)$ for a smooth, compactly supported, normalized function $\varphi.$ The condition on $\uit$ for validity 
of this local relation is unchanged, but the condition on $\ell$ becomes $L_\varphi(t)\gg \ell\gg \ell_c,$ where $L_\varphi(t)$ 
is the spatial dimension of the support of $\varphi(\cdot,t),$ which must be held fixed as first $Do\to\infty$ and then $\ell,\uit\to 0.$
Although this relation is ``space-time local in the sense of distributions'',  the spatial average here is over many increment lengths 
$\ell$ (which in turn must be much larger than $\ell_c$). In particular, the result does not help to justify a ``refined similarity 
hypothesis'' of the type \eqref{C19}, which involves on the lefthand side an average of $\varsigma$ over a region of extent $\ell$ 
in space and $\uit$ in velocity.    

A discontented reader might wonder why the kinetic 4/5th-law that we present does not make use 
of the simple ``point-splitting'' argument employed in standard derivations for incompressible fluids \cite{uriel1995turbulence}
or for gyrokinetics \cite{plunk2009theory,plunk2010two}. The problem is that there is no obvious ``point-splitting'' of the phase-space entropy 
density $\sit[f_s]=-f_s \log f_s$ for which one can show as $Do\to\infty $ that:  (i) all terms in the point-split relation 
remain finite even as $\grad_\bx$-gradients and $\grad_\bv$-gradients diverge, and (ii) the contribution of the 
Landau collision integral can furthermore be neglected. Standard verifications of (i) use essentially the fact that 
kinetic energy per volume $(1/2)|\bu|^2$ for incompressible fluids and free-energy per phase-volume $g^2/2F_0$
for gyrokinetics (see \cite{plunk2009theory}, eq.(4.16)) are quadratic in the solution fields.  Careful derivations of the analogue of (ii) 
for incompressible fluids (e.g. \cite{uriel1995turbulence}, eq.(6.47), or \cite{eyink2010notes}, p.5, Ch.II.B) 
use the simple form $\nu \triangle\bu$ of viscous diffusion.  None of these standard arguments obviously carries over to the 
entropy density and the Vlasov-Maxwell-Landau kinetic equations, whereas our coarse-graining regularization  
(\ref{IV1}) in the main text trivially guarantees (i) and has been shown also to yield (ii). It is worth remarking that the standard 
point-splitting argument does guarantee (i) for the quadratic quantity $(1/2)f_s^2,$ which is an ideal invariant for smooth 
Vlasov-Maxwell solutions. On the other hand, this quadratic quantity satisfies no $H$-theorem for the VML equations 
and it is also not obvious how to justify (ii) for a point-splitting of this quantity. 

\bibliography{bibliography.bib}

\end{document}